\documentclass[11pt,a4paper]{article}
\def\AmSTeX{\leavevmode\hbox{$\mathcal A\kern-.2em\lower.376ex%
        \hbox{$\mathcal M$}\kern-.2em\mathcal S$-\TeX}}
\usepackage{graphicx}
\usepackage[numbers,sort&compress]{natbib}
\usepackage{amsmath}
\usepackage{amssymb}
\usepackage{subfigure}
\usepackage{mathrsfs}
\usepackage{bm}
\usepackage[amsmath,thmmarks,hyperref]{ntheorem}
\ifx\pdfoutput\undefined \pdffalse \fi \ifx\pdfoutput\relax
\pdffalse \fi
\ifx\texonly\undefined\let\texonly\relax\fi
\ifx\endtexonly\undefined\let\endtexonly\relax\fi \texonly
  \let\htmlonly\iffalse
  \let\endhtmlonly\fi
\endtexonly

\htmlonly
  
\endhtmlonly
\texonly
  \usepackage{makeidx}

  \usepackage{multicol}
  

\endtexonly
\title{}
\author{\thanks{}}
\date{}
\begin{document}

\title{The Study of Rare $B_c\rightarrow D^{(*)}_{s,d}l\bar{l}$~Decays}

\author{Wan-Li~Ju\footnote{wl\_ju\_hit@163.com}~$^1$, Guo-Li Wang\footnote{gl\_wang@hit.edu.cn}~$^1$, Hui-Feng Fu\footnote{huifeng\_fu@163.com}~$^2$, Tian-Hong Wang\footnote{thwang.hit@gmail.com}~$^1$ and Yue Jiang\footnote{jiangure@hit.edu.cn}~$^1$\\
{\it \small  $^1$Department of Physics, Harbin Institute of
Technology,
Harbin, 150001, China}\\
{ \small $^2$\it Department of Physics, Tsinghua University, Beijing, 100084, China} }


\maketitle

\baselineskip=20pt
\begin{abstract}

In this paper, we study rare decays $B_c\rightarrow
D^{(*)}_{s,d}l\bar{l}$ within the Standard Model. The penguin, box,
annihilation, color-favored cascade and color-suppressed cascade
contributions are included. Based on our calculation, the
annihilation and color-favored cascade diagrams play important
roles in the differential branching fractions, forward-backward
asymmetries, longitudinal polarizations of the final vector mesons and
leptonic longitudinal polarization asymmetries. More importantly,
color-favored cascade decays largely enhance the resonance cascade
contributions. To avoid the resonance cascade
contribution pollution, new cutting regions are put forward.

\end{abstract}

\section{Introduction}

 The $B_c$ meson is the ground state of the bottom-charm bound system. Its first observation was
at the Fermilab Tevatron by the CDF Collaboration in 1998 through
the cascade decay $B_c\rightarrow J/\psi\bar{l}\nu$ and
$J/\psi\rightarrow \mu\bar{\mu}$~\cite{CDF}.

Since its discovery, the $B_c$ meson has
attracted more and more attention. First, because of
explicit flavors and the relationship $M(B_c)<M(B)+M(D)$, the strong and electromagnetic decay channels are forbidden, while the weak
decay channels are allowed. Thus, the $B_c$ decay offers us an ideal field for studying the weak interaction. Second, the $B_c$ meson consists of two heavy quarks $c$ and
$b$. They can decay independently, or both of them participate in one
process. Thus, compared with $B$ and $D$ mesons, the $B_c$ meson owns more
decay channels and a larger final phase space. Third, the
mass spectra of $\bar cc$ and
$\bar bb$ bound states have been extensively studied, both theoretically and experimentally. The
double-heavy explicit flavored meson spectrum will offer a
new circumstance to research the low energy QCD and examine the
previous comprehension of the meson structure. In the aspect of experiment, the Large Hadron Collider~(LHC) has circulated the
proton beams since 2009. The number of the produced $B_c$ mesons is
expected to be $5\times 10^{10}$~\cite{123,Ho and Ji} per year when
it runs at the highest luminosity $10^{34}$ cm$^{-2}$s$^{-1}$ and
highest energy $14$ TeV. So most $B_c$ decay channels will be
observed with this huge number. It is believed that the
predictions about the $B_c$ meson in theory will be checked by the
experiment data in the near future.

Among the channels of the $B_c$ meson transitions, the rare
decays $b\rightarrow s(d)l\bar{l}$ and $b\rightarrow s(d)\gamma$
have been emphasized recently. These decays are the single-quark flavor-changing neutral current~(FCNC) processes.
Based on the Glashow-Iliopoulos-Maiani~(GIM)~\cite{GIM} mechanism in the Standard Model (SM), they are forbidden at tree level but induced by electro-weak
loop diagrams. It implies that the SM contribution is greatly
suppressed. And, as the SM contribution is suppressed, the new physics (NP) effects may become important. So studying the rare decays will be helpful to
test the SM stringently and detect the NP effects indirectly. Among these rare decay channels, the semi-leptonic processes $b\rightarrow
s(d)l\bar{l}$ are favored because of their abundant observables, for instance, the differential branching fraction, the forward-backward asymmetry, the longitudinal polarization of the final
vector meson, the leptonic longitudinal polarization asymmetry and so
on. More information on the Wilson coefficients of the rare decays can be extracted from these observables.

In existing studies, the $b\rightarrow sl\bar{l}$ processes have
been discussed widely through the $B\rightarrow K^{(*)}l\bar{l}$
decays. In the $B\rightarrow K^{(*)}l\bar{l}$ processes, the
$b\rightarrow sl\bar{l}$ contribution is dominant, while the
annihilation diagrams are CKM suppressed
$|V^*_{ub}V_{us}|/|V^*_{ts}V_{tb}|\sim \lambda^2$ and the spectator
scattering effect is at the next $\alpha_s$ order. As shown
in~\cite{D.Ebert,LHCbBK1,LHCbBK2}, by including only the
$b\rightarrow sl\bar{l}$ contribution in the SM, the differential
branching fractions, the forward-backward asymmetries and the
longitudinal polarizations of the final vector mesons are in agreement
with the experimental data. However, in $B_c\rightarrow
D^{(*)}_{s,d}l\bar{l}$ rare decays, the situation is different. The
annihilation diagrams in $B_c\rightarrow
D^{(*)}_{s,d}l\bar{l}$ rare decays are CKM allowed
$|V^*_{cb}V_{cs(d)}|/|V^*_{ts(d)}V_{tb}|\sim 1$ and enhanced by a 3
times larger color factor. Thus, the annihilation contributions
should be considered seriously. In this way, the study of the $B_c\rightarrow
D^{(*)}_{s,d}l\bar{l}$ rare decays provides outstanding features.
First, the NP effects will be more distinct by considering the
annihilation processes. As pointed out in~\cite{pakistan2,pakistan3},
with the help of the interference between the $b\rightarrow
s(d)l\bar{l}$ contribution and the SM annihilation contribution, the
predictions of SM with a fourth generation~(SM4) and the Super-Symmetric~(SUSY) Models will deviate stronger from the SM
predictions in the processes $B_c\rightarrow D^{*}_{s}l\bar{l}$ than
in the processes $B\rightarrow K^{*} l\bar{l}$. Furthermore, in
$B_c$ rare decays, the particles beyond SM may contribute not
only to the $b\rightarrow sl\bar{l}$ transitions but also to the
annihilation diagrams. If so, this double-contribution mechanism may
make the NP effect more evident. Second, the processes
$B_c\rightarrow D_{s,d}l\bar{l}$ can help us comprehend the large
isospin asymmetry phenomenon in the $B\rightarrow K \mu \bar{\mu}$
decay. Recently, LHCb \cite{isospinexp}~has reported that the
isospin asymmetry~$A_I$ of the $B\rightarrow K^{*} \mu\bar{\mu}$
process is in agreement with the naive SM expectation of a vanishing
asymmetry. However, in the $B\rightarrow K \mu\bar{\mu}$ process,
there is a $4\sigma$ deviation from zero for $A_I$. Even if
calculations~\cite{isospin1,isospin2,isospin3} including the
annihilation and spectator-scattering diagrams are carried out, the
isospin asymmetry of the process $B\rightarrow K \mu\bar{\mu}$ is
also less than $0.05$. One may infer that studying the spectator
effects should be emphasized. In the decays $B_c\rightarrow
D_{s,d}l\bar{l}$, spectator effects will be obvious based on the
particular annihilation mechanism. So the rare $B_c$
decays can offer a new and helpful field to reveal the reason of the
large isospin asymmetry in the $B\rightarrow K \mu \bar{\mu}$ decay.

In previous
works~\cite{D.Ebert,AFaessler,Gengchaoqiang,Ho}, which include the $b\rightarrow s(d)l\bar{l}$ short-distance contribution
and the $b\rightarrow \bar{c}cs(d)\rightarrow s(d)\bar{l}l$
long-distance diagrams, the $B_c\rightarrow D_{s,d}l\bar{l}$ and
$B_c\rightarrow D^{*}_{s,d}l\bar{l}$ processes~have been calculated, while in Refs.~\cite{TengWang,KAzizi1,KAzizi3} only the $b\rightarrow s(d)l\bar{l}$ short-distance contribution is considered. Recently,
using the annihilation form-factors of $B_c\rightarrow
D^{*}_{s}\gamma$~\cite{KAzizigamma,HYCheng,DDU}, the $B_c\rightarrow
D^{*}_{s}l\bar{l}$ transitions combined with the annihilation effects
have been analysed in Refs.~\cite{pakistan1,pakistan2,pakistan3}. Actually, the
relationship $(P_i-P_f)^2=0~\textmd{GeV}^2$ holds only for the
$B_c\rightarrow D^{*}_{s,d}\gamma$ processes. If the $B_c\rightarrow
D^{(*)}_{s,d}l\bar{l}$ annihilation diagrams are considered, more
form-factors will appear. In the aspect of
the long-distance interaction,
the papers~\cite{D.Ebert,AFaessler,Gengchaoqiang,Ho,pakistan2}
have calculated the color-suppressed cascade diagrams. The
color-suppressed cascade long-distance contribution dominates the $B\rightarrow K^{(*)}J/\psi(\psi(2S))\rightarrow K^{(*)}\bar{l}l$~processes. However, both
color-suppressed and color-favored long-distance diagrams contribute
to $B_c\rightarrow D_{s(d)}^{(*)}J/\psi(\psi(2S))\rightarrow D_{s(d)}
^{(*)}\bar{l}l$ processes. Furthermore, according to our pervious calculations
on the non-leptonic $B_c\rightarrow D_{s(d)}^{(*)}J/\psi$ decays~\cite{huifengfu}, the color-favored
contributions are almost 3 times larger than the color-suppressed ones.

So besides including the $b\rightarrow s(d)l\bar{l}$ effect, it is
motivated to investigate the $B_c\rightarrow D^{(*)}_{s,d}l\bar{l}$ decays:~(i)~using the annihilation form-factors of the $B_c\rightarrow
D^{(*)}_{s,d}l\bar{l}$ decays instead of the $B_c\rightarrow
D^{*}_{s,d}\gamma$ processes;~(ii)~with both the color-suppressed and
color-favored long-distance contributions.

To investigate the $B_c\rightarrow D^{(*)}_{s,d}l\bar{l}$ decays, the Operator Production Expansion~(OPE) mechanism and
factorization ansatz can be employed. In this method, the
amplitude can be separated into two parts: the short-distance
Wilson coefficients~(If the decays involve the resonance cascade processes, extra long-distance terms will appear.) and the long-distance
hadronic matrix elements which are operators sandwiched by the initial and
the final states. In the framework of Re-normalization Group
Equations~(RGE), the Wilson coefficients can be obtained at next-to-leading~(NL)~order or next-to-next-to-leading~(NNL)~order QCD
corrections~\cite{WILLSON,twoloop} perturbatively. But calculating the hadronic matrix elements is a
non-perturbative problem and a model dependent method has to be
chosen to do it.

Till now, the hadronic matrix elements have been investigated in several
approaches: the relativistic constituent quark model
\cite{AFaessler,Gengchaoqiang}, the light-cone quark model
\cite{Gengchaoqiang,Ho,TengWang}, the three point QCD sum rules
\cite{KAzizi1,KAzizi3} and the QCD-motivated relativistic
quark model \cite{D.Ebert}. In this paper, we choose the
Mandelstam Formalism~(MF) \cite{S. Mandelstam} and the
Bethe-Salpeter (BS) equation \cite{E1,E2} to investigate the hadronic
matrix elements. This method has several particular features. First,
the BS equation, based on the quantum field theory, is a relativistic
equation to describe a two-body bound state. Using this method, the Gaussian wave
function is abandoned and a relativistic form of the wave function
has been introduced \cite{guoli wang,guoliwang1-,guoliwang0-}. In
this way, the wave function has the same parity and charge parity as
the bound state. Different forms of wave functions denote
different states, e.g., the wave function of a pseudoscalar meson
is much different from the one of a vector meson. Only in the non-relativistic limit, they are
similar \cite{chao-hsi chang}. Second, in the decays
$B_c\rightarrow D^{(*)}_{s,d} l\bar{l}$, the mass of initial meson
$B_c$ is much larger than that of the final meson $D^{(*)}_{s,d}$. As a
result, the final meson recoil can be large and the relativistic
effect should be taken seriously. Based on the Mandelstam
Formalism, our method keeps the relativistic effect not only
from the relativistic wave functions but also from the
kinematics~\cite{chao-hsi chang1}. At last, the weak
annihilation amplitude involving the heavy-light meson $B$ can
be calculated within the Heavy Quark Effective
Theory~(HQET)~\cite{HQET1,HQET2} under the $m_{u(d)}/m_b\sim0$
limit. However, unlike the $B$ and $D$ meson, the $B_c$ meson consists of two
heavy quarks and the relationship $m_c/m_b\sim m_{u(d)}/m_b\sim0$ is not suitable. So in this paper, we calculate the annihilation
hadronic matrix elements based on the BS method. In our
calculation, dynamics of both heavy quarks is considered. And our
method can be used to deal with not only the double-heavy quark meson but
also the heavy-light meson. Besides, the
weak-annihilation hadronic currents satisfy the gauge-invariance
at the $(P_i-P_f)^2=0~\textmd{GeV}^2$.

This paper is organized as follows. In section 2, the theoretical
details of the effective hamiltonian, hadronic transition matrix elements and the observables of the $B_c\rightarrow D^{(*)}_{s,d}l\bar{l}$ processes are given. In section 3, the
numerical results and discussions are presented. In section 4, we
summarize this paper.

\section{Theoretical Details}
\subsection{Effective Hamiltonian}

In this paper, we calculate the $B_c\rightarrow
D^{(*)}_{s,d}l\bar{l}$ decays including the short-distance~(SD) and
long-distance~(LD)~contributions in the SM. The SD diagrams contain
the Penguin, Box~(PB) diagrams and annihilation~(Ann) graphs as shown in~Figs.~1~(a,~b,~c).
The LD parts include the
color-suppressed~(CS) and the color-favored~(CF) contributions as
shown in Figs.~1~(d,~e). We have the effective Hamiltonian for
annihilation, color-suppressed cascade and color-favored cascade
diagrams:
\begin{equation}
\begin{split}
\mathcal{H}^{eff}_{1}=\frac{G_F}{\sqrt{2}}V_{cb}V^*_{cs(d)}\left(C_1(\mu)Q_1(\mu)+C_2(\mu)Q_2(\mu)\right),\\
\end{split}
\end{equation}
where $G_F$ is the Fermi constant and $C_1,~C_2$ are willson
coefficients. In this paper, when the annihilation, color-suppressed and color-favored cascade
diagrams are calculated, the effects of $Q_{3-6}$ are neglected because of their small Wilson coefficients.

The penguin and box diagrams' effective
Hamiltonian, that is, the $b\rightarrow s(d)\bar{l}l$ effective
Hamiltonian, is given by:
\begin{equation}
\begin{split}
\mathcal{H}^{eff}_{2}=\frac{G_F}{\sqrt{2}}\left\{-V_{tb}V^*_{ts(d)}\underset {i = 1} {\overset {10} {\Sigma}}C_i(\mu)Q_i(\mu)\right\},\\
\end{split}
\end{equation}
where $V_{cb},~V_{cs(d)},~V_{tb}$ and~$V_{ts(d)}$ are the CKM
matrix elements. The set of local operators is \cite{AFaessler,WILLSON}:
\begin{equation}
\begin{split}
Q_1&=(\bar{s}_ic_j)_{V-A}(\bar{c}_jb_i)_{V-A},~~~~~~~~~~~~~~~~~~~~~Q_2=(\bar{s}c)_{V-A}(\bar{c}b)_{V-A},\\
Q_3&=(\bar{s}b)_{V-A}\sum_q(\bar{q}q)_{V-A},~~~~~~~~~~~~~~~~~~~~Q_4=(\bar{s}_ib_j)_{V-A}\sum_q(\bar{q}_jq_i)_{V-A},\\
Q_5&=(\bar{s}b)_{V-A}\sum_q(\bar{q}q)_{V+A},~~~~~~~~~~~~~~~~~~~~Q_6=(\bar{s}_ib_j)_{V-A}\sum_q(\bar{q}_jq_i)_{V+A},\\
Q_7&=\frac{e}{8\pi^2}m_b\left(\bar{s}\sigma^{\mu\nu}(1+\gamma_5)b\right)F_{\mu\nu},~~~~~~~~~~~Q_8=\frac{g}{8\pi^2}m_b\left(\bar{s}_i\sigma^{\mu\nu}(1+\gamma_5)T_{ij}b_j\right)G_{\mu\nu},\\
Q_9&=\frac{e^2}{8\pi^2}(\bar{s}b)_{V-A}(\bar{l}l)_{V},~~~~~~~~~~~~~~~~~~~~~~~~Q_{10}=\frac{e^2}{8\pi^2}(\bar{s}b)_{V-A}(\bar{l}l)_{A}.
\end{split}
\end{equation}
For $b\rightarrow s(d)\bar{l}l$ processes, with the help of
RGE~\cite{AJBuras} approach, $C_{7,9}(m_b)$ are linear combinations
of according $C_{7,9}(M_W)$ and $C_{1-6}(M_W)$. The operators
$Q_{1-6}$ themselves have matrix elements that contribute to $b \to
sl\bar{l}$ and the perturbative parts can be included into $C^{eff}_{
7,9}$. In our calculation, considering that the effect of $Q_8$ belongs to the higher $\alpha_s$ order, we do not include its contribution. Thus, the amplitude induced by effective Hamiltonian is shown as:
\begin{equation}
\mathcal{M}_{b\rightarrow
s(d)l^+l^-}=i\frac{G_{F}\alpha_{em}}{2\sqrt{2}\pi}V_{tb}V_{ts(d)}^{*}\left\{\left[C^{eff}_{9}W_{\mu}-\frac{2m_{b}}{(P_i-P_f)^2}C^{eff}_{7}W_{\mu}^{T}\right]\bar{l}\gamma^{\mu}l+C_{10}W_{\mu}\bar{l}\gamma^{\mu}\gamma_5l\right\}.
\end{equation}
In the above equation, we have defined the hadronic matrix elements $W_{\mu}$ and $W_{\mu}^{T}$:
$$W_{\mu}=\langle D^{(*)}_{s(d)}|
\bar{s}(\bar{d})\gamma_{\mu}(1-\gamma_{5})b| B_c\rangle ,
~~~~~W_{\mu}^{T}=\langle
D^{(*)}_{s(d)}|\bar{s}(\bar{d})i\sigma_{\mu\nu
}(P_i-P_f)^{\nu}(1+\gamma_5) b| B_c\rangle, $$ where $P_i$ and $P_f$ are momenta of the initial and final mesons, respectively. The
antisymmetric tensor is defined as $\sigma_{\mu\nu }=\frac{i}{2}[
\gamma_{\mu},\gamma_{\nu}]$ and $m_b$ is the mass of $b$ quark.

The effective Wilson
coefficients $C^{eff}_7$ and $C^{eff}_9$ are defined as:
\begin{equation}
\begin{split}
C^{eff}_7=&C_7-C_5/3-C_6,\\
C_9^{eff}=&C_9+(3C_1+C_2+3C_3+C_4+3C_5+C_6) h(\hat{m_c},s)\\
-&\frac{1}{2}h(1,s)(4C_3+4C_4+3C_5+C_6)-\frac{1}{2}h(0,s)(C_3+3C_4)\\
+&\frac{2}{9}(3C_3+C_4+3C_5+C_6), \\
\end{split}
\label{eq3}
\end{equation}
where $h(\hat{m_c},s)$ ($\hat{m_c}=m_c/m_b$), $h(1,s)$ and $h(0,s)$
describe the contributions of the four-fermi operator $Q_{1-6}$
loops. In Refs~\cite{AFaessler,AJBuras,MMisiak}, with the help of RGE, the Wilson coefficients $C_i~(i=1-10)$ are evolved
from $M_W$ scale to the $m_b$ scale. In this
paper, we adopt the same expressions of $h$ functions and the same numerical values of
the Wilson
coefficients as those in Ref.~\cite{AFaessler}.

For the annihilation diagrams, according to the naive factorization, we write the amplitude as:
\begin{equation}
\begin{split}
\mathcal{M}_{Ann}=&-iV_{cb}V^*_{cs(d)}\langle D^{(*)}_{s(d)}\bar{l}l|\mathcal{J}_{em}\mathcal{H}_1|B_c\rangle\\
=&V_{cb}V^{*}_{cs(d)}\frac{ i
\alpha_{em}}{(P_i-P_f)^2}\frac{G_{F}}{2\sqrt{2}\pi}\left(\frac{C_1}{N_c}+C_2\right)W_{ann}^{\mu}\bar{l}\gamma_{\mu}l,
\end{split}
\end{equation}
where in the second equality the Fierz-arrangement identity is employed. The annihilation hadronic matrix elements can be expressed as:$$W^{\mu}_{ann}=W^{\mu}_1+W^{\mu}_2+W^{\mu}_3+W^{\mu}_4.$$
In the above equation, $W^{\mu}_{1,2,3,4}$ are defined as:
\begin{equation}
\begin{split}
W_1^{\mu}=&(-8\pi^2)\langle D^{(*)}_{s(d)}|\bar{s}(\bar{d})
\gamma_{\alpha}(1-\gamma_5)c|0\rangle\langle0|\bar{c}\gamma^{\alpha}(1-\gamma_5)
\frac{1}{\not\!{p}_{q_1}-m_{q_1}+i \epsilon}(-\frac{1}{3})\gamma^{\mu}b| B_c\rangle,\\
W_2^{\mu}=&(-8\pi^2)\langle D^{(*)}_{s(d)}|\bar{s}(\bar{d})
\gamma_{\alpha}(1-\gamma_5)c|0\rangle\langle0|\bar{c}(\frac{2}{3})\gamma^{\mu}
\frac{1}{\not\!{p}_{q_2}-m_{q_2}+i \epsilon}\gamma^{\alpha}(1-\gamma_5)b| B_c\rangle,\\
W_3^{\mu}=&(-8\pi^2)\langle D^{(*)}_{s(d)}|\bar{s}(\bar{d}) (-\frac{1}{3})\gamma^{\mu}\frac{1}{\not\!{p}_{q_4}-m_{q_4}+i  \epsilon}\gamma_{\alpha}(1-\gamma_5)c|0\rangle\langle0|\bar{c}\gamma^{\alpha}(1-\gamma_5)b| B_c\rangle,\\
W_4^{\mu}=&(-8\pi^2)\langle D^{(*)}_{s(d)}|\bar{s}(\bar{d})\gamma_{\alpha}(1-\gamma_5)\frac{1}{\not\!{p}_{q_3}-m_{q_3}+i  \epsilon}(\frac{2}{3})\gamma^{\mu}c|0\rangle\langle0|\bar{c}\gamma^{\alpha}(1-\gamma_5)b| B_c\rangle,\\
\end{split}
\end{equation}
where $p_{q_{1(2-4)}}$ and $m_{q_{1(2-4)}}$ are momenta and
masses of the propagated quarks, respectively.

 The LD processes considered in this paper are those that are induced by the resonance cascade decays, i.e, $B_c\rightarrow D^{(*)}_{s(d)}V\rightarrow D^{(*)}_{s(d)}\bar{l}l$, whose contributions can be described by the relationship $Br(B_c\rightarrow D^{(*)}_{s(d)}\bar{l}l)_{cascade}\sim Br(B_c\rightarrow D^{(*)}_{s(d)}V)\times Br(V\rightarrow\bar{l}l)$ approximately. (The relationship written here is not used to compute any observables in this paper but just employed to make the following sentences transparent.) The resonances $V$ denote $J^{PC}=1^{--}$ mesons which could be the $\bar{u}u$, $\bar{d}d$, $\bar{s}s$ and $\bar{c}c$ bound states. 
 During our calculation, we ignore the effects of the $B_c\rightarrow D^{(*)}_{s(d)}\rho(\omega,\phi)\rightarrow D^{(*)}_{s(d)}\bar{l}l$ cascade decays. On one hand, the strong decays allowed by Okubo-Zweig-Iizuka (OZI) rules are the dominant channels of the $\rho$, $\omega$ and $\phi$ mesons, while the strong decays of $J/\psi(\psi(2S))$ are suppressed by OZI rules. So the processes $\rho(\omega,\phi)\rightarrow l^{+}l^{-}$ which are induced by electromagnetic interaction have much smaller branching fractions than $J/\psi(\psi(2S))\rightarrow l^{+}l^{-}$. On the other hand, considering the small CKM matrix elements $V_{ub}$ and $V_{us}$, the branching fractions of $B_c\rightarrow D^{(*)}_{s}\rho(\omega)$ processes will be suppressed. And because of the small Wilson coefficients $C_{3-6}$, the numerical values of $Br(B_c\rightarrow D^{(*)}_{s,d}\phi)$ are small. So we calculate only the $B_c\rightarrow D^{(*)}_{s(d)}J/\psi(\psi(2S))\rightarrow D^{(*)}_{s(d)}\bar{l}l$ processes in this paper.

 As pointed out in Refs.~\cite{RESchaohsi,RESBKgamma}, the propagator of $V$ can be written in the Breit-Wigner form. So, from the naive factorization, the amplitude can be written as:
\begin{equation}
\begin{split}
\mathcal{M}_{LD}=&\underset{V}{\Sigma}\langle \bar{l}l|\mathcal{J}^{(2)}_{em}|V\rangle\frac{i}{Q^2-M_V^2+i\Gamma_VM_V}\langle D^{(*)}_{s(d)}V|(-i)\mathcal{H}_1|B_c\rangle e^{i\varphi},\\
\end{split}
\end{equation}
where $\mathcal{J}^{(2)}_{em}\equiv
\bar{l}\gamma_{\mu}l(-i4\pi\alpha_{em}/M_V^2)\bar{c}\frac{2}{3}\gamma^{\mu}c$. $M_V$ and $\Gamma_V$ are the mass and width of the resonance
meson, respectively. From Fierz-arrangement identity and naive factorization, we write the hadronic matrix element
$\langle D^{(*)}_{s(d)}V|\mathcal{H}_1|B_c\rangle $ as:
\begin{equation}
\begin{split}
\langle D^{(*)}_{s(d)}V|\mathcal{H}_1|B_c\rangle
=&V_{cb}V^*_{cs(d)}\frac{G_F}{\sqrt{2}}\left(C_1+\frac{C_2}{N_c}\right)\langle
D^{(*)}_{s(d)}|\bar{s}(\bar{d})\gamma^{\nu}(1-\gamma_5)b|B_c\rangle\langle
V|\bar{c}\gamma_{\nu}(1-
\gamma_5)c|0\rangle  \\
+&V_{cb}V^*_{cs(d)}\frac{G_F}{\sqrt{2}}\left(C_2+\frac{C_1}{N_c}\right)\langle
V|\bar{c}\gamma^{\nu}(1-\gamma_5)b|B_c\rangle\langle
D^{(*)}_{s(d)}|\bar{s}(\bar{d})\gamma_{\nu}(1- \gamma_5)c|0\rangle,
\end{split}
\end{equation}
where on the right side of the above equation, the first term is from the CS diagram and the
second term gives the CF contribution. If only the
CS and $b\rightarrow s(d)l\bar{l}$ contributions are included, our amplitude
$\mathcal{M}_{b\rightarrow s(d)l^+l^-}+\mathcal{M}_{LD}$ will be
similar with those in Refs.~\cite{D.Ebert,AFaessler,CSlim2,NGDeshpande,PJODonnell}.
Taking account of the unitarity condition on the elastic Breit-Wigner
resonance amplitude~\cite{PJODonnell}, we take $\varphi=0$ in this paper.

To simplify the calculation, we define:
\begin{equation}
\begin{split}
\mathcal{M}^{CS}_{LD}\equiv&\underset{V}{\Sigma}\langle \bar{l}l|\mathcal{J}^{(2)}_{em}|V\rangle\frac{i}{Q^2-M_V^2+i\Gamma_VM_V}V_{cb}V^*_{cs(d)}\frac{G_F}{\sqrt{2}}\left(C_1+\frac{C_2}{N_c}\right)\\
&\langle
D^{(*)}_{s(d)}|\bar{s}(\bar{d})\gamma^{\nu}(1-\gamma_5)b|B_c\rangle\langle
V|\bar{c}\gamma_{\nu}(1-
\gamma_5)c|0\rangle \\
=&i\frac{G_{F}\alpha_{em}}{2\sqrt{2}\pi}V_{tb}V_{ts(d)}^{*}C_9^{CS}W_{\mu}\bar{l}\gamma^{\mu}l ,\\
\end{split}
\end{equation}
where in the second step the relationship $\Gamma(V\rightarrow\bar{l}l)=\pi\alpha^2_{em}16f^2_{V}/(27M_V)$ is used. And the
$C_9^{CS}$ is given by:$$C_9^{CS}\equiv\frac{V_{cb}V_{cs(d)}^{*}}{V_{tb}V_{ts(d)}^{*}}\frac{9\pi}{\alpha^2_{em}}\left(C_1+\frac{C_2}{N_c}\right)\underset{V}{\sum}\frac{\Gamma(V\rightarrow\bar{l}l)M_V}{Q^2-M_V^2+i\Gamma_VM_V}.$$

Similarly, we also define:
\begin{equation}
\begin{split}
\mathcal{M}^{CF}_{LD}\equiv&\underset{V}{\Sigma}\langle \bar{l}l|\mathcal{J}^{(2)}_{em}|V\rangle\frac{i}{Q^2-M_V^2+i\Gamma_VM_V}V_{cb}V^*_{cs(d)}\frac{G_F}{\sqrt{2}}\left(C_2+\frac{C_1}{N_c}\right)\\
&\langle
V|\bar{c}\gamma^{\nu}(1-\gamma_5)b|B_c\rangle\langle
D^{(*)}_{s(d)}|\bar{s}(\bar{d})\gamma_{\nu}(1- \gamma_5)c|0\rangle\\
\equiv&V_{cb}V^{*}_{cs(d)} i
\alpha_{em}\frac{G_{F}}{2\sqrt{2}\pi}\left(C_2+\frac{C_1}{N_c}\right)W_{CF}^{\mu}\bar{l}\gamma_{\mu}l.
\end{split}
\end{equation}

By including the BP, Ann, CF cascade and CS cascade diagrams, the amplitude calculated in this paper can be written as:
$$\mathcal{M}_{B_c\rightarrow D^{(*)}_{s(d)}\bar{l}l}=\mathcal{M}_{b\rightarrow
s(d)l\bar{l}}+\mathcal{M}_{LD}+\mathcal{M}_{Ann}.$$

Considering the Lorentz and parity symmetries, $W_{\mu}$ and
$W_{\mu}^T$ could be written in terms of the form factors:
\begin{equation}
\begin{split}
\langle D_{s(d)}(P_f)|\bar{s}\bar{(d)}\gamma^{\mu}(1-\gamma_{5})b| B_c(P_i)\rangle=&F_z \left(P_+^{\mu}-\frac{P_+\cdot Q}{Q^2}Q^{\mu}\right)+F_0 \frac{P_+\cdot Q}{Q^2}Q^{\mu},\\
\langle D_{s(d)}(P_f)|\bar{s}\bar{(d)}i\sigma^{\mu\nu}Q_{\nu}(1+\gamma_{5})b|B_c(P_i)\rangle=&\frac{-F_T }{M_i+M_f}\left\{Q^2P_+^{\mu}-(P_+\cdot Q)Q^{\mu}\right\},\\
\langle D^{*}_{s(d)}(P_f,\epsilon_{f})|\bar{s}\bar{(d)}\gamma^{\mu}(1-\gamma_{5})b| B_c(P_i)\rangle=&\frac{iV }{M_i+M_f}\epsilon^{\mu\epsilon_{f} Q P_+}-2M_fA_0 \frac{\epsilon_f \cdot Q}{Q^2}Q^{\mu}\\
&-(M_i+M_f)A_1 \left(\epsilon_f^{\mu}-\frac{\epsilon_f\cdot Q}{Q^2}Q^{\mu}\right)\\
+&A_2 \frac{\epsilon_f\cdot Q}{M_i+M_f}\left\{P_+^{\mu}-\frac{P_+\cdot Q}{Q^2}Q^{\mu}\right\},\\
\langle D^{*}_{s(d)}(P_f,\epsilon_{f})|\bar{s}\bar{(d)}i\sigma^{\mu\nu}Q_{\nu}(1+\gamma_{5})b| B_c(P_i)\rangle=&-iT_1 \epsilon^{\mu\epsilon_f Q P_+}\\ &+T_2 \left\{P_+\cdot Q \epsilon_f^{\mu}-(\epsilon_f \cdot Q)P_+^{\mu}\right\}\\
&+T_3 \left(\epsilon_f\cdot Q\right)\left\{Q^{\mu}-\frac{Q^2}{P_+\cdot Q}P_+^{\mu}\right\}.\\
\end{split}
\end{equation}
The form-factors of annihilation and color-favored diagrams are defined as:
\begin{equation}
\begin{split}
W^{\mu}_{ann}(B_c\rightarrow D_{s,d})=&(M_i-M_f)^2\left\{\frac{1}{2} B_1  \left(P_+^{\mu }-Q^{\mu }\right)+\frac{1}{2} B_2  \left(Q^{\mu }+P_+^{\mu }\right)\right\},\\
W^{\mu}_{ann}(B_c\rightarrow D^{*}_{s,d})=&(M_i-M_f)\left\{T_{1ann}  ~M_i^2\epsilon_f^{\mu  }+\frac{1}{2} T_{2ann} \left(Q\cdot \epsilon_f\right) \left(Q^{\mu }+P_+^{\mu }\right)\right.\\
+&\left.\frac{1}{2} T_{3ann} \left(Q\cdot\epsilon_f\right) \left(P_+^{\mu }-Q^{\mu }\right)
+\frac{1}{2} i V_{ann} ~ \epsilon ^{\mu \epsilon_f
QP_+}\right\},\\
W^{\mu}_{CF}(B_c\rightarrow D_{s,d})=&(M_i-M_f)^2\left\{\frac{1}{2} B_{1CF}  \left(P_+^{\mu }-Q^{\mu }\right)+\frac{1}{2} B_{2CF}  \left(Q^{\mu }+P_+^{\mu }\right)\right\},\\
W^{\mu}_{CF}(B_c\rightarrow D^{*}_{s,d})=&(M_i-M_f)\left\{T_{1CF}  M_i^2\epsilon_f^{\mu  }+\frac{1}{2} T_{2CF}  \left(Q\cdot \epsilon_f\right) \left(Q^{\mu }+P_+^{\mu }\right)\right.\\+
&\left.\frac{1}{2} T_{3CF}  \left(Q\cdot\epsilon_f\right) \left(P_+^{\mu }-Q^{\mu }\right)
+\frac{1}{2} i V_{CF} ~ \epsilon ^{\mu \epsilon_f
QP_+}\right\},
\end{split}
\end{equation}
where $P_+=P_i+P_f$ and $Q=P_i-P_f$ have been used. $F_{z}$,
$F_0$, $F_T$, $V$, $A_1$, $A_2$, $A_0$, $T_1$, $T_2$, $T_3$, $B_1$, $B_2$, $B_{1CF}$, $B_{2CF}$, $T_{1ann}$, $T_{2ann}$, $T_{3ann}$, $V_{ann}$, $T_{1CF}$, $T_{2CF}$, $T_{3CF}$ and $V_{CF}$ are the
form factors.

\subsection{Hadronic Transition Matrix Elements in the Bethe-Salpeter Method}

In this subsection, we show the details of how to calculate the
hadronic matrix elements in Eqs.~(12,~13). Within the Bethe-Salpeter method, a meson is considered as a bound state of the quark and the anti-quark.
For a meson with mass $M$, momentum $P$ and relative momentum
$q$, we can define:
$$q=\alpha_2p_1-\alpha_1p_2, ~~~~~P=p_1+p_2, ~~~~~\alpha_1=m_1/(m_1+m_2),
~~~~~\alpha_2=m_2/(m_1+m_2),$$ where $p_1$ ($p_2$) and $m_1$ ($m_2$)
are the momentum and mass of the constituent quark(anti-quark), respectively. To
describe the bound state, the wave
function is essential. In this paper, we
obtain it by solving the BS equation:
$$(\not\!{p_1}-m_1)\chi(q)(\not\!{p_2}+m_2)=i\int\frac{d^4k}{(2\pi)^4}V(P,k,q)\chi(k),$$
where $\chi(q)$ is the BS wave function and $V(P,k,q)$ is the
interaction kernel.

To solve the equation, we need to reduce it to its instantaneous version, the Salpeter equation \cite{guoli wang}. Then we need a proper interaction kernel and the forms of the wave functions. In our method, the Cornell potential, which is one of the most successful and widely used potentials in describing heavy mesons, is adopted as the interaction kernel.

The behavior of a bound state is determined by its $J^P$
quantum number, so we give the form of the wave function based on the meson's $J^P$ quantum number. As a result, the wave functions of the different
mesons fulfill different BS equations~\cite{chao-hsi
chang,guoli wang}. For a pseudoscalar meson, $J^P=0^-$, the positive energy wave function can be written as \cite{guoli wang}:
\begin{equation}
\varphi^{++}_{0^-}(P,q_{_{P_{\bot}}})=a_1\left[a_2+\frac{\not\!P}{M}+\not\!q_{_{P_{\bot}}}a_3+\frac{\not\!q_{_{P_{\bot}}}\not\!P-\not\!P\not\!q_{_{P_{\bot}}}}{2M}a_4\right]\gamma_5.
\end{equation}

One can check that the wave function has the correct $J^P$ quantum
number. In Eq.~(14), the parameters are
defined as:
\begin{equation}
\begin{split}
a_1&=\frac{M}{2}\left\{\varphi_1(q^2_{_{P_{\bot}}})+\varphi_2(q^2_{_{P_{\bot}}})\frac{m_1+m_2}{\omega_1+\omega_2}\right\},\\
a_2&=\frac{\omega_1+\omega_2}{m_1+m_2},\\
a_3&=-\frac{m_1-m_2}{m_1\omega_2+m_2\omega_1},\\
a_4&=\frac{\omega_1+\omega_2}{m_1\omega_2+m_2\omega_1},\\
\end{split}
\end{equation}
where $\omega_i$ is defined as
$\omega_i\equiv\sqrt{(m^2_i-q^2_{_{P_{\bot}}})}$ ($i=1(2)$ denotes
the (anti-)quark. For $B_c^-$ meson, $i=1$ denotes the $b$
quark and $i=2$ denotes the $\bar{c}$ quark. For $D_{s(d)}^-$ meson, $i=1$ denotes the $s(d)$ quark and $i=2$ denotes the $\bar{c}$
quark.). We also use the definition $q^{\mu}_{_{P_{\bot}}}\equiv q^{\mu}-(P\cdot q/M^2)P^{\mu}$ where
$P$ and $M$ are the momentum and mass of the bound
state, respectively. In the Center-of-Mass-System~(CMS) of the meson,
$q^{\mu}_{_{P_{\bot}}}=(0,\vec{q})$ is obvious. Using instantaneous approximation, the numerical
values of $\varphi_1(q^2_{_{P_{\bot}}})$ and
$\varphi_2(q^2_{_{P_{\bot}}})$ are obtained by solving the full
Salpeter equations \cite{guoli wang}.

For vector mesons, $J^P=1^-$, the positive energy wave function can be
written as \cite{guoliwang1-}:
\begin{equation}
\begin{split}
\varphi^{++}_{1^-}(P,q_{_{P_{\bot}}},\epsilon)=&b_1\not\!\epsilon +b_2\not\!\epsilon \not\!P+b_3(\not\!q_{_{P_{\bot}}}\not\!\epsilon -q_{_{P_{\bot}}}\cdot
\epsilon )
+b_4(\not\!P\not\!\epsilon \not\!q_{_{P_{\bot}}}-\\
&\not\!P q_{_{P_{\bot}}}\cdot
\epsilon )+q_{_{P_{\bot}}}\cdot
\epsilon [b_5+b_6\not\!P+b_7\not\!q_{_{P_{\bot}}}+
\frac{b_8}{2}(\not\!P\not\!q_{_{P_{\bot}}}-\not\!q_{_{P_{\bot}}}\not\!P)],
\end{split}
\end{equation}
where the coefficients are shown as:
\begin{equation}
\begin{split}
b_1=&\frac{M}{2}\left(\varphi_5-\varphi_6\frac{\omega_1+\omega_2}{m_1+m_2}\right),\\
b_2=&-\frac{m_1+m_2}{2(\omega_1+\omega_2)}\left(\varphi_5-\varphi_6\frac{\omega_1+\omega_2}{m_1+m_2}\right),\\
b_3=&\frac{M(m_2\omega_1-m_1\omega_2)}{-2q_{_{P_{\bot}}}^2(\omega_1+\omega_2)}\left(\varphi_5-\varphi_6\frac{\omega_1+\omega_2}{m_1+m_2}\right),\\
b_4=&\frac{\omega_1+\omega_2}{2(\omega_1\omega_2+m_1m_2-q_{_{P_{\bot}}}^2)}\left(\varphi_5-\varphi_6\frac{\omega_1+\omega_2}{m_1+m_2}\right),\\
b_5=&\frac{m_1+m_2}{2M(\omega_1\omega_2+m_1m_2+q_{_{P_{\bot}}}^2)}\left[M^2\left(\varphi_5-\varphi_6\frac{m_1+m_2}{\omega_1+\omega_2}\right)+q_{_{P_{\bot}}}^2\left(\varphi_3+\varphi_4\frac{m_1+m_2}{\omega_1+\omega_2}\right)\right],\\
b_6=&\frac{\omega_1-\omega_2}{2M^2(\omega_1\omega_2+m_1m_2+q_{_{P_{\bot}}}^2)}\left[M^2\left(\varphi_5-\varphi_6\frac{m_1+m_2}{\omega_1+\omega_2}\right)+q_{_{P_{\bot}}}^2\left(\varphi_3+\varphi_4\frac{m_1+m_2}{\omega_1+\omega_2}\right)\right],\\
b_7=&\frac{1}{2M}\left(\varphi_3+\varphi_4\frac{m_1+m_2}{\omega_1+\omega_2}\right)-\varphi_6\frac{M}{m_1\omega_2+m_2\omega_1},\\
b_8=&\frac{1}{2M^2}\frac{\omega_1+\omega_2}{m_1+m_2}\left[\left(\varphi_3+\varphi_4\frac{m_1+m_2}{\omega_1+\omega_2}\right)-\varphi_5\frac{2M^2}{\omega_1\omega_2+m_1m_2-q_{_{P_{\bot}}}^2}\right].
\end{split}
\end{equation}
In Eqs.~(16,~17), the same definitions of $P$, $M$, $\omega_i$, $m_i$
and $q_{_{P_{\bot}}}$ as for the pseudo-scalar are used. For $J/\psi$
and $\psi(2S)$ mesons, $i=1$ denotes the $c$ quark and $i=2$ denotes
the $\bar{c}$ quark. For $D_{s(d)}^{*-}$, $i=1$ denotes
$s(d)$ quark and $i=2$ denotes $\bar{c}$ quark. The numerical
values of wave functions $\varphi_3(q_{_{P_{\bot}}})$,
$\varphi_4(q_{_{P_{\bot}}})$, $\varphi_5(q_{_{P_{\bot}}})$ and
$\varphi_6(q_{_{P_{\bot}}})$ can be obtained by solving the full
Salpeter equations for a vector meson. And the details can be found in Ref. \cite{guoliwang1-}.

Using the MF \cite{S. Mandelstam}, the hadronic
matrix elements $W^{\mu}$ and $W^{\mu}_T$ can be expressed as an
overlapping integral over the initial and final meson wave functions
\cite{chao-hsi chang}. We find that the contribution from the
positive energy wave function is dominant. Thus, the other parts have been
ignored. After integrating over $q^0_{_{P_{\bot}}}$, the
hadronic matrix elements can be obtained:
\begin{equation}
W_{\mu}=-\int\frac{d^3\vec{q}}{(2\pi)^3}\mathrm{Tr}\left\{\frac{\not\!{P_i}}{M_i}\bar{\varphi}^{++}_{f}\gamma_{\mu}\left(1-\gamma_5\right)\varphi^{++}_{i}\right\},
\end{equation}
\begin{equation}
W_{\mu}^T=-\int\frac{d^3\vec{q}}{(2\pi)^3}\mathrm{Tr}\left\{\frac{\not\!{P_i}}{M_i}\bar{\varphi}^{++}_{f}i\sigma_{\mu\nu}(P_i-P_f)^{\nu}\left(1+\gamma_5\right)\varphi^{++}_{i}\right\},
\end{equation}
where
 $P_i$ and $M_i$ are the momentum and mass of the initial meson, respectively. $\varphi^{++}_{i}$ and $\varphi^{++}_{f}$ are the positive energy wave functions of the initial and final mesons, respectively. 

Similarly, as to the annihilation hadronic currents, we have:
\begin{equation}
\begin{split}
W^{\beta}_1=&3(-8\pi^2)\int\frac{d^3\vec{q}_{_{P_{f
\bot}}}}{(2\pi)^3}\mathrm{Tr}\left\{\bar{\varphi_f}^{++}
\gamma_{\alpha}(1-\gamma_5)\right\}\\
&\int\frac{d^3\vec{q}_{_{P_{i\bot}}}}{(2\pi)^3}\mathrm{Tr}\left\{\gamma^{\alpha}(1-\gamma_5)
\frac{1}{\not\!{p}_{q_1}-m_{q_1}+i \epsilon}(-\frac{1}{3})\gamma^{\beta}\varphi^{++}_{i}\right\},\\
\end{split}
\end{equation}
where the cases of $W_2,~W_3,~W_4$ are similar. In our method, the pseudo-scalar meson wave functions $\varphi_1,~\varphi_2$ and the vector meson wave
functions $\varphi_3,~\varphi_4,~\varphi_5,~\varphi_6$ have the
approximate relationship
$$\varphi_1-\varphi_2=\varphi_3+\varphi_4=\varphi_5+\varphi_6=0.$$
Using this relationship, the calculations can be highly simplified
but a tiny deviation will emerge.

For the color-favored long-distance hadronic matrix element $W^{\mu}_{CF}$ as shown in Eq.~(11), it is proportional to the product of the transition matrix element $\langle
V|\bar{c}\gamma^{\nu}(1-\gamma_5)b|B_c\rangle$ multiplied by decay constant $\langle
D^{(*)}_{s(d)}(V)|\bar{s}(\bar{d})\gamma_{\nu}(1-
\gamma_5)c|0\rangle$. Using Eq.~(18), $\langle
V|\bar{c}\gamma^{\nu}(1-\gamma_5)b|B_c\rangle$ can be obtained. And $\langle
D^{(*)}_{s(d)}(V)|\bar{s}(\bar{d})\gamma_{\nu}(1-
\gamma_5)c|0\rangle$ has been discussed in our previous
paper~\cite{guoliwang1-,guoliwang0-}.

\subsection{Observables of the $B_c\rightarrow D^{(*)}_{s,d}l\bar{l}$ processes }

As pointed out in Refs.~\cite{AFaessler,D.Ebert}, during the calculation of the observables, it is convenient to project the hadronic matrix elements to
the 4-Helicity-Basis $\varepsilon_H^{\dag\mu}(\pm,0,t)$. With this projection, both the hadronic and leptonic currents can be calculated in
their own C.M.S separately. And the leptonic
angular distributions can be easily expressed in the terms of helicity
amplitudes. The helicity amplitudes in this paper are shown
as:

$(1) ~B_c\rightarrow D^{*}_{s,d}l^{\pm}l^{\mp} ~~~\text{transition}$
\begin{equation}
\begin{split}
H_{\pm}^{(1)}=&   -\left(M_f+M_i\right) \left\{ A_1 (C^{eff}_9+C^{CS}_9)+\frac{2 C^{eff}_7 m_b}{Q^2} \left(-M_f+M_i\right) T_2\right\}\\
&+ R_{\text{PBAnn}}M_i^2(M_i-M_f) (T_{1\text{ann}}+Q^2T_{1\text{CF}})\\
& \pm\sqrt{\lambda } \left[ \frac{V}{M_f+M_i} (C_9^{eff}+C_9^{CS})+ \frac{2 C_7^{eff}m_b}{Q^2}T_1\right.\\
&\left.+ \frac{1}{2}R_{\text{PBAnn}} (M_i- M_f)(V_{\text{ann}}+Q^2V_{\text{CF}})\right]  ,\\
H_{\pm}^{(2)}=&-C_{10} \left(\mp \frac{V}{M_f+M_i} \sqrt{\lambda }+A_1 \left(M_f+M_i\right)\right),\\
H_t^{(1)}=&+\frac{1}{4 \sqrt{Q^2} M_f} \sqrt{\lambda } R_{\text{PBAnn}}(M_i-M_f) \left\{2 M_i^2(T_{1\text{ann}}+Q^2T_{1\text{CF}})\right.\\
&\left.+\left(Q^2-M_f^2+M_i^2\right) (T_{2 \text{ann}}+Q^2T_{2 \text{CF}})-\left(Q^2+M_f^2-M_i^2\right) (T_{3 \text{ann}}+Q^2T_{3 \text{CF}})\right\}\\
&-\frac{ \sqrt{\lambda } A_0 (C_9^{eff}+C_9^{CS})}{\sqrt{Q^2}},\\
H_t^{(2)}=&-\frac{ \sqrt{\lambda } A_0 C_{10}}{\sqrt{Q^2}},\\
H_0^{(1)}=&\frac{1}{4 \sqrt{Q^2} M_f\left(M_f+M_i\right)} \left\{2(C_9^{eff}+C_9^{CS})\left[ A_2 \lambda   + A_1  \left(M_f\right.\right.\right.\\
&\left.\left.+M_i\right) {}^2 \left(M_f^2-M_i^2+Q^2\right)\right]+
 R_{\text{PBAnn}} \left(M_f^2-M_i^2\right)\left[2 M_i^2\left(T_{1\text{ann}}\right.\right.\\
 &\left.\left.\left.+Q^2T_{1\text{CF}}\right) \left(M_f^2-M_i^2+Q^2\right)-\lambda  \left(T_{2 \text{ann}}+T_{3 \text{ann}}+Q^2T_{2 \text{CF}}+Q^2T_{3 \text{CF}}\right)\right]\right\}\\
 &+ \frac{C_7^{eff} m_b}{\sqrt{Q^2}M_f} \left\{-T_2 \left(3 M_f^2+M_i^2-Q^2\right)-\lambda   \frac{T_3}{M_f^2-M_i^2}\right\},\\
H_0^{(2)}=&\frac{C_{10}}{2 \sqrt{Q^2} M_f } \left\{ \lambda  \frac{A_2}{\left(M_f+M_i\right)}+ A_1 \left(M_f+M_i\right) \left(Q^2+M_f^2-M_i^2\right)\right\},\\
\end{split}
\end{equation}
where $\lambda=(M_i^2-M_f^2){}^2+Q^2(Q^2-2M_i^2-2M_f^2)$
and $R_{\text{PBAnn}}=\frac{1}{Q^2}\frac{V_{cb}V^*_{cs(d)}}{V_{(tb)}V^*_{ts(d)}}\left(\frac{C_1}{N_c}+C_2 \right)$.

(2)~$ B_c\rightarrow D_{s,d}l^{\pm}l^{\mp} ~~~\text{transition}$
\begin{equation}
\begin{split}
H_{\pm}^{(1)}=&0,\\
H_{\pm}^{(2)}=&0,\\
H_t^{(1)}=&\frac{(M_i-M_f)^2R_{\text{BPAnn}}}{2 \sqrt{Q^2}}\left\{\left(B_2+Q^2B_{2CF}-B_1-Q^2B_{1CF}\right) Q^2 +\left(M_i^2-\right.\right.\\
&\left.\left.M_f^2\right)  \left[\left(B_1+B_2+Q^2B_{1CF}+Q^2B_{2CF}\right) \right]\right\}+\frac{M_i^2-M_f^2}{\sqrt{Q^2}} (C^{eff}_9+C^{CS}_9) F_0,\\
H_t^{(2)}=&\frac{C_{10} F_0}{\sqrt{Q^2}}\left( M_i^2-M_f^2\right),\\
H_0^{(1)}=&\frac{2 C^{eff}_7 m_b
F_T\sqrt{\lambda }}{ \sqrt{Q^2} \left(M_f+M_i\right)}   +\frac{\sqrt{\lambda }}{2 \sqrt{Q^2} } \left[
\left(B_1+B_2+Q^2B_{1CF}+Q^2B_{2CF}\right)\right.\\
&\left.(M_i-M_f)^2 R_{\text{BPAnn}}+2 (C^{eff}_9+C_9^{CS}) F_z\right] ,\\
H_0^{(2)}=&\frac{C_{10} F_z \sqrt{\lambda }}{\sqrt{Q^2}},\\
\end{split}
\end{equation}
where because the final meson $D_{s(d)}$ is a pseudo-scalar meson and has no
polarization direction, the transverse helicity amplitudes of the
$B_c\rightarrow D_{s(d)}l^{\pm}l^{\mp}$ processes are $0$ (Of course, if new physics operators contribute, they will be in a different situation.).

Based on the calculations in Ref.~\cite{AFaessler}, the differential
branching fraction is:
$$
\frac{\mathrm{d}Br}{\mathrm{d}Q^2}=\frac{G^2_F}{(2\pi)^3\Gamma_{B_c}}\left(\frac{\alpha_{em}|V^*_{ts(d)}V_{tb}|}{2\pi}\right)^2\frac{\lambda^{1/2}Q^2}{48M^3_{B_c}}\sqrt{1-\frac{4m_l^2}{Q^2}}\mathcal{M}^2_H,
$$
where $\mathcal{M}_H$ is defined as:
\begin{equation}
\begin{split}
\mathcal{M}^2_H=& \left(H^{(1)}_+H^{\dag(1)}_+
+H^{(1)}_-H^{\dag(1)}_-+H^{(1)}_0H^{(1)\dag}_0\right)\left(1+\frac{2m^2_l}{Q^2}\right)+\\
&\left(H^{(2)}_+H^{\dag(2)}_++H^{(2)}_-H^{\dag(2)}_-
+H^{(2)}_0H^{(2)\dag}_0 \right)
\left(1-\frac{4m^2_l}{Q^2}\right)+\frac{2m_l^2}{Q^2}3H^{(2)}_tH^{\dag(2)}_t.
\end{split}
\end{equation}

Besides that, we also compute the forward-backward asymmetries
$A_{FB}$ and the longitudinal polarizations $P_L$ of the final vector mesons
in the decays $B_c\rightarrow D^{*}_{s,d}\bar{l}l$. In the investigation of the
$B\rightarrow K^*\bar{l}l$ processes, $A_{FB}$ and
$P_L$ have been widely paid attention to, both theoretically and experimentally. We hope to obtain more information on the Wilson coefficients through studying these observables. From the
results in Ref.~\cite{AFaessler}, we have:
\begin{equation}
\begin{split}
A_{FB}=&\frac{3}{4}\sqrt{1-\frac{4m^2_l}{Q^2}}\frac{2}{\mathcal{M}^2_H}\left\{\mathrm{Re}\left(H_+^{(1)}H_+^{\dag(2)}\right)-\mathrm{Re}\left(H_-^{(1)}H_-^{\dag(2)}\right)\right\},\\
P_L=&\frac{1}{\mathcal{M}^2_H} \left\{
H^{(1)}_0H^{\dag(1)}_0\left(1+\frac{2m^2_l}{Q^2}\right)+H^{(2)}_0H^{\dag(2)}_0
\left(1-\frac{4m^2_l}{Q^2}\right)+\frac{2m_l^2}{Q^2}3H^{(2)}_tH^{\dag(2)}_t
\right\}.
\end{split}
\end{equation}
In this paper, only the longitudinal polarizations $P_L$ of the final
vector mesons are investigated. The transverse polarizations $P_T$ of
the final vector mesons can be obtained from the relationship $P_T=1-P_L$ obviously. Furthermore, we also study the leptonic longitudinal polarization
asymmetry $A_{LPL}$, which is defined as:
\begin{equation}
\begin{split}
A_{LPL}&\equiv\frac{dBr_{h=-1}/dQ^2-dBr_{h=1}/dQ^2}{dBr_{h=-1}/dQ^2+dBr_{h=1}/dQ^2}\\
&=\sqrt{1-\frac{4m^2_l}{Q^2}}\frac{2}{\mathcal{M}^2_H}\left\{\mathrm{Re}\left(H_+^{(1)}H_+^{\dag(2)}\right)+\mathrm{Re}\left(H_-^{(1)}H_-^{\dag(2)}\right)+\mathrm{Re}\left(H_0^{(1)}H_0^{\dag(2)}\right)\right\},
\end{split}
\end{equation}
where $h=+1(-1)$ denotes right~(left) handed $l^-$. And the second step followed the derivation in Ref~\cite{AFaessler}.

\section{Numerical Results and Discussions}

 In this section, the parameters, the form factors, the differential branching fractions, forward-backward asymmetries, longitudinal polarizations of the final vector mesons and leptonic longitudinal polarization asymmetries are presented.
\subsection{Parameters and Theoretical Uncertainties}

In our calculation, BS-model dependent parameters are employed, which include the masses of the constituent quarks and the Cornell-potential-parameters $\Lambda_{BS}$,
$\alpha_{BS}$, $V_{BS}$ and $\lambda_{BS}$. The masses of the constituent quarks are taken as $m_b=4.96$ GeV, $m_c=1.62$ GeV, $m_s=0.5$ GeV and $m_d=0.311$ GeV.
As shown in Ref. \cite{110changchaohsi}, the mass spectra with these
input constituent quark masses are in good agreement with the
experimental data. Thus, this set of numerical values is adopted in
this paper. The numerical values of Cornell-potential-parameters are adopted as Ref.~\cite{huifengfu}.

Moreover, the masses and the lifetimes of
$B_c,~D^{(*)}_{s(d)},~J/\psi,~\psi(2S)$ are taken from
Particle Data Group (PDG) \cite{amsler c}, as the values of $\alpha_{em},~G_F$
and $V_{CKM}$.

In this paper, the theoretical uncertainties are estimated including two aspects. On one hand, theoretical uncertainties caused by BS-model
dependent parameters are studied. This kind of errors is calculated by changing the numerical values of our model
dependent parameters by $\pm5\%$, which
include the masses of constituent quarks and the Cornell-potential-parameters. We find that, in BS method, the observables are more sensitive to $\Lambda_{BS}$ and $\lambda_{BS}$ than the other BS-model
dependent parameters.

On the other hand, the systematic uncertainties which the naive factorisation hypothesis arouses are investigated. In the investigation of $B\rightarrow K
^{(*)}\bar{l}l$ processes, to include non-factorisable contributions of LD cascade processes $B\rightarrow K^{(*)} J/\psi(\psi(2S))\rightarrow K
^{(*)}\bar{l}l$, the $\kappa$ factor which is determined by comparing the experimental data with the theoretical results is introduced frequently~\cite{kappa,AFaessler,D.Ebert}. However, the situation in $B_c\rightarrow D_{s,d}^{(*)}J/\psi(\psi(2S))\rightarrow D_{s,d}^{(*)}\bar{l}l$ is different. The cascade processes $B\rightarrow K^{(*)} J/\psi(\psi(2S))\rightarrow K
^{(*)}\bar{l}l$ include only CS contributions and the $\kappa$ factor can be used to include the non-factorisable effects of CS diagrams, but transitions $B_c\rightarrow D_{s,d}^{(*)}J/\psi(\psi(2S))\rightarrow D_{s,d}^{(*)}\bar{l}l$ contain both CS and CF diagrams. Thus, another way is adopted in this paper. In the previous calculations on the non-leptonic decays of $B$ meson~\cite{NF1,NF2,NF3,NF4}, to contain the non-factorisable contributions, the number of colors $N_c$ in the expression $(C_1/N_c+C_2)$ or $(C_1+C_2/N_c)$ is treated as parameter which should be obtained by fitting the experimental data, i.e, the effective number of colors.
However, the $B_c$ meson which consists two heavy quarks is much different from $B$ meson.
So the parameters constrained by $B$ experimental data are not chosen to use in this paper. And considering that at present the data on $B_c$ non-leptonic decays are still lacking, we calculate the observables with $N_c=3$ but for estimating the systematic uncertainties which are aroused by non-factorisable contributions in the the processes $B_c\rightarrow D_{s,d}^{(*)}J/\psi(\psi(2S))\rightarrow D_{s,d}^{(*)}\bar{l}l$, we change the numerical values of $N_c$ in Eq.~(9) within the region $[2,\infty]$. Besides, the systematic uncertainties aroused by non-factorisable weak annihilation contributions are dealt with using the similar way in which the numerical value of $N_c$ in Eq.~(6) is changed within the region $[2,\infty]$.

During the calculation of theoretical uncertainties, it is found that the CS contributions are very sensitive to $N_c$. For instance, as shown in emerald area of Fig.~11 (e), the error band of $P_L$ that includes PB and CS diagrams is quite large, which is brought about dominantly by the $N_c$. And if there are sufficient experimental data on $B_c$ non-leptonic decays, the number values of $N_c$ can be obtained through fitting the experimental data and will be constrained within a small region. As a result, the error bands which include CS cascade contributions can be much narrower.

From the error bands as shown in Figs.~6-13, one may find that the error bands of $A_{LPL}$, $P_L$ and $A_{FB}$ which include only the PB diagrams are fairly narrow.
For instance, as shown in Figs.~6~(c), $A_{LPL}$ including only the PB diagrams has tiny theoretical uncertainty. This can be understood from the Eq.~(22) and Eq.~(25). If only PB contributions are considered, we have the relationships $H^{(1)}_{\pm}=H^{(2)}_{\pm}=0$ and $H_0^{(1)}\propto\left\{4C_7^{eff}m_b F_T+2\left(M_f+M_i\right)C_9^{eff}F_z\right\}$. Considering that $C^{eff}_7$ has a small value, we can write the relationship $H_0^{(1)}\propto\left\{2\left(M_f+M_i\right)C_9^{eff}F_z\right\}\propto F_z$ approximately. If $m_{\mu}\sim0~\text{GeV}$ is considered, as shown in Eq.~(25), both the numerator and denominator of $A_{LPL}$ are proportional to $F_z$, while $F_T$ and $F_0$ are suppressed by $C_7^{eff}$ and $m_{\mu}$ respectively. So the ratio, that is, $A_{LPL}$, has tiny theoretical uncertainty. If Ann, CS and CF contributions are added, $H_0^{(1)}$ will be affected by not only the numerical values of $N_c$ but also the Ann and CF form-factors, which can not be neglected. So there are wide error bands in Figs.~6~(d). The cases of the other figures are similar.

\subsection{Form Factors of the Decays $B^{-}_c\rightarrow D^{-(*)}_{s(d)}
l\bar{l}$}

The form factors of the decays $B^{-}_c\rightarrow D^{-(*)}_{s(d)}
l\bar{l}$ are shown in Figs.~2-5. As seen from Figs.~2-5, the
form-factors of $W^{\mu}$ and $W_T^{\mu}$ have different features
from those of $W_{ann}^{\mu}$. In our model, $W_T^{\mu}$ and
$W^{\mu}$ depend on the overlapping region of
initial and final wave functions. In the low $Q^2 $ region, the final
meson $D^{-(*)}_{s(d)}$ has a large recoil momentum. As a
result, the overlapping region of initial and final wave functions is small and the form factors of $W_{\mu}$ and $W^T_{\mu}$ will
be suppressed around $Q^2\sim0~\textmd{GeV}^2$. In the high $Q^2
$ region, the final meson $D^{-(*)}_{s(d)}$ has a tiny
recoil momentum and the overlapping region of initial and final wave
functions will be enhanced. So we find that the form factors of
$W_{\mu}$ and $W^T_{\mu}$ become larger with $Q^2$ increasing.

However, the calculation of $W_{ann}^{\mu}$ is quite different. As shown in Eqs.~(7,~20), it is a
product of the initial~(final) meson decay constant multiplied by the final~(initial) annihilation hadronic matrix element. The initial and final wave functions are integrated independently. To do the
integral, we use the following relationship,
 \begin{equation} \frac{1}{x-x_0 \pm
i\epsilon}=P\left(\frac{1}{x-x_0} \right)\mp i \pi \delta(x-x_0).
\end{equation}

In this way, the form-factors of $W_{ann}^{\mu}$ are complex~(The same scenario has also been reported by Ref.~\cite{CSKIM} in analysis of the t-channel annihilation diagrams in the $B^0\rightarrow D^0 \mu \bar{\mu}$ process by pQCD method.) and we define them as:
\begin{equation}
\begin{split}
B&_1\equiv B_{1Aann}+iB_{1Bann},~~~~~~~~~~B_2\equiv B_{2Aann}+iB_{2Bann},\\
V&_{ann}\equiv V_{Aann}+iV_{Bann},~~~~~~~~~~T_{1ann}\equiv T_{1Aann}+iT_{1Bann},\\
T&_{2ann}\equiv T_{2Aann}+iT_{2Bann},~~~~~~~~T_{3ann}\equiv
T_{3Aann}+iT_{3Bann}.
\end{split}
\end{equation}

\subsection{Observables and Discussions}

With the form-factors above, the differential branching fractions
$dBr/dQ^2$, forward-backward asymmetries $A_{FB}$, leptonic longitudinal
polarization asymmetries $A_{LPL}$ and longitudinal polarizations $P_L$
of the final vector mesons are calculated using Eqs.~(21-25) and shown in
~Figs.~6-13. In our calculation, if only $Q_7$ and Ann~contributions are considered, we find that the
relationship $dBr^{Q_7+Ann}/dQ^2>dBr^{Ann}/dQ^2>dBr^{Q_7}/dQ^2$ is
established in the $Q^2\sim0~\text{GeV}^2$ region. In calculation of
$B_c\rightarrow D_{s,d}^{*}\gamma$~\cite{KAzizigamma,HYCheng,DDU},
the same relationship also holds.

We plot the leptonic longitudinal polarization asymmetries $A_{LPL}$ in
Figs.~6~(c,~d), Figs.~8~(c,~d), Figs.~10~(g,~h) and Figs.~12~(g,~h) for
the $B_c\rightarrow D_{s(d)}^{(*)}\mu\bar{\mu}$ processes and in
Figs.~7~(c,~d), Figs.~9~(c,~d), Figs.~11~(g,~h) and Figs.~13~(g,~h) for
the $B_c\rightarrow D_{s(d)}^{(*)}\tau\bar{\tau}$ processes.
If only the PB and CS contributions are considered, in the area
far away from $Q^2=0~\text{GeV}^2$ and the resonance regions around $J/\psi$ and
$\psi(2S)$, considering the relationship
$C_9^{eff}\sim-C_{10}\gg2m_bC_7/(M_i+M_f)$, the leptonic term $\bar{l}\gamma^{\mu}(1-\gamma_5) l$ contributes to $A_{LPL}$ dominantly. For
$m_{\mu}\sim0$~GeV, the leptonic
spin-flip contribution can be neglected. As a result, $A_{LPL}(B_c\rightarrow D_{s(d)}^{(*)}\mu\bar{\mu})\sim-1$ which includes PB and CS diagrams is shown. For
$B_c\rightarrow D_{s(d)}^{(*)}\tau\bar{\tau}$ processes, the
approximation $m_{\tau}\sim0$~GeV is not fulfilled and the leptonic
spin-flip contribution is significant. Consequently, $A_{LPL}(B_c\rightarrow D_{s(d)}^{(*)}\tau\bar{\tau})$ which includes PB and CS contributions
has an evident deviation from $-1$. If the Ann diagrams are added in the processes $B_c\rightarrow D_{s(d)}^{(*)}\mu\bar{\mu}$,
both $\bar{l}\gamma_{\mu}(1-\gamma_5)l$ and
$\bar{l}\gamma_{\mu}(1+\gamma_5)l$ terms contribute, which can not be
neglected. Thus, $A_{LPL}(B_c\rightarrow D_{s(d)}^{(*)}\mu\bar{\mu})$ which are far from $-1$ are shown in the dash-dot-dash lines of
Fig.~6~(d), Fig.~8~(d), Fig.~10~(h) and Fig.~13~(h). If both CS and CF cascade processes are
added in the processes $B_c\rightarrow D_{s(d)}^{(*)}\mu\bar{\mu}$, $A_{LPL}(B_c\rightarrow D_{s(d)}^{(*)}\mu\bar{\mu})$ will be suppressed around the $Q^2\sim0~\text{GeV}^2$ region, as shown in the solid lines of
Fig.~6~(d), Fig.~8~(d), Fig.~10~(h) and Fig.~13~(h).

The forward-backward asymmetries $A_{FB}$ of
$B_c\rightarrow D^*_{s(d)}\mu\bar{\mu}$ are plotted in Figs.~10~(c,~d)
and Figs.~12~(c,~d). As shown in the dash-dot-dash lines of Fig.~10~(c) and Fig.~12~(c), if PB diagrams
are considered, because of the $\gamma$~penguin effect,
forward-backward asymmetries $A_{FB}$ will be negative around $Q^2\sim0~\text{GeV}^2$. If the annihilation contribution is
included, as shown in the dash-dot-dash lines of Fig.~10~(d) and Fig.~12~(d), the negative deviation from zero will be larger. For the
$B_c\rightarrow D^*_{s(d)}\tau\bar{\tau}$ processes, the
forward-backward asymmetries are given in Figs.~11~(c,~d) and
Figs.~13~(c,~d). Actually, the forward-backward asymmetry $A_{FB}$ depends on the transverse CP-odd helicity amplitudes. Because there
are no transverse helicity amplitudes in decay $B_c\rightarrow
D_{s(d)}l\bar{l}$, $A_{FB}$ will be zero. In this paper, they are not plotted.

In the Figs.~10-13~(e,~f), the longitudinal polarizations
$P_L$ of the final vector mesons are plotted. As shown in the Figs.~10~(e,~f) and Figs.~12~(e,~f), $P_L$ are almost proportional to $Q^2$ within the low $Q^2$ region but inversely related to $Q^2$ in the high $Q^2$ area. In the $Q^2\sim Q^2_{MAX}$ region, the final
meson has a tiny recoil momentum and the polarization is almost
averaged. So $P_L\sim\frac{1}{3}$ is found near the $Q^2_{MAX}$ point.
In
the $Q^2\sim0~\text{GeV}^2$ region, the $Q_7$~and annihilation
diagrams play important parts in $P_L$. Considering that the real $\gamma$ has no longitudinal
polarization, in a physical sense, $P_L$ should be towards $0$ nearby the $Q^2= 0~\text{GeV}^2$ point. This feature is in agreement
with the one in Ref.~\cite{D.Ebert} but has obvious difference
from those in Refs.~\cite{pakistan1,pakistan2}. Beside that, as shown in Fig.~10 (c) and Fig.~12 (c), $P_L$ with only PB processes differ slightly from those which include both PB and CS contributions.

In Figs.~6-13~(a,~b), the differential branching
fractions of $B_c\rightarrow D_{s(d)}^{(*)}\bar{l}l$ are plotted. From the solid lines of Fig.~6~(b), Fig.~8~(b), Fig.~10~(b) and Fig.~12~(b), it is found that the differential branching fractions near the resonance region of $J/\psi$ are a lot larger than
the ones around the resonance region of $\psi(2S)$. The cases of the Fig.~6~(a), Fig.~8~(a), Fig.~10~(a) and Fig.~12~(a) are similar. This phenomenon can be understood from the wave functions of $J/\psi$ and $\psi(2S)$. In BS method, $J/\psi$ is the ground state of $^3S_1$ $\bar{c}c$ bound system and there is no node in the wave function of $J/\psi$. But $\psi(2S)$ is the first radial excited state of $^3S_1$ $\bar{c}c$ bound system and one node appears in the wave function of $\psi(2S)$, which causes large cancelation in the integration over the wave functions. So compared with decays $B_c\rightarrow D_{s,d}^{(*)}J/\psi\rightarrow D_{s(d)}^{(*)}\bar{l}l$, the $B_c\rightarrow D_{s,d}^{(*)}\psi(2S)\rightarrow D_{s(d)}^{(*)}\bar{l}l$ processes contribute less.
Besides, if only SD contribution
is considered, as shown in dash-dot-dash lines of Figs.~6-13 (a,~b), the
annihilation diagrams will enhance the differential branching
fractions obviously around the $Q^2\sim[12,~18]~\text{GeV}^2$ region.
This is caused by the large imaginary-parts of the annihilation
form-factors as shown in Figs.~2-5.

To avoid the pollution of resonance cascade decays $B_c\rightarrow D^{(*)}_{s(d)}V\rightarrow D^{(*)}_{s(d)}
\bar{l}l$, when the branching fractions
of $B_c\rightarrow D_{s(d)}^{(*)}\bar{l}l$ are calculated, the cutting regions on the di-muon mass around the $J/\psi$ and $\psi(2S)$ masses are required. In Refs.~\cite{Gengchaoqiang,Ho}, through analysing the contributions of the
PB and CS cascade diagrams, the experimental cutting regions were defined. However, based on our differential branching
fractions in Figs.~6-13~(a,~b), the CF cascade diagrams affect the differential branching fractions substantially. So new experimental cutting regions will be necessary. In this paper, we attempt to define the experimental cutting
regions: $5~\text{GeV}^2<Q^2<15~\text{GeV}^2$ and the regions of interest are given as:
\begin{equation}
\begin{split}
\text{Region}&~(1): 0.2~\text{GeV}^2<Q^2<5~\text{GeV}^2,\\
\text{Region}&~(2): 15~\text{GeV}^2<Q^2<Q^2_{MAX}.
\end{split}
\end{equation}
In Table.~1, the branching fractions including penguin-box,
annihilation, color-suppressed cascade and color-favored cascade
diagrams within Region~(1) and Region~(2) are shown. From the
Table.~1, the $B_c\rightarrow D_s^*(2112)\mu\bar{\mu}$ process has
the largest branching fraction among the decays $B_c\rightarrow D_{s,d}^{(*)}\bar{l}l$, which is $1.24\times10^{-7}$.

\section{Conclusion}

In summery, the processes $B_c\rightarrow D^{(*)}_{(s,d)}l\bar{l}$
induced by the $b\rightarrow s(d)\bar{l}l$, the annihilation,
color-favored resonance cascade and color-suppressed resonance
cascade contributions are analysed within the SM. Our results show that:~(1) the annihilation contribution which could be ignored in
$B\rightarrow K^{(*)}\bar{l}l$ should be emphasized in $B_c\rightarrow D^{(*)}_{(s,d)}l\bar{l}$.
Especially in the $Q^2\sim0~ \text{GeV}^2$ region, the
forward-backward asymmetries~$A_{FB}$ and leptonic longitudinal
polarization asymmetries~$A_{LPL}$ have sensitive values to the
annihilation effects; (2) considering color-favored resonance cascade
effects, the branching fractions using the previous experimental cutting regions
are seriously polluted by the $J/\psi$ and $\psi(2S)$ cascade
decays. So the new experimental cutting regions are
proposed.

\section*{Acknowledgments}

 This work was supported in part by the National Natural Science
Foundation of China (NSFC) under Grant No. 11175051 and Program for Innovation Research of Science in Harbin Institute of Technology.

\begin{table}[ph]
\caption{Branching ratios including penguin-box, annhilation,
color-favored and color-suppressed cascade effects in unit of
$10^{-8}$, where the regions of interest are defined as: $\text{Region}~(1):
0.2~\text{GeV}^2<Q^2<5~\text{GeV}^2;~~ \text{Region}~(2):
15~\text{GeV}^2<Q^2<Q^2_{MAX}$ }
\begin{center}
{\begin{tabular}{|c|c|c|c|c|c|}
\hline decay&Region~(1)&Region~(2)\\
\hline $Br(B_c\rightarrow D_s\mu \bar{\mu})$&$0.81^{+0.45}_{-0.29}$&$0.86^{+0.47}_{-0.23}$\\
\hline $Br(B_c\rightarrow D_s\tau \bar{\tau})$&0&$1.5^{+0.4}_{-0.3}$\\
\hline $Br(B_c\rightarrow D_s^*(2112)\mu \bar{\mu})$&$5.5^{+1.7}_{-2.1}$&$6.9^{+3.7}_{-1.7}$\\
\hline $Br(B_c\rightarrow D_s^*(2112)\tau \bar{\tau})$&0&$2.5^{+2.3}_{-1.0}$\\
\hline $Br(B_c\rightarrow D\mu \bar{\mu})$&$0.014^{+0.008}_{-0.006}$&$0.039^{+0.025}_{-0.011}$\\
\hline $Br(B_c\rightarrow D\tau \bar{\tau})$&0.0&$0.059^{+0.016}_{-0.012}$\\
\hline $Br(B_c\rightarrow D^*(2010)\mu \bar{\mu})$&$0.15^{+0.05}_{-0.06}$&$0.31^{+0.16}_{-0.08}$\\
\hline $Br(B_c\rightarrow D^*(2010)\tau \bar{\tau})$&0.0&$0.12^{+0.10}_{-0.04}$\\
\hline
\end{tabular} }
\end{center}
\end{table}

\begin{figure}[htbp]
\centering
{\includegraphics[width = 0.99\textwidth,height=0.8\textheight]{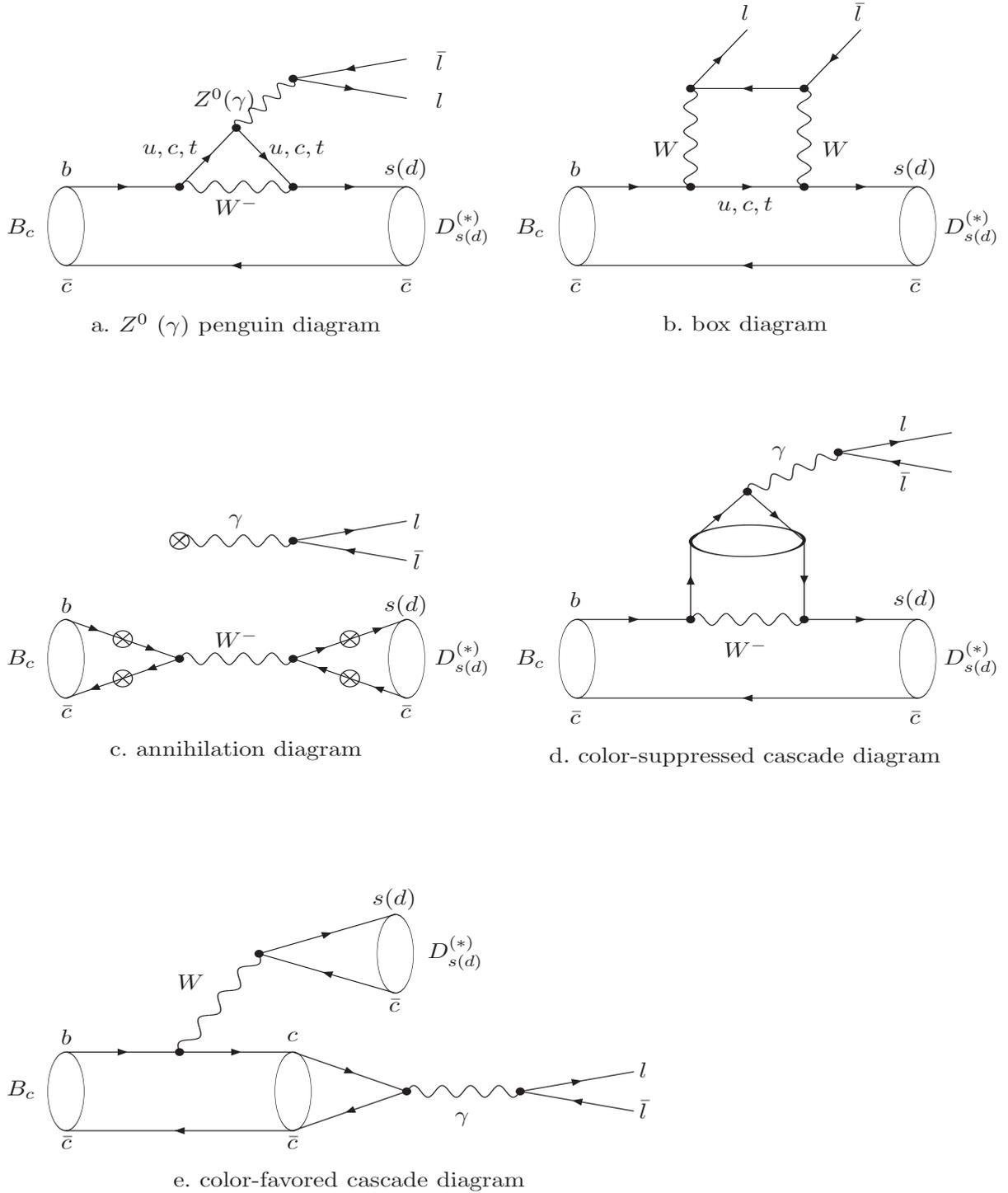}}
\caption{Diagrams of $B_c\rightarrow D^{(*)}_{s,d} l\bar{l}$. In annihilation diagrams c) the photon can be emitted from each quark, denoted by $\bigotimes$, and decays to the lepton pair.}
\end{figure}

\begin{figure}[htbp]
\centering \subfigure[Form-Factors of $W^{\mu}$ and $W^{\mu}_T$
induced by penguin and box diagrams]{\includegraphics[width =
0.49\textwidth,height=0.4\textheight]{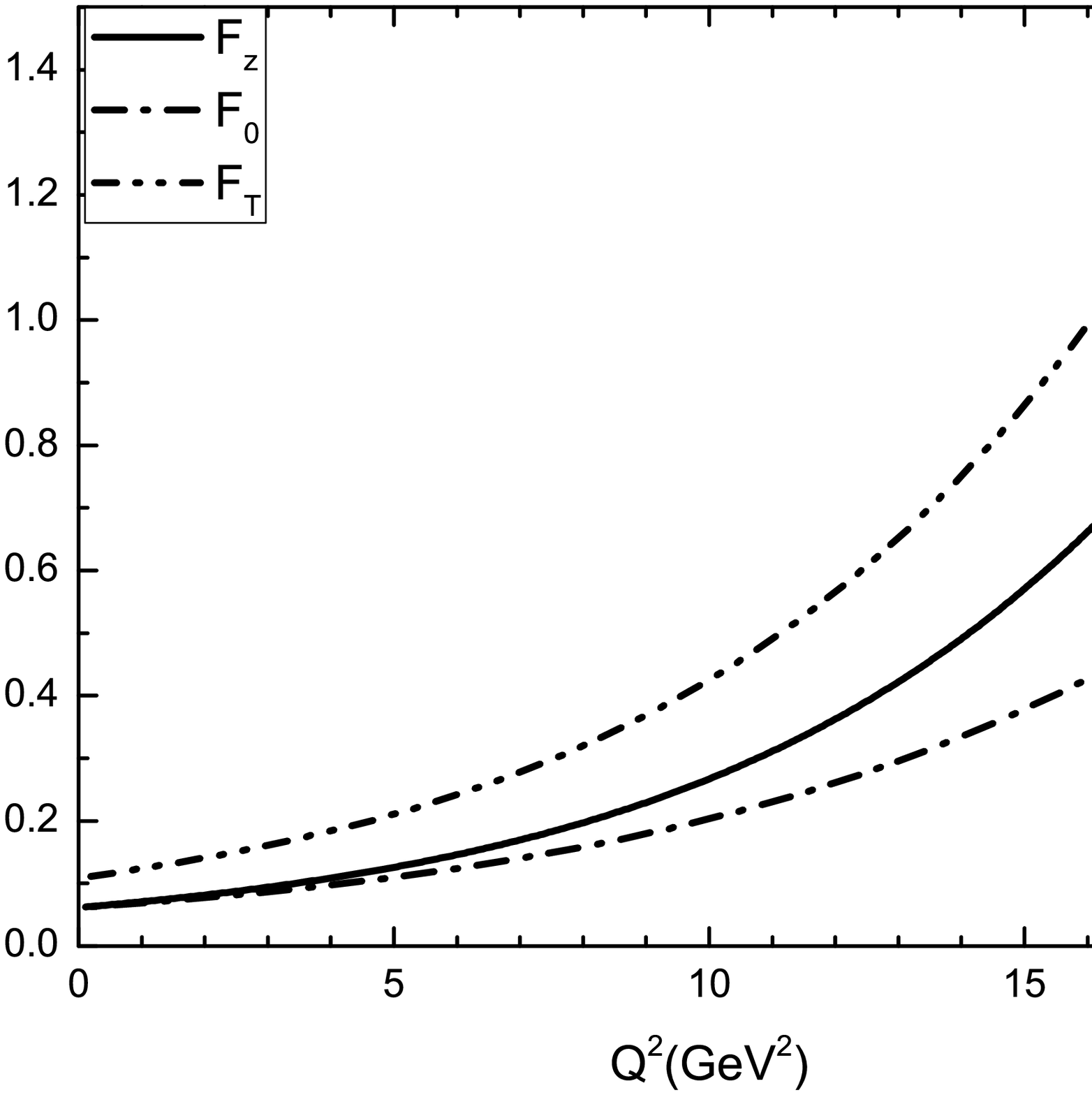}}
\subfigure[Form-Factors of $W^{\mu}_{ann}$ induced by annihilation
diagrams]{\includegraphics[width =
0.49\textwidth,height=0.4\textheight]{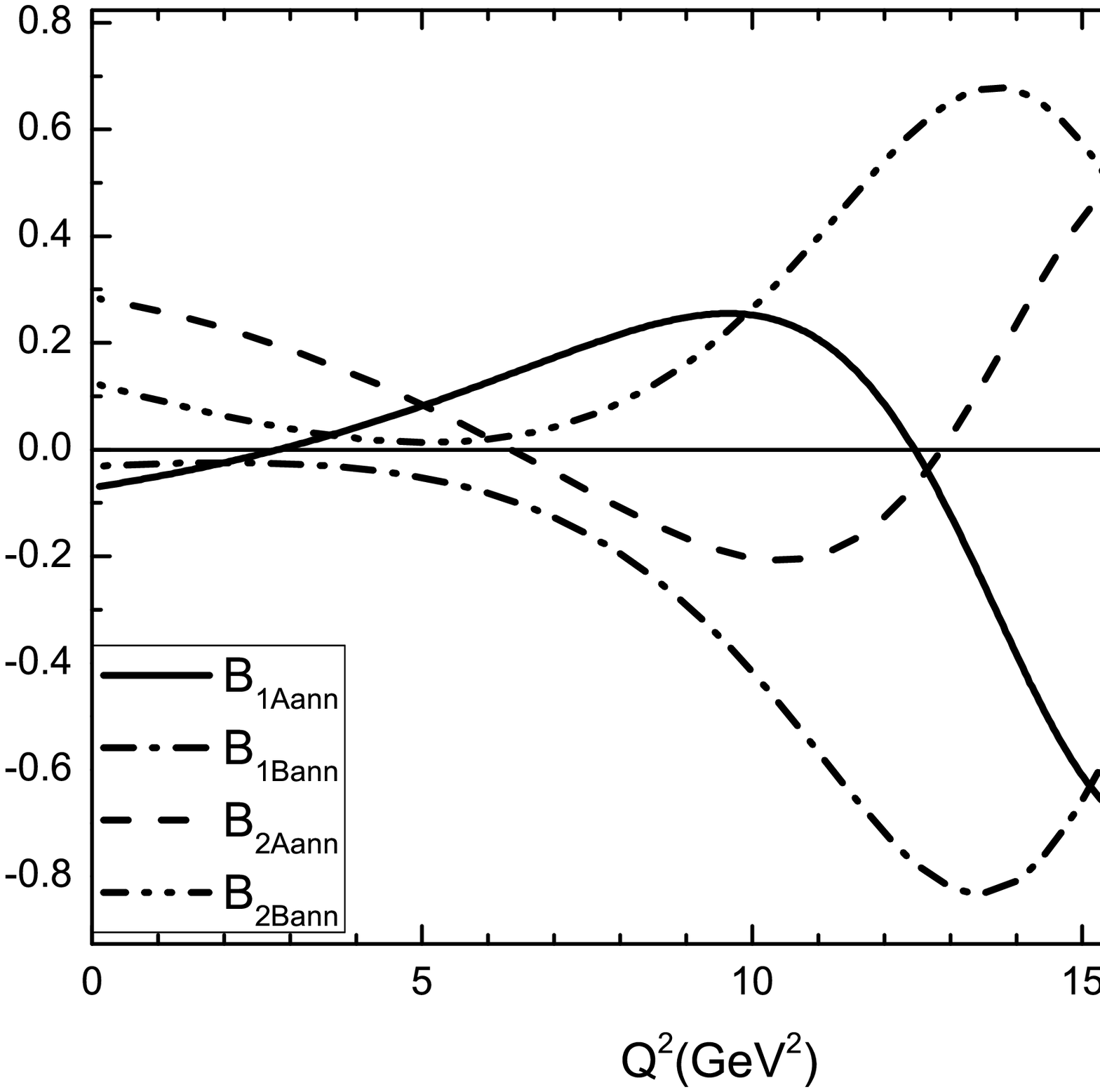}}
\caption{Form-Factors of $B_c\rightarrow D l\bar{l}$.}
\end{figure}

\begin{figure}[htbp]
\centering \subfigure[Form-Factors of $W^{\mu}$ and $W^{\mu}_T$
induced by penguin and box diagrams]{\includegraphics[width =
0.49\textwidth,height=0.4\textheight]{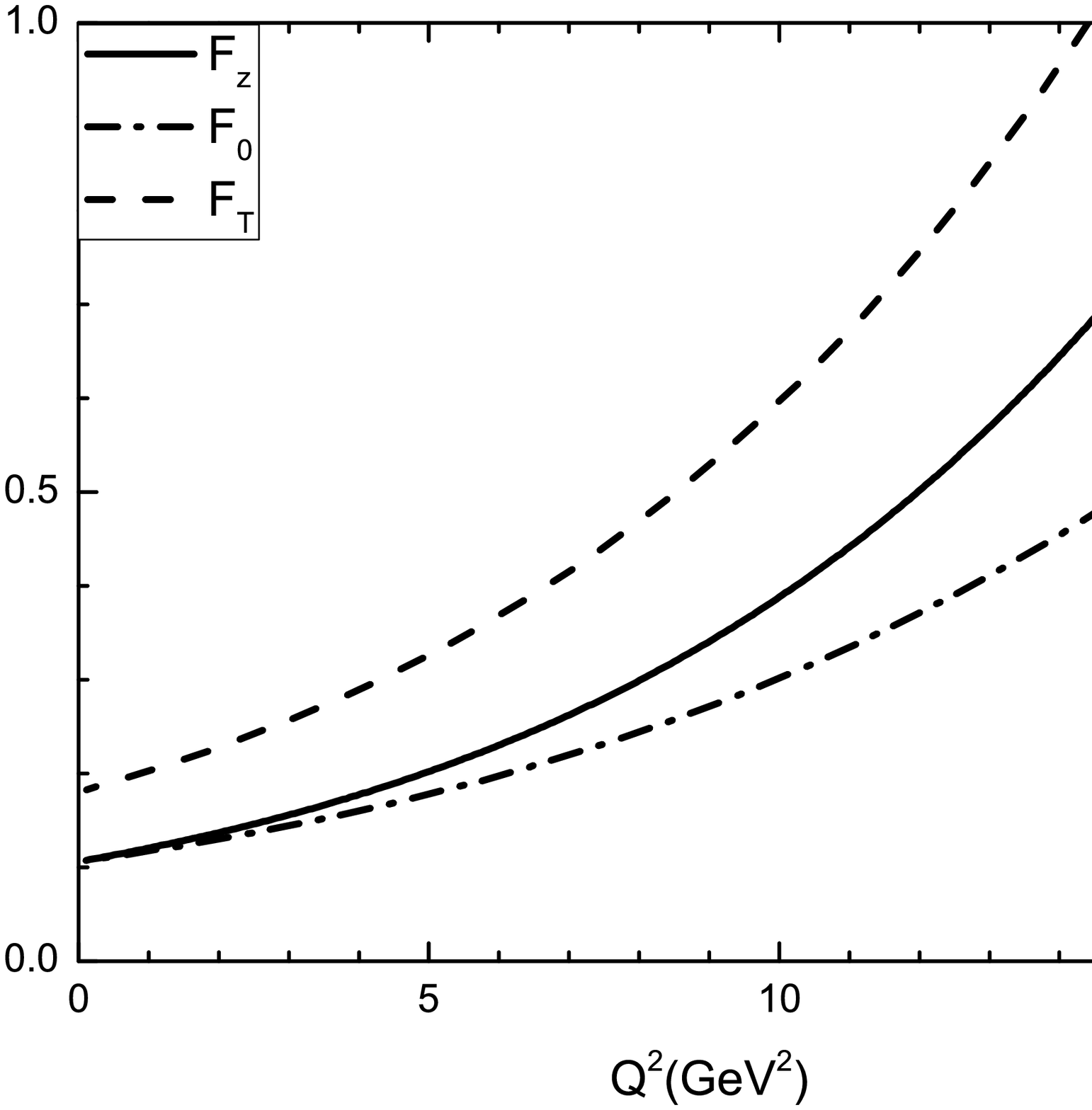}}
\subfigure[Form-Factors of $W^{\mu}_{ann}$ induced by annihilation
diagrams]{\includegraphics[width =
0.49\textwidth,height=0.4\textheight]{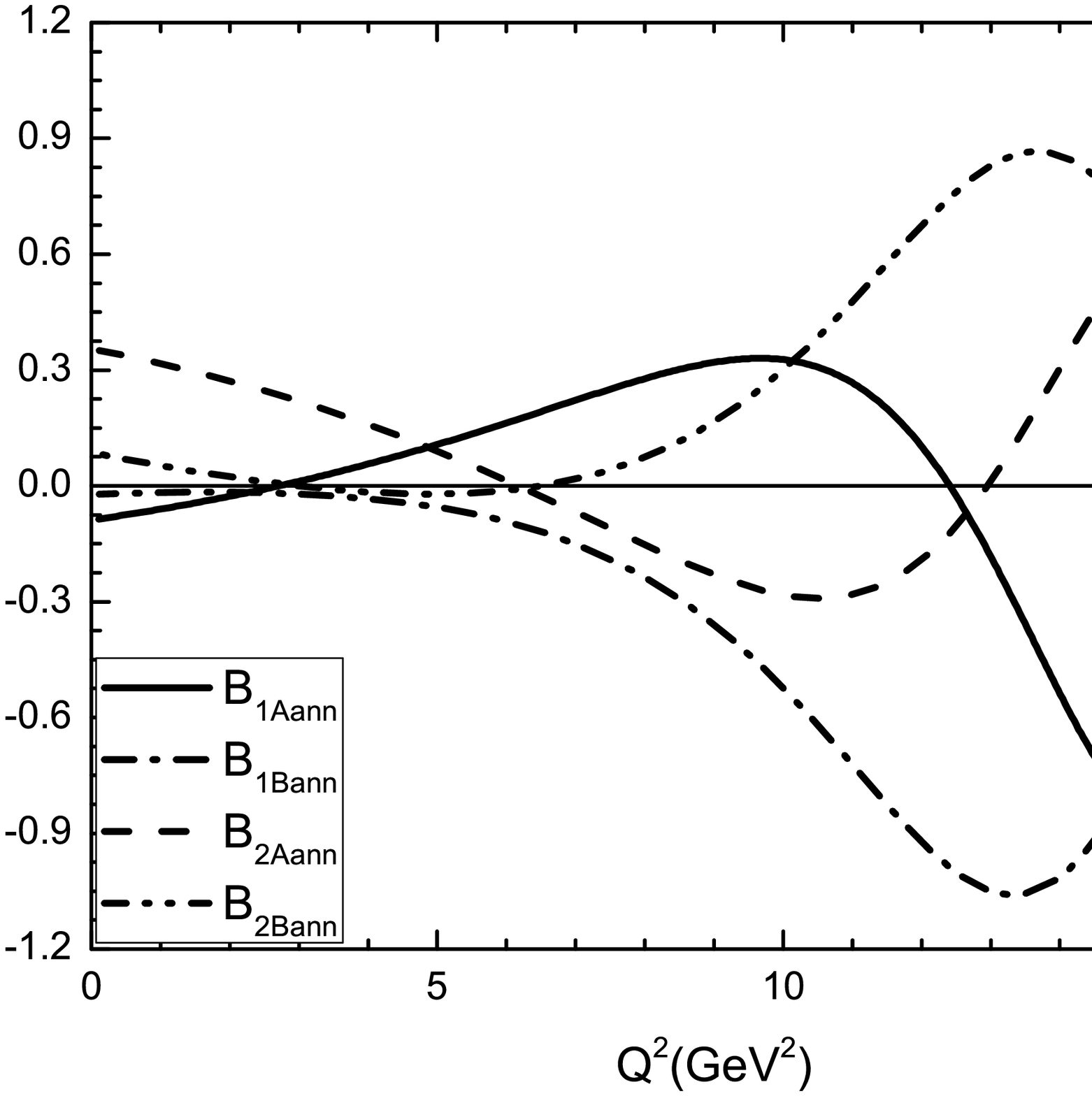}}
\caption{Form-Factors of $B_c\rightarrow D_s l\bar{l}$.}
\end{figure}

\begin{figure}[htbp]
\centering \subfigure[Form-factors of $W^{\mu}$ induced by $Z^0$
penguin and box diagrams]{\includegraphics[width =
0.49\textwidth,height=0.4\textheight]{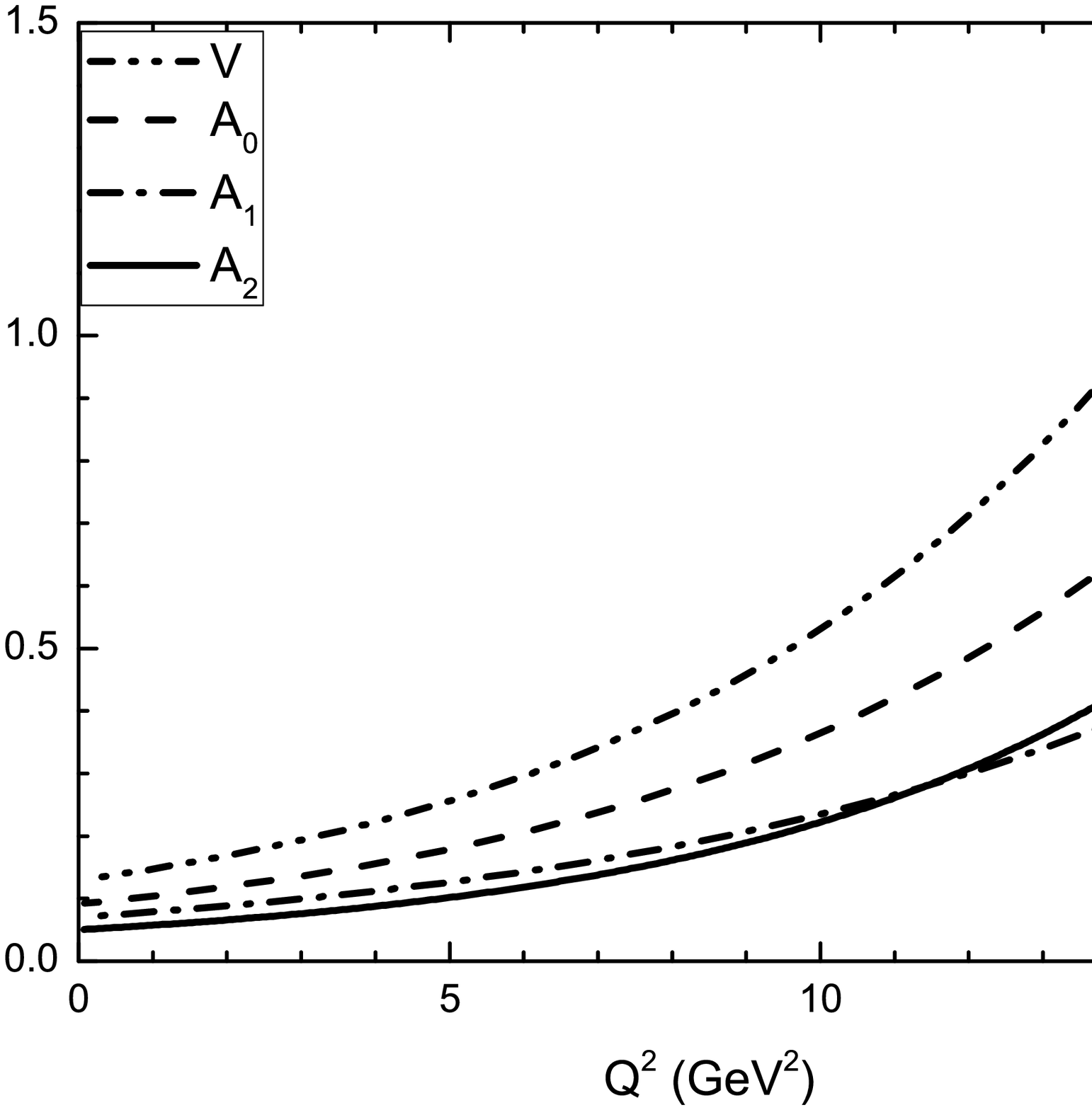}}
\subfigure[Form-factors of $W^{\mu}_T$ induced by $\gamma$ penguin diagrams]{\includegraphics[width = 0.49\textwidth,height=0.4\textheight]{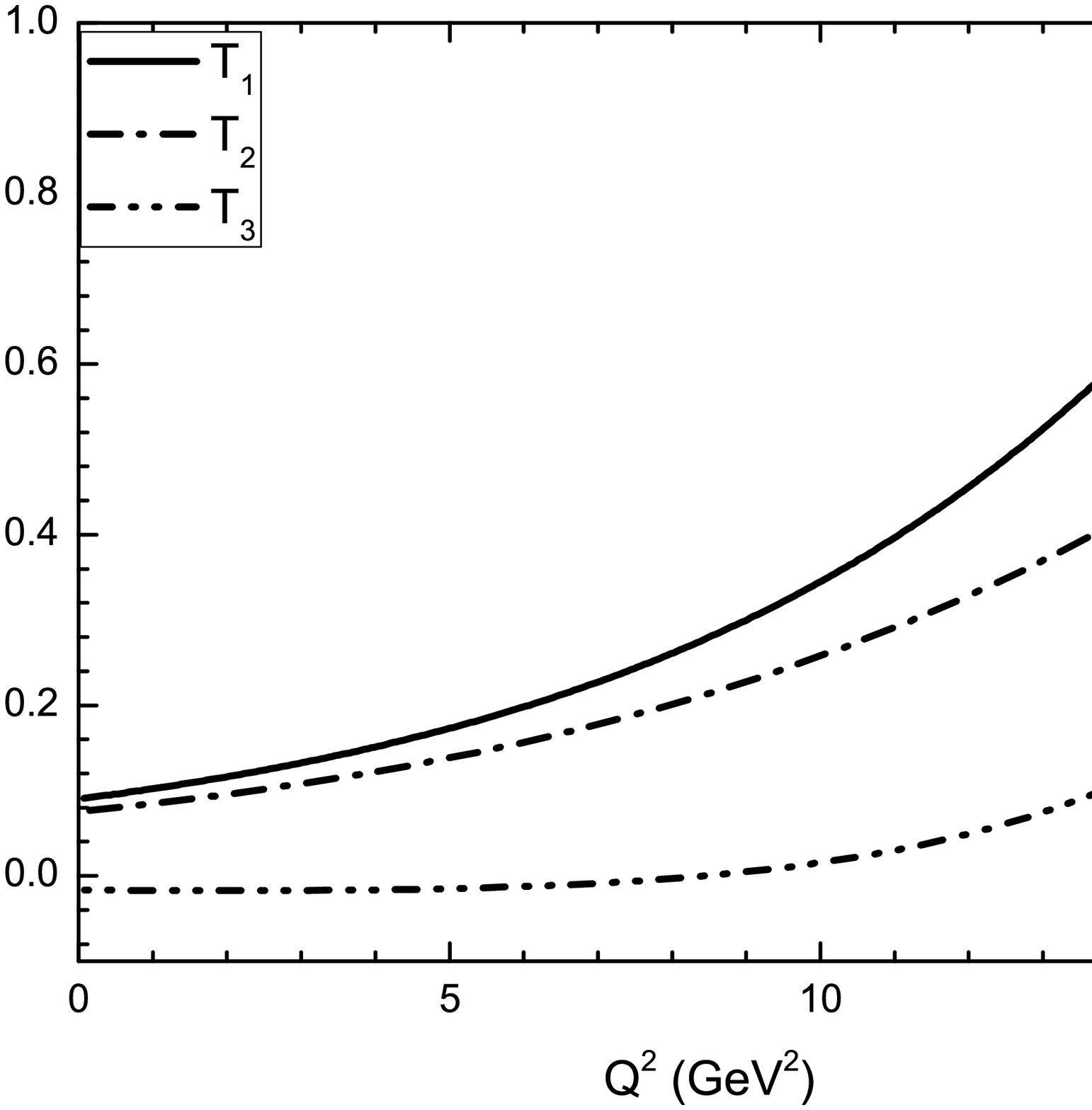}}\\
\subfigure[Real parts of $W^{\mu}_{ann}$ Form-Factors induced by
annihilation diagrams]{\includegraphics[width =
0.49\textwidth,height=0.4\textheight]{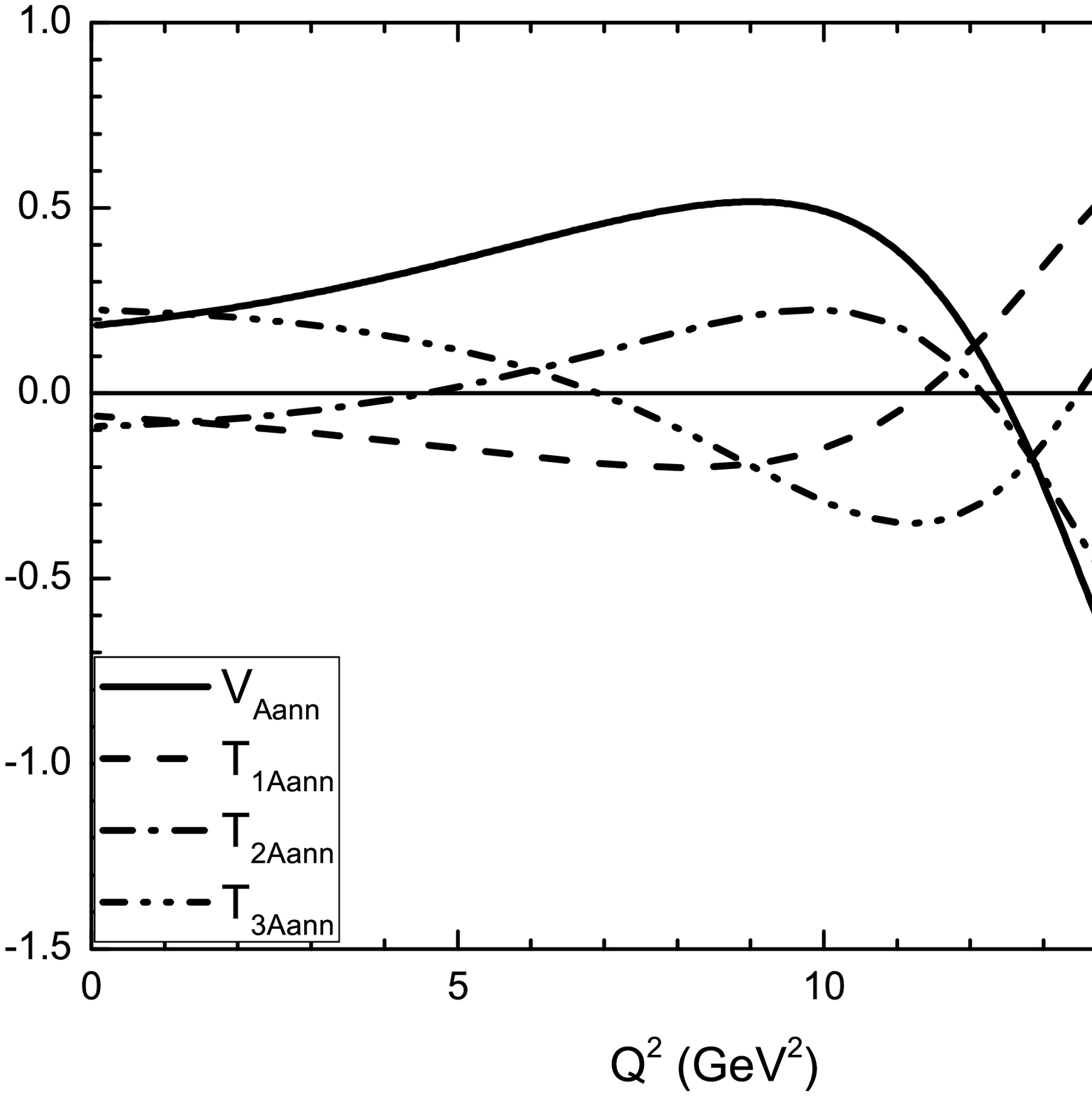}}
\subfigure[Imaginary parts of $W^{\mu}_{ann}$ Form-Factors induced
by annihilation
diagrams]{\includegraphics[width = 0.49\textwidth,height=0.4\textheight]{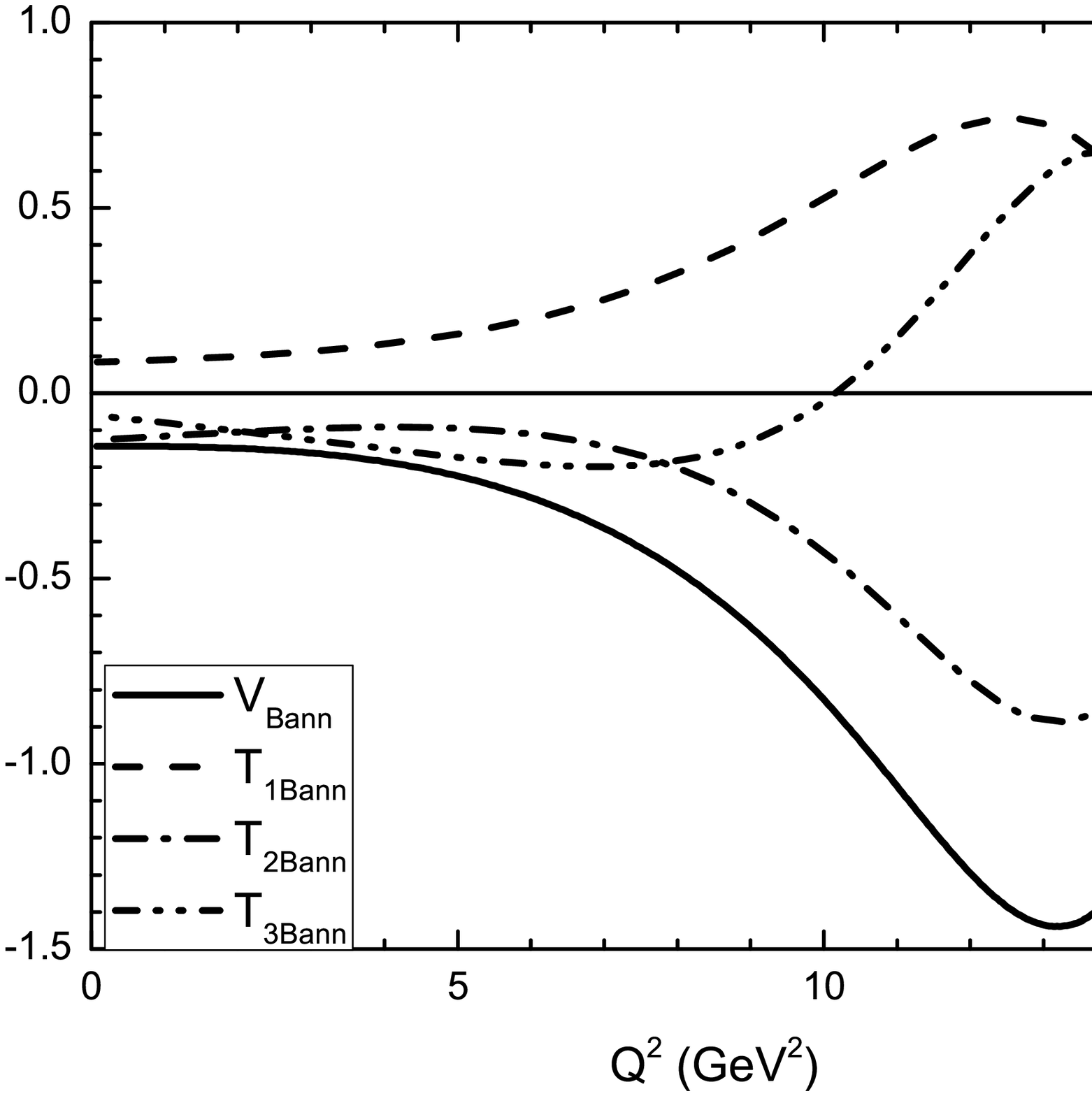}}\\
\caption{Form-factors of $B_c\rightarrow D^* l\bar{l}$.}
\end{figure}
\begin{figure}[htbp]
\centering \subfigure[Form-factors of $W^{\mu}$ induced by $Z^0$
penguin and box diagrams]{\includegraphics[width =
0.49\textwidth,height=0.4\textheight]{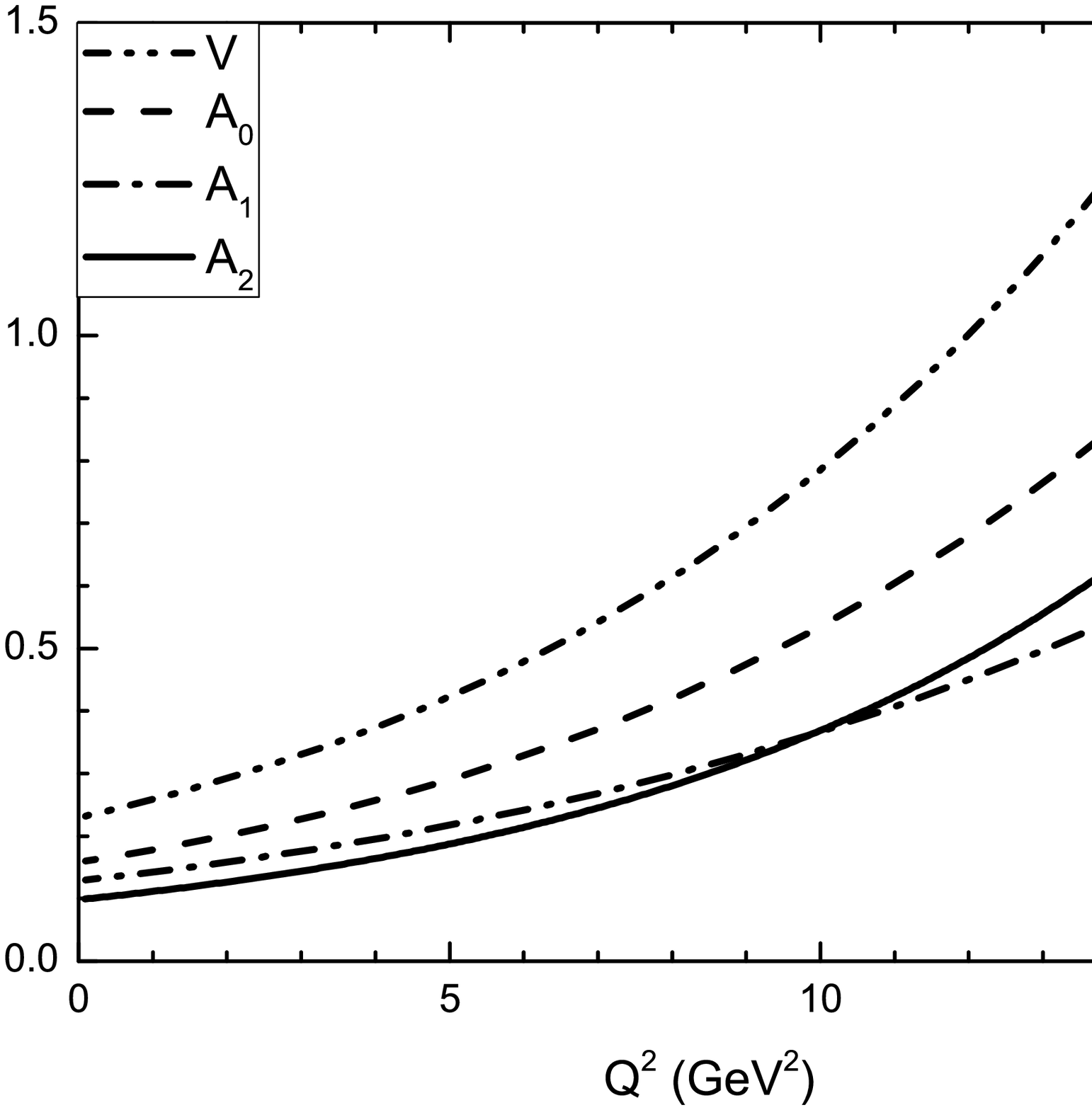}}
\subfigure[Form-factors of $W^{\mu}_T$ induced by $\gamma$ penguin diagrams]{\includegraphics[width = 0.49\textwidth,height=0.4\textheight]{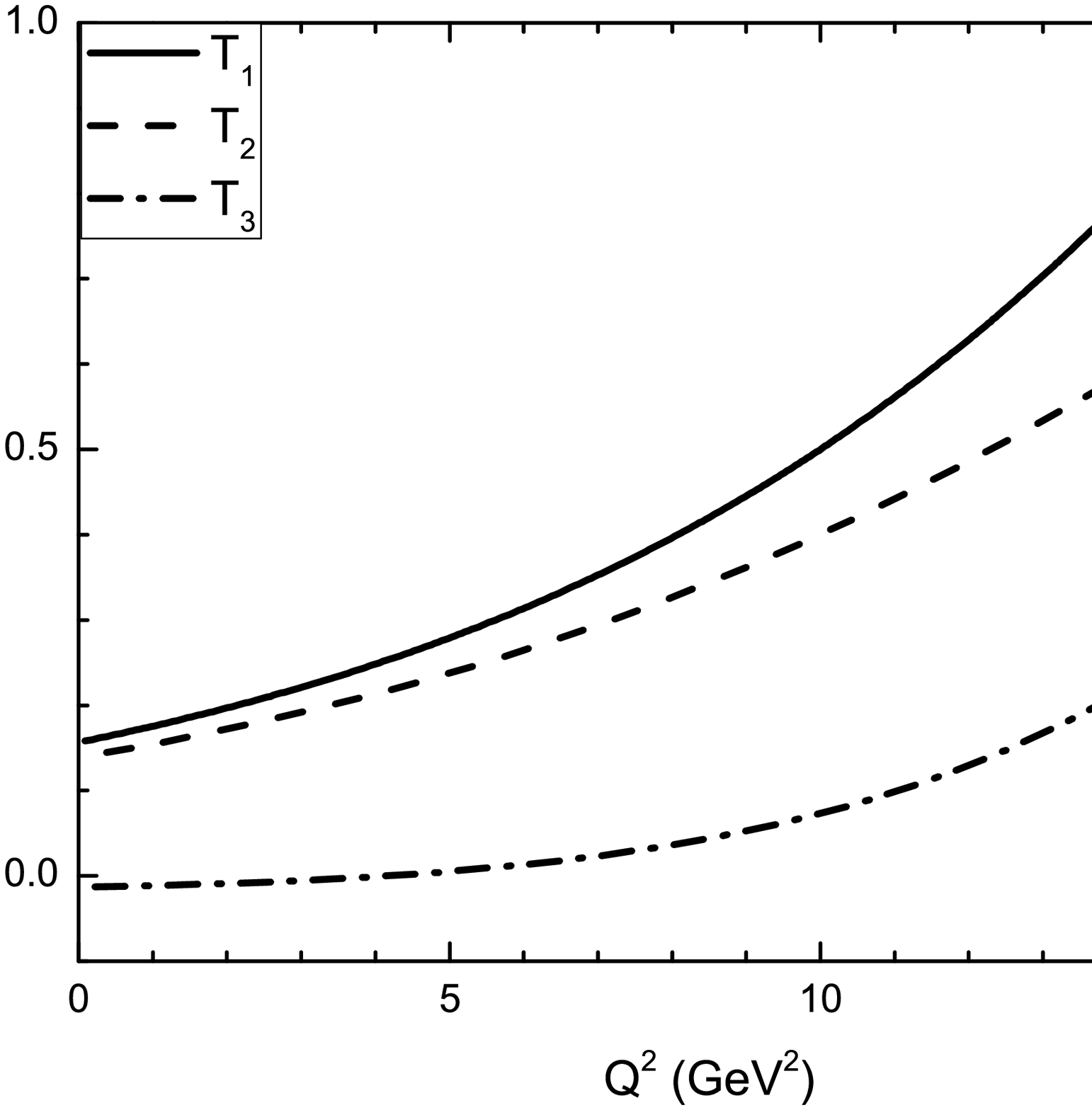}}\\
\subfigure[Real parts of $W^{\mu}_{ann}$ form-factors induced by
annihilation diagrams]{\includegraphics[width =
0.49\textwidth,height=0.4\textheight]{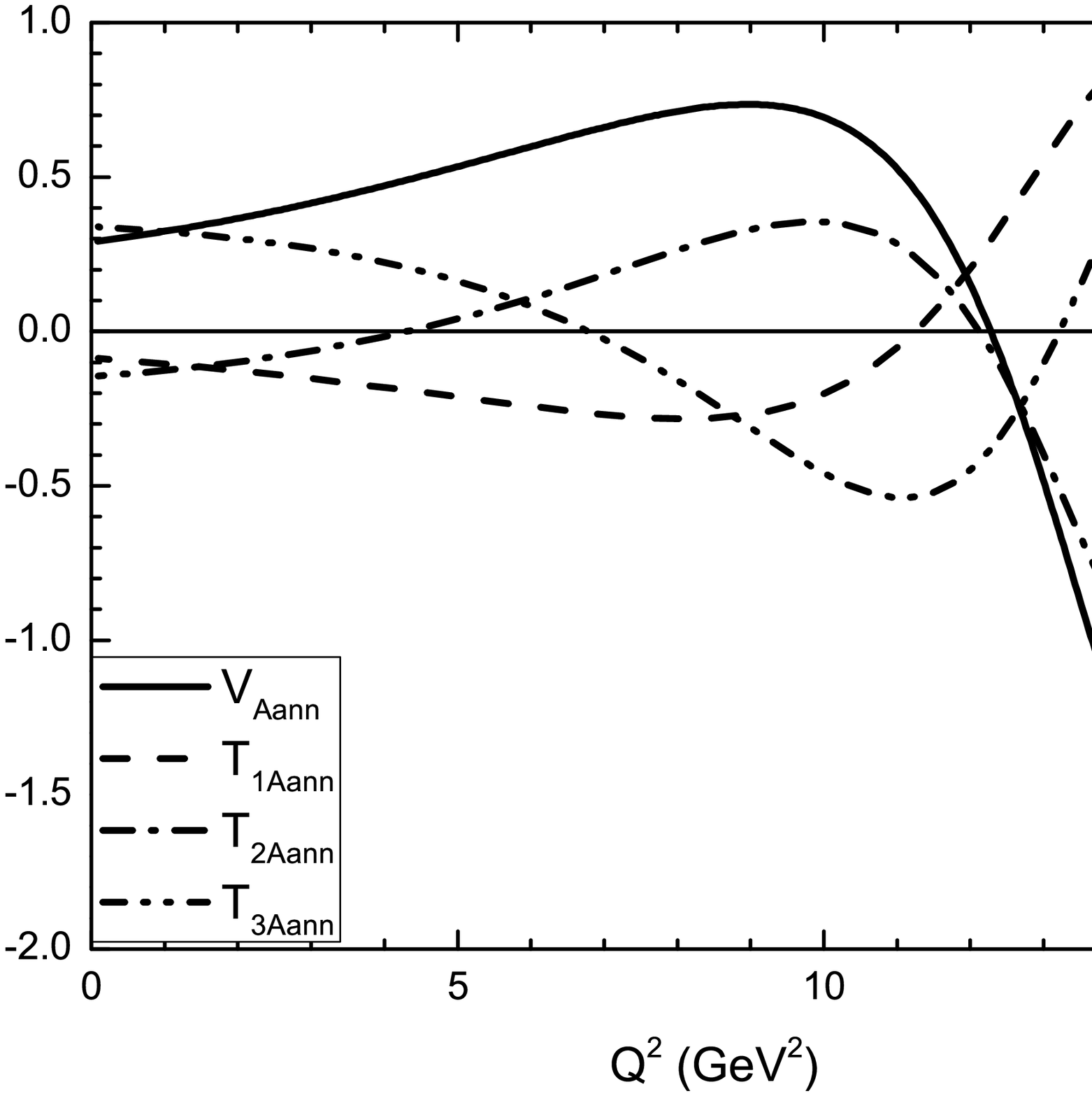}}
\subfigure[Imaginary parts of $W^{\mu}_{ann}$ form-factors induced
by annihilation
diagrams]{\includegraphics[width = 0.49\textwidth,height=0.4\textheight]{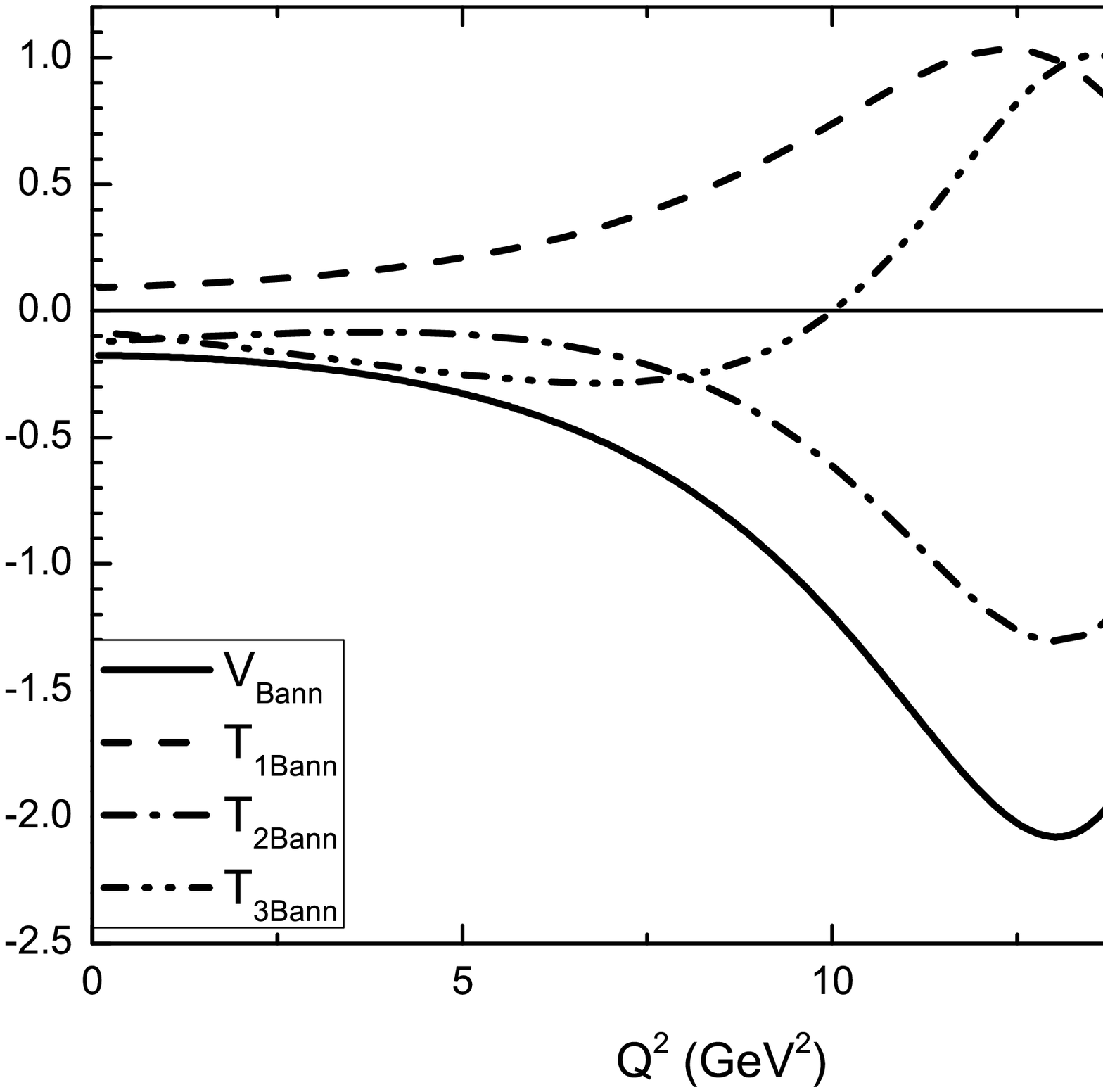}}\\
\caption{Form-factors of $B_c\rightarrow D_s^* l\bar{l}$.}
\end{figure}

\begin{figure}[htbp]
\centering
\subfigure[Differential branching fraction]{\includegraphics[width = 0.49\textwidth,height=0.4\textheight]{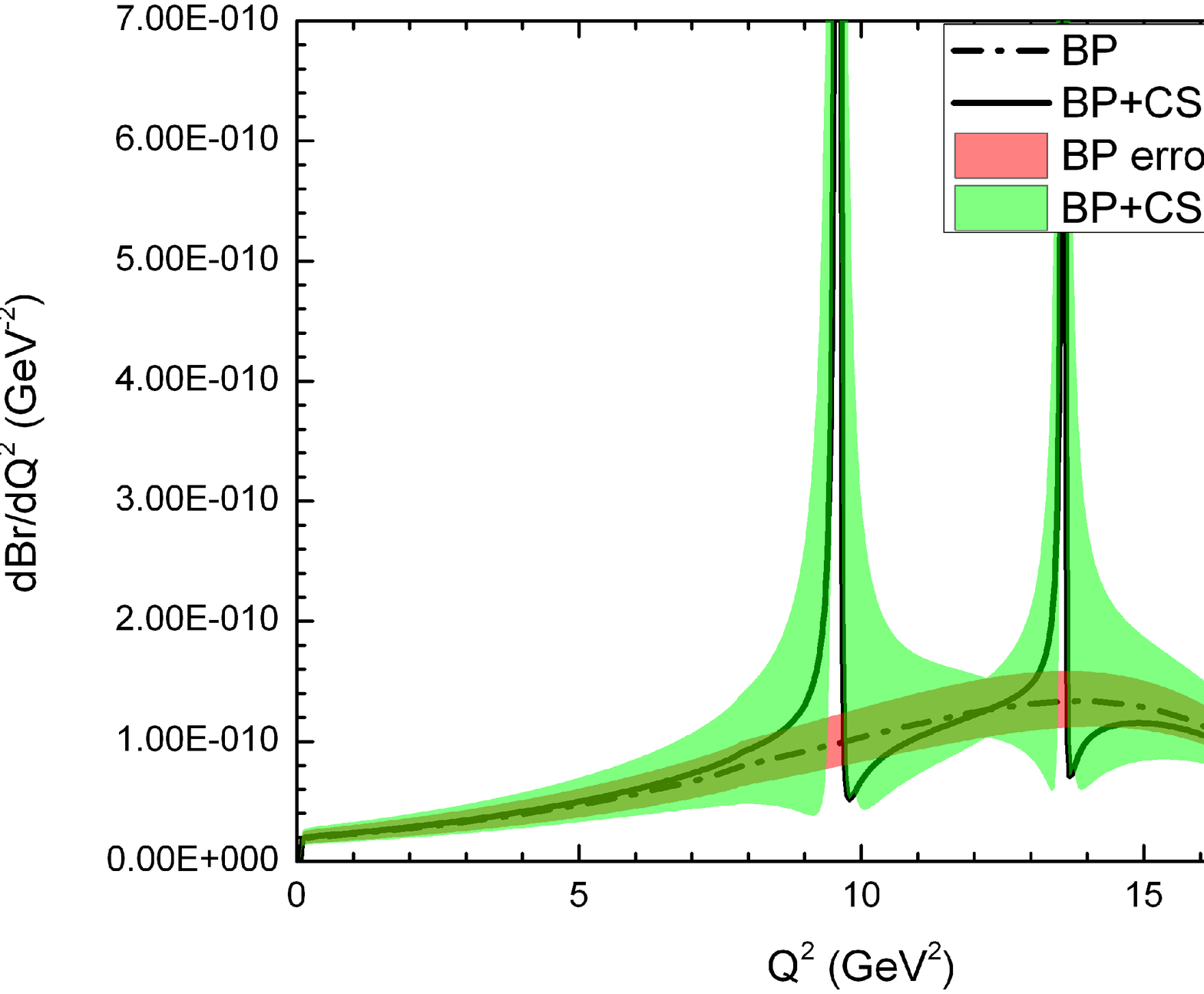}}
\subfigure[Differential branching fraction]{\includegraphics[width = 0.49\textwidth,height=0.4\textheight]{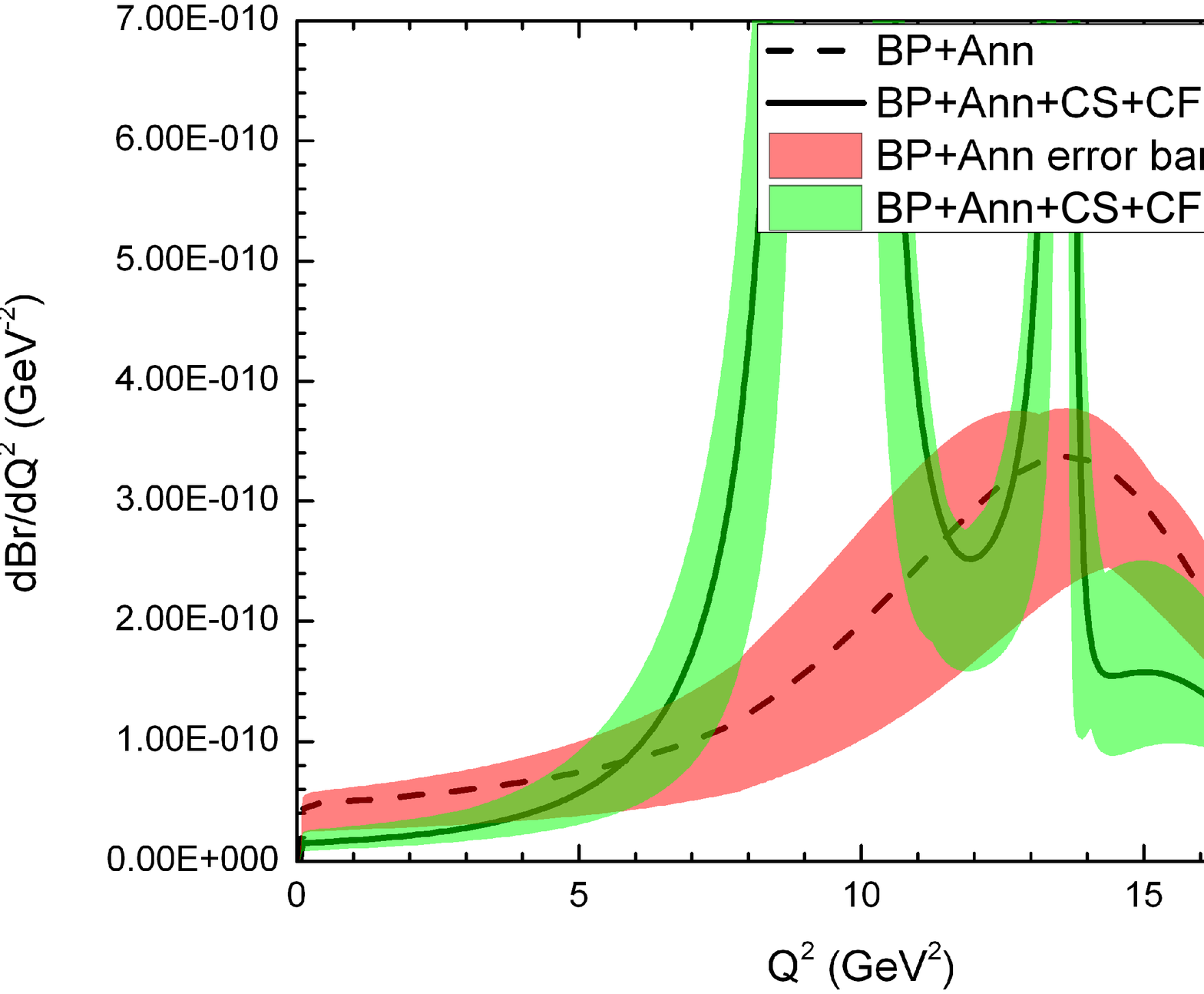}}\\
\subfigure[Leptonic longitudinal polarization asymmetry]{\includegraphics[width = 0.49\textwidth,height=0.4\textheight]{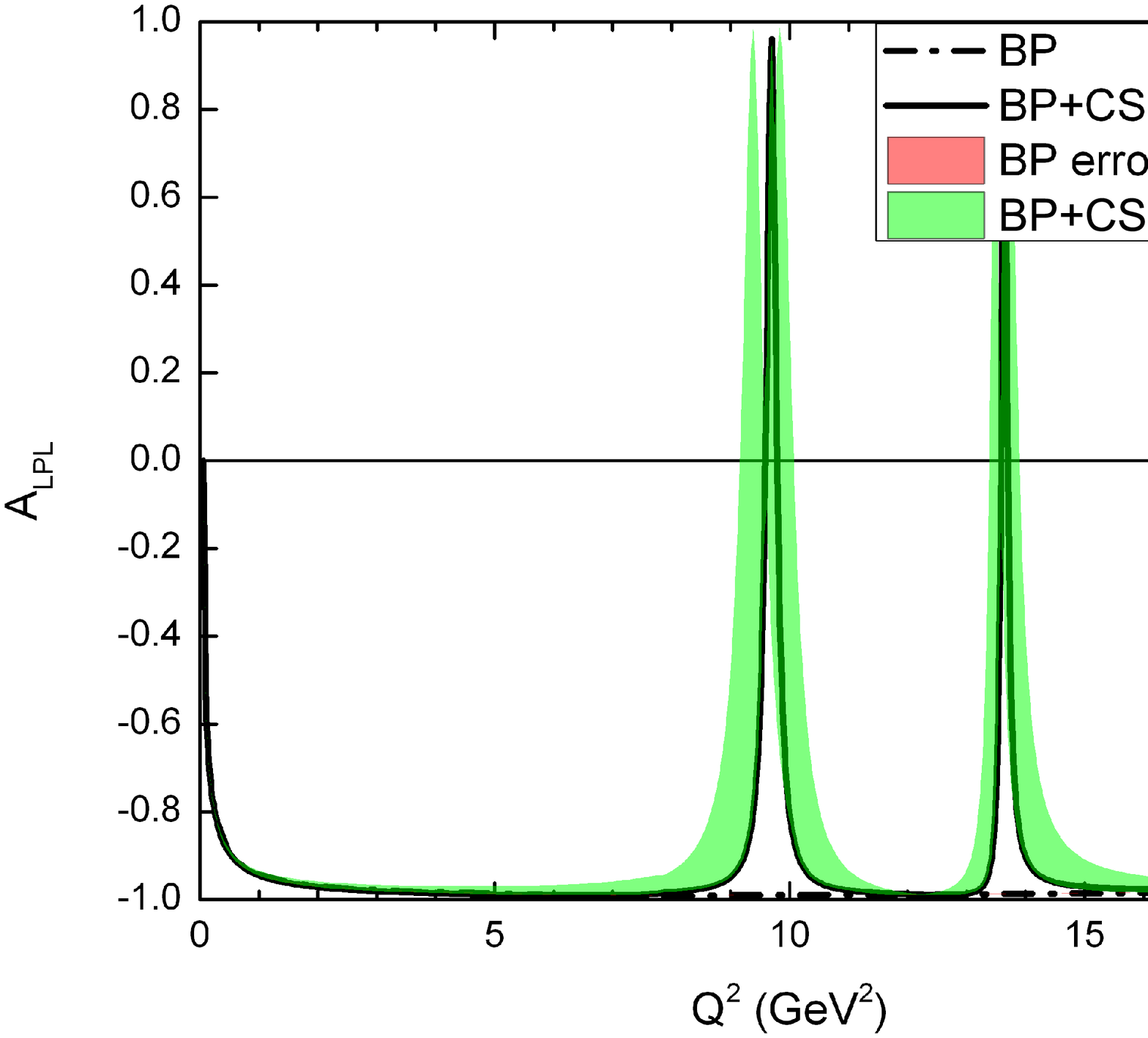}}
\subfigure[Leptonic longitudinal polarization asymmetry]{\includegraphics[width = 0.49\textwidth,height=0.4\textheight]{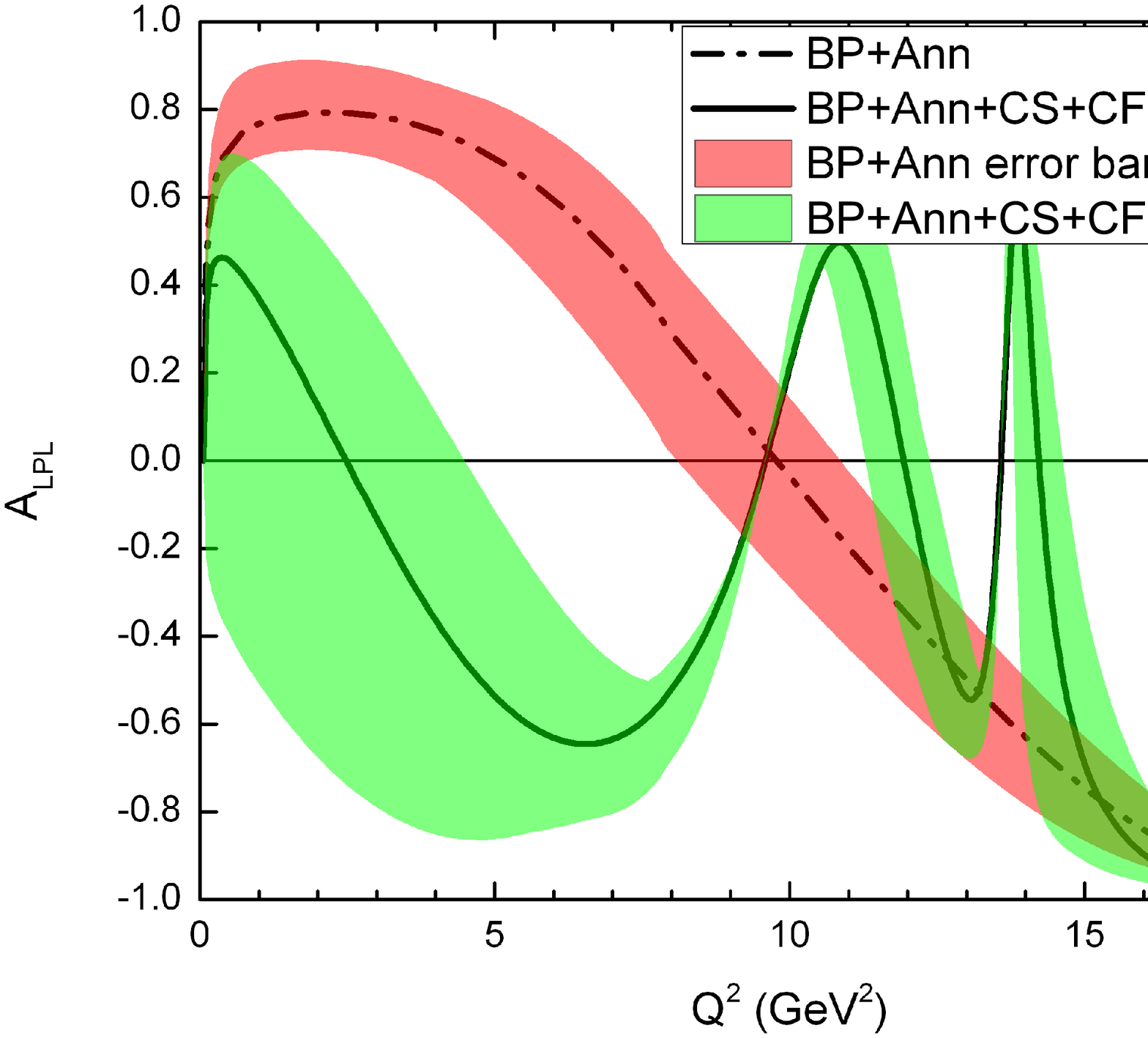}}
\caption{Observables of $B_c\rightarrow D \mu\bar{\mu}$. } 
\end{figure}

\begin{figure}[htbp]
\centering
\subfigure[Differential branching fraction]{\includegraphics[width = 0.49\textwidth,height=0.4\textheight]{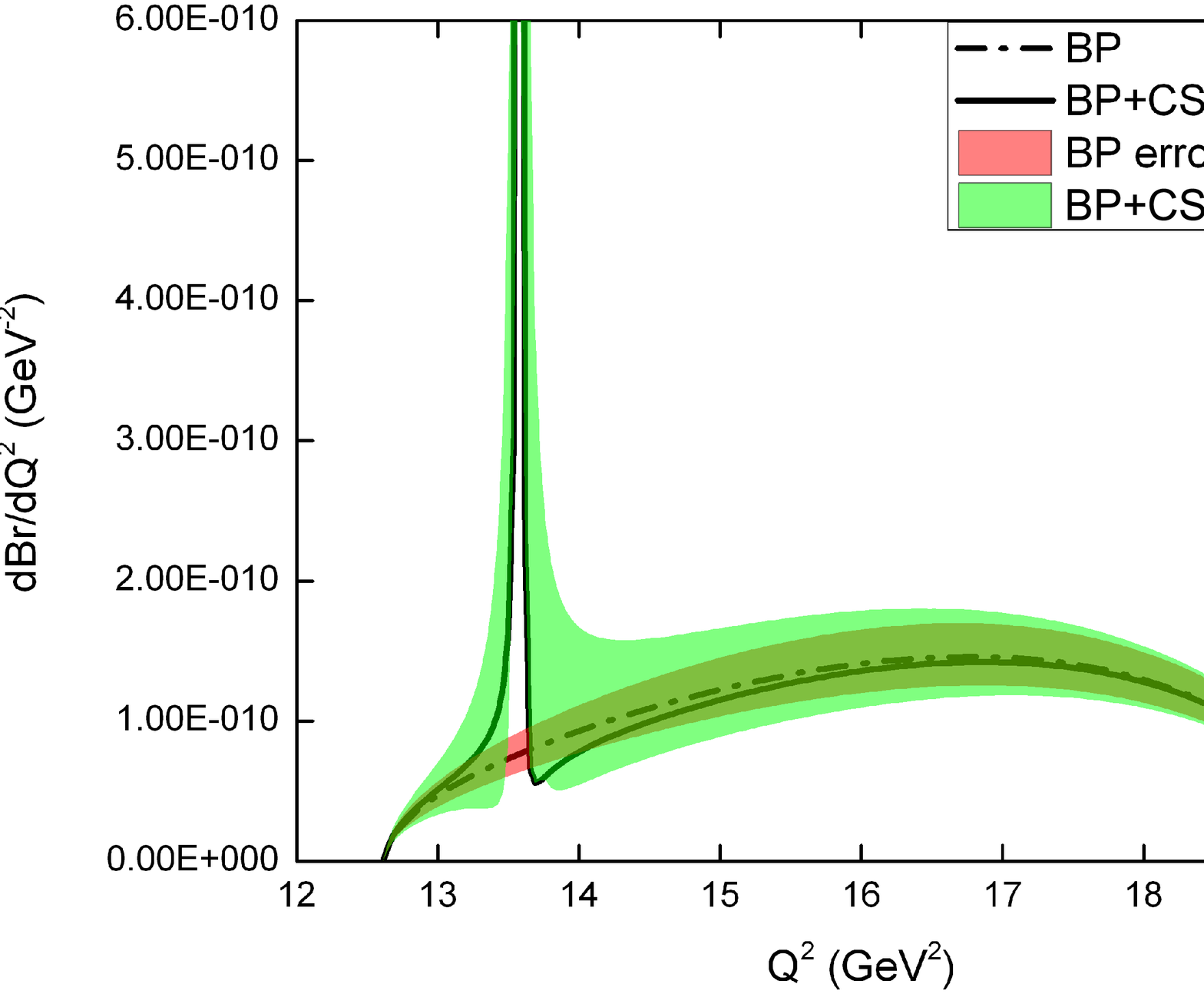}}
\subfigure[Differential branching fraction]{\includegraphics[width = 0.49\textwidth,height=0.4\textheight]{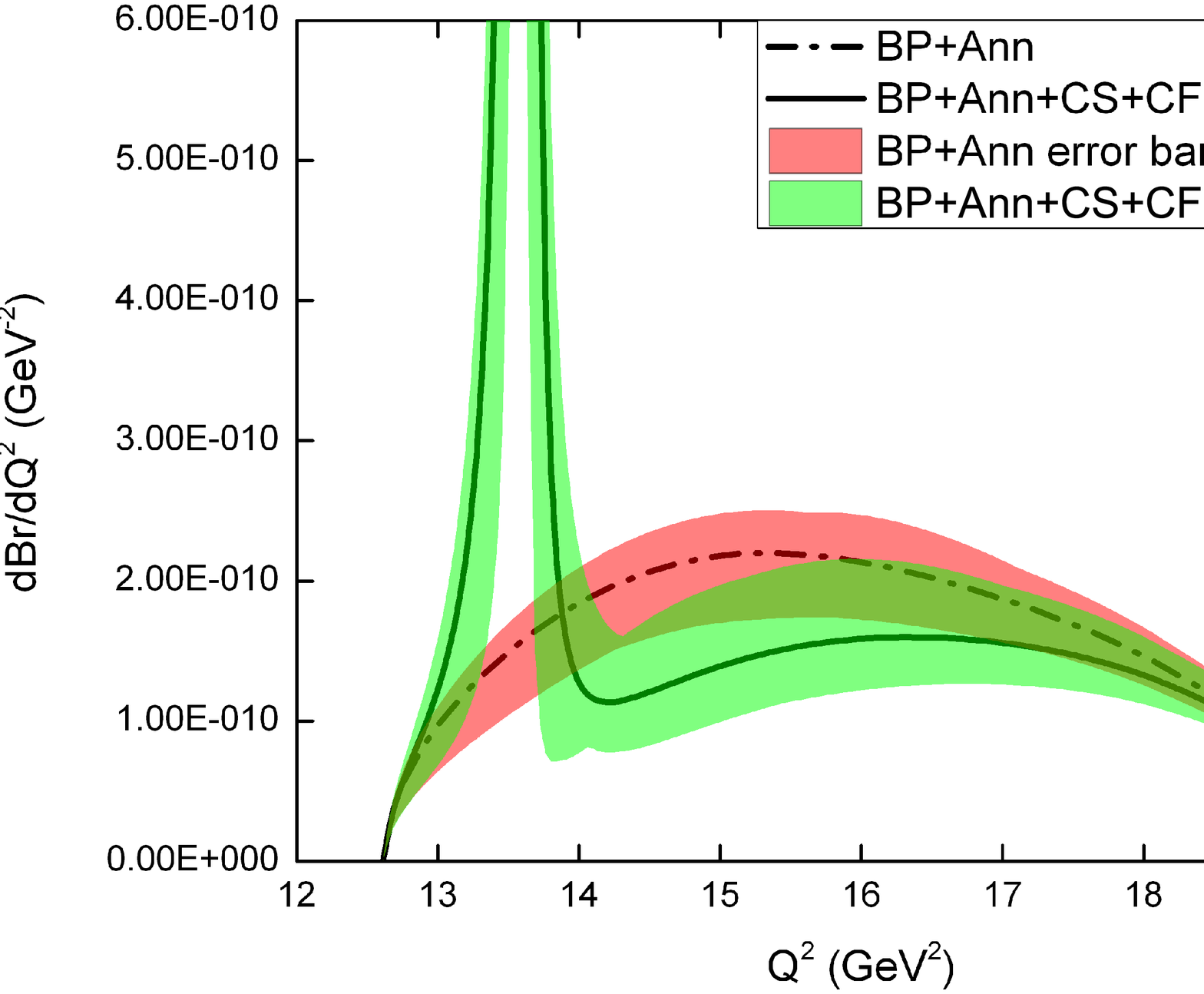}}\\
\subfigure[Leptonic longitudinal polarization asymmetry]{\includegraphics[width = 0.49\textwidth,height=0.4\textheight]{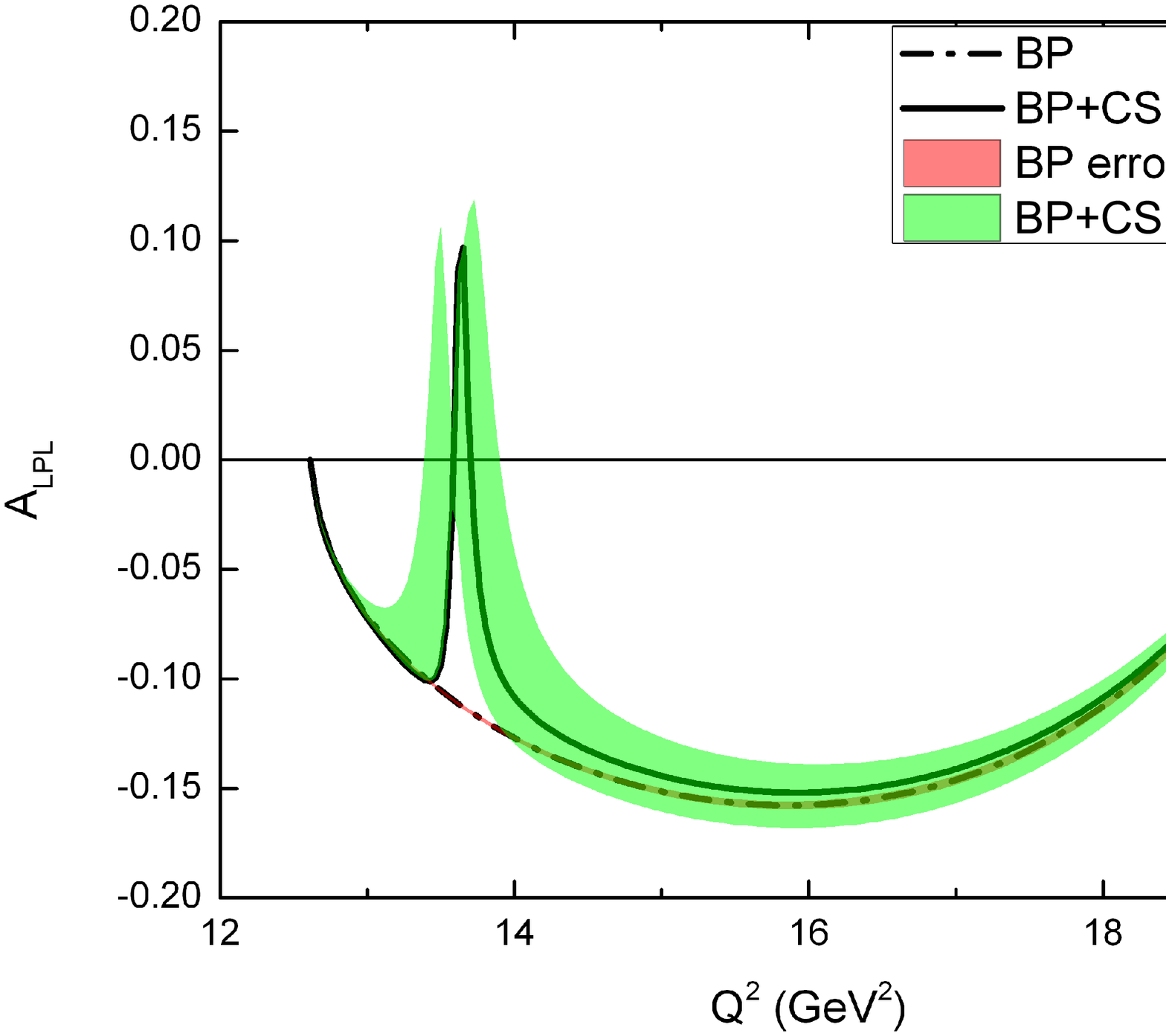}}
\subfigure[Leptonic longitudinal polarization asymmetry]{\includegraphics[width = 0.49\textwidth,height=0.4\textheight]{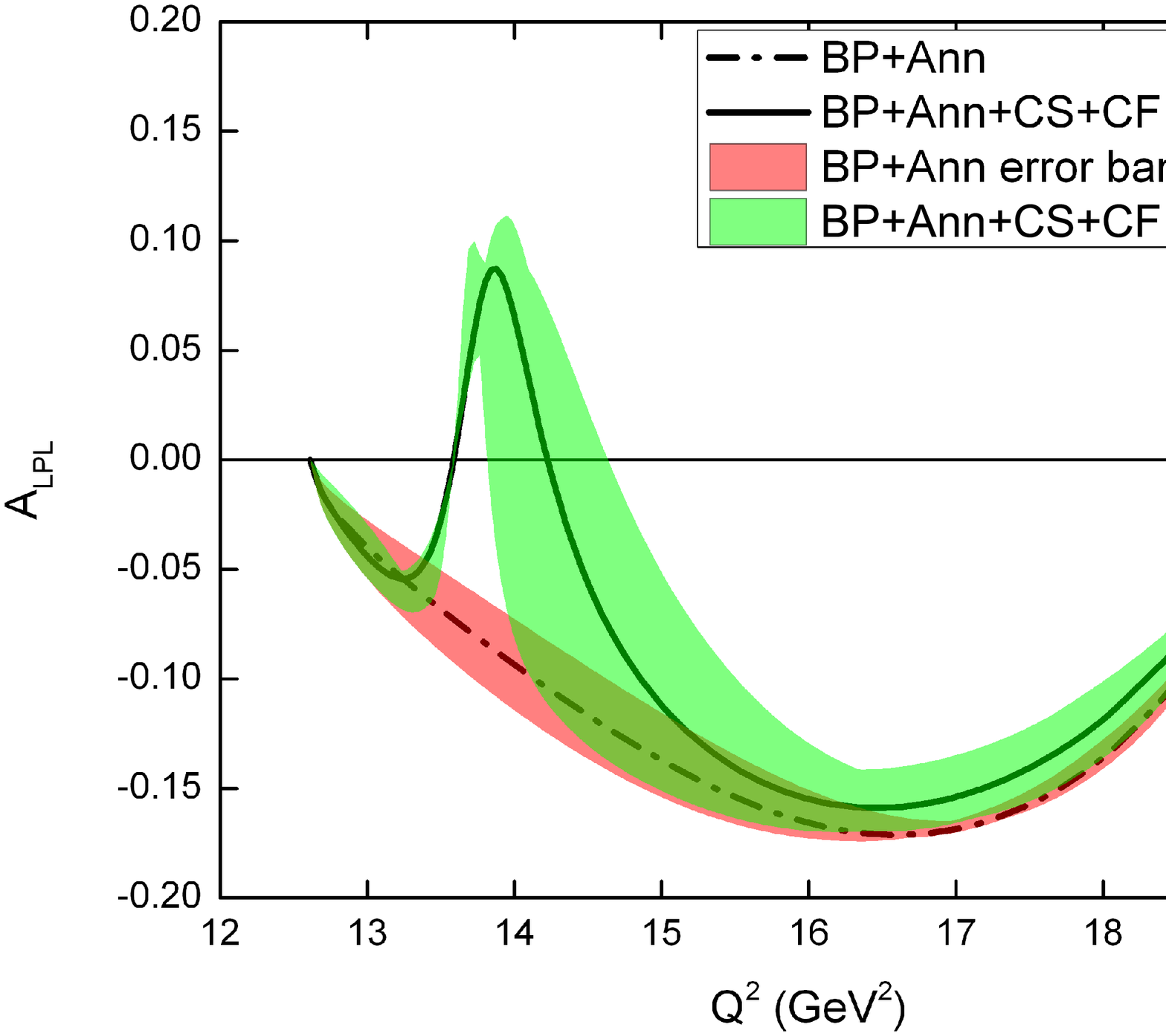}}
\caption{Observables of $B_c\rightarrow D \tau\bar{\tau}$. } 
\end{figure}

\begin{figure}[htbp]
\centering
\subfigure[Differential branching fraction]{\includegraphics[width = 0.49\textwidth,height=0.4\textheight]{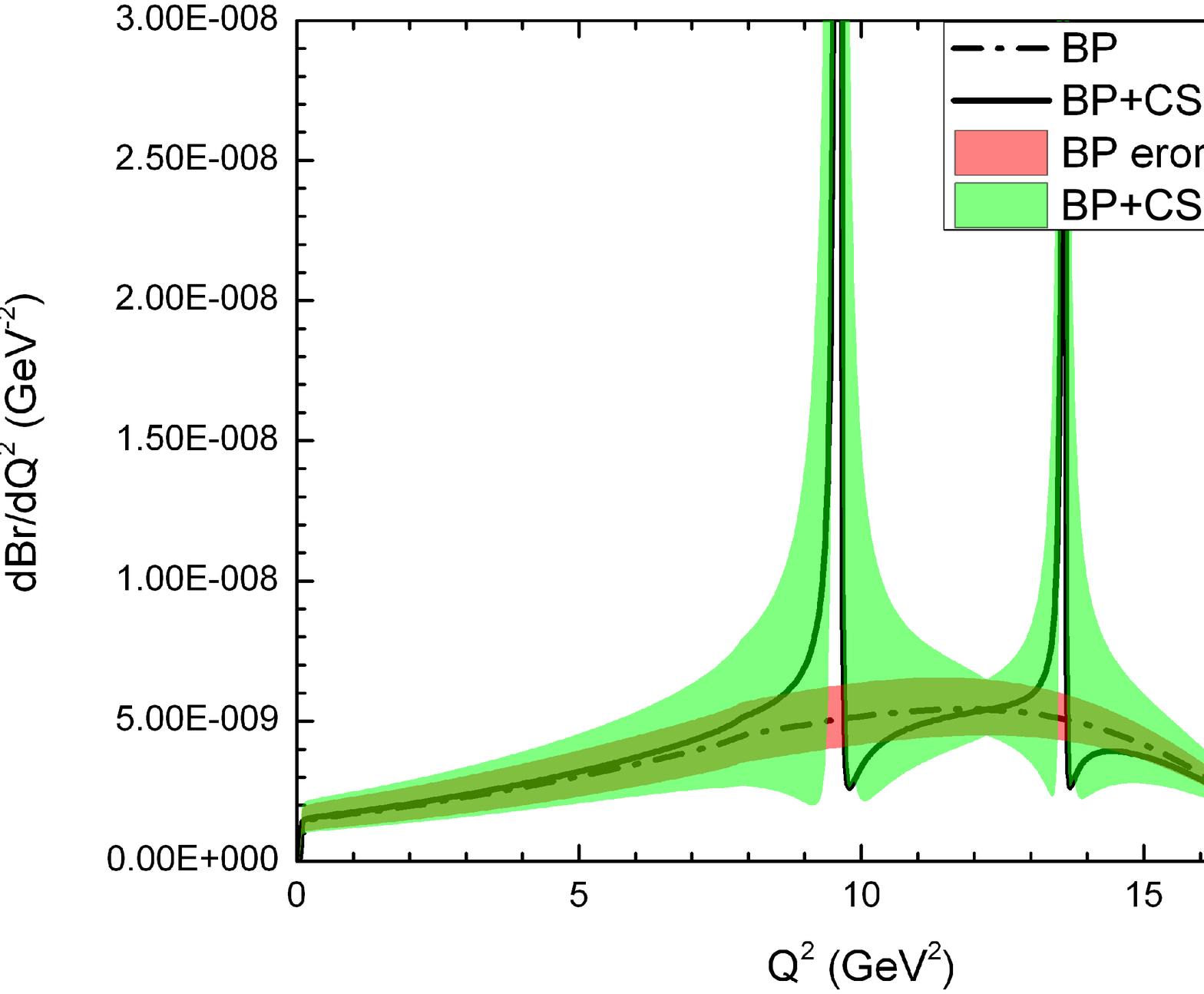}}
\subfigure[Differential branching fraction]{\includegraphics[width = 0.49\textwidth,height=0.4\textheight]{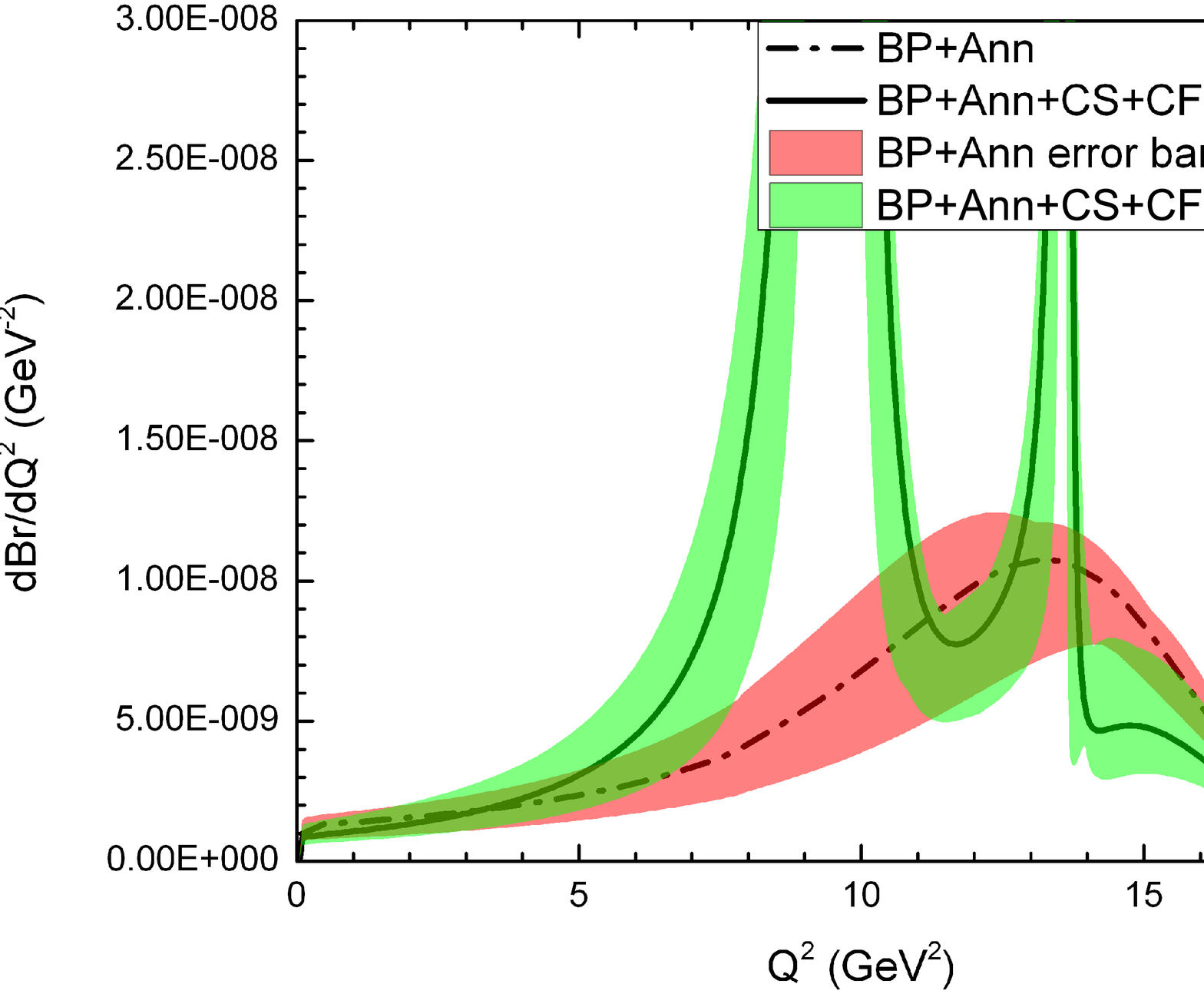}}\\
\subfigure[Leptonic longitudinal polarization asymmetry]{\includegraphics[width = 0.49\textwidth,height=0.4\textheight]{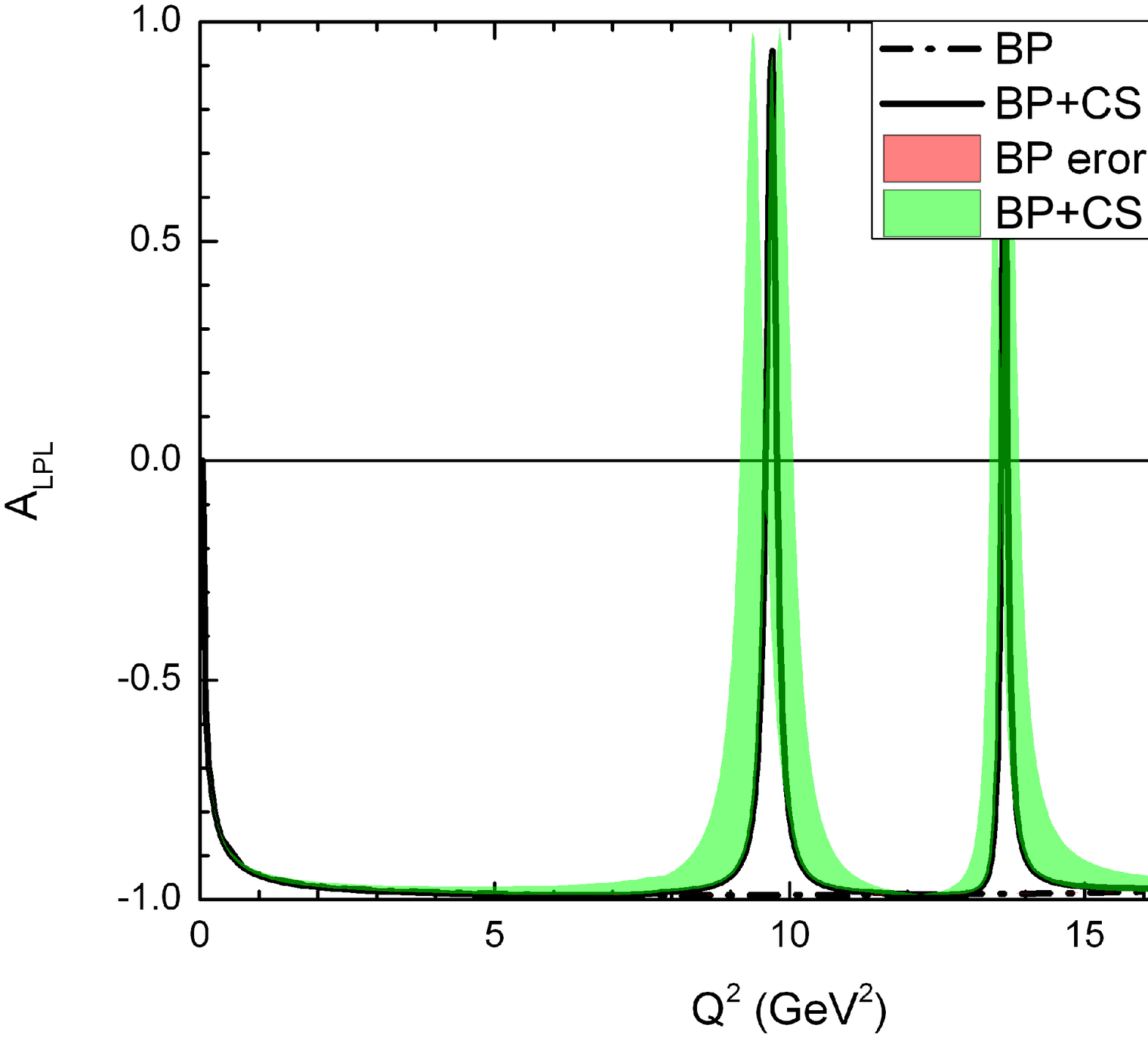}}
\subfigure[Leptonic longitudinal polarization asymmetry]{\includegraphics[width = 0.49\textwidth,height=0.4\textheight]{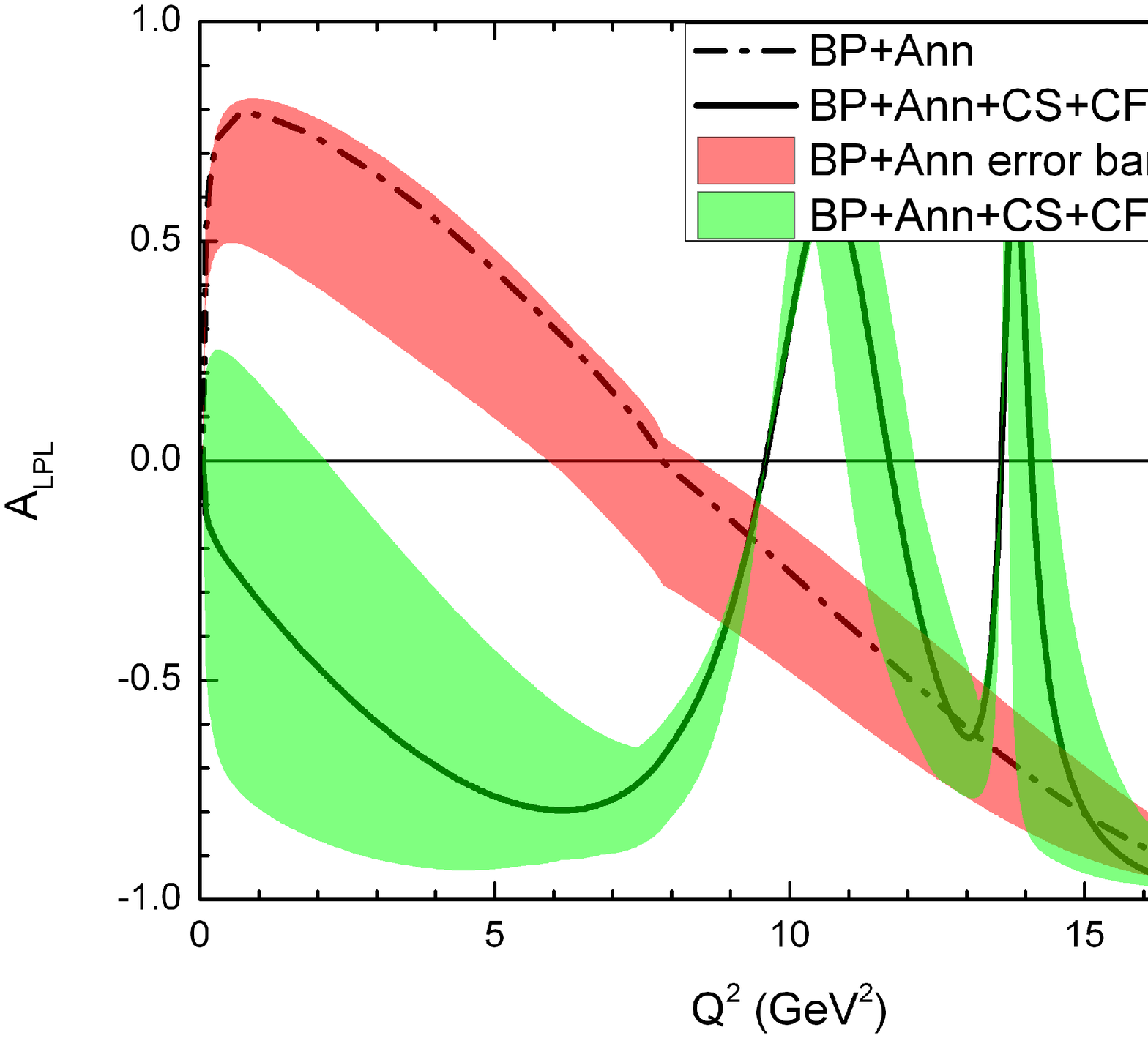}}
\caption{Observables of $B_c\rightarrow D_s \mu\bar{\mu}$. } 
\end{figure}

\begin{figure}[htbp]
\centering
\subfigure[Differential branching fraction]{\includegraphics[width = 0.49\textwidth,height=0.4\textheight]{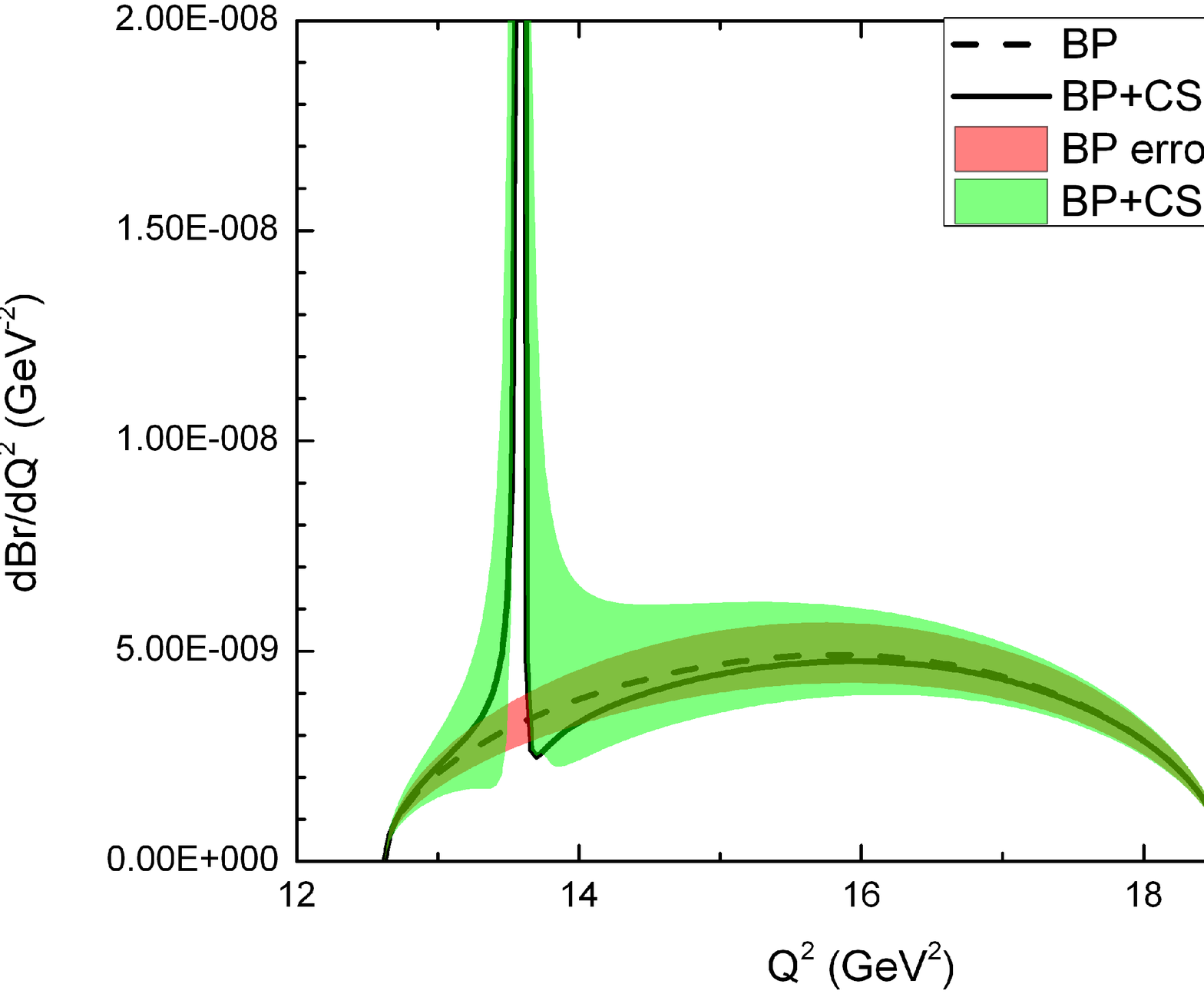}}
\subfigure[Differential branching fraction]{\includegraphics[width = 0.49\textwidth,height=0.4\textheight]{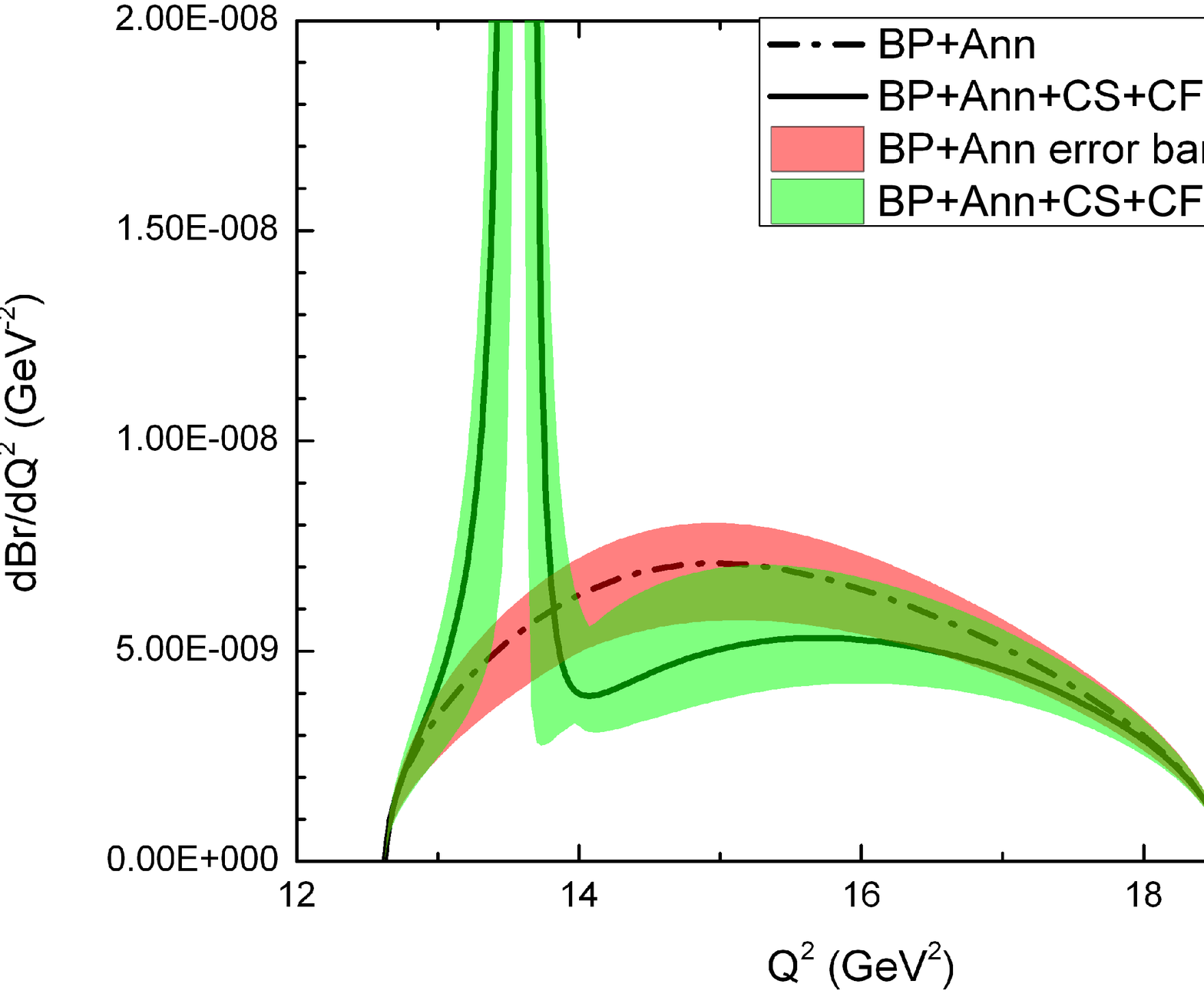}}\\
\subfigure[Leptonic longitudinal polarization asymmetry]{\includegraphics[width = 0.49\textwidth,height=0.4\textheight]{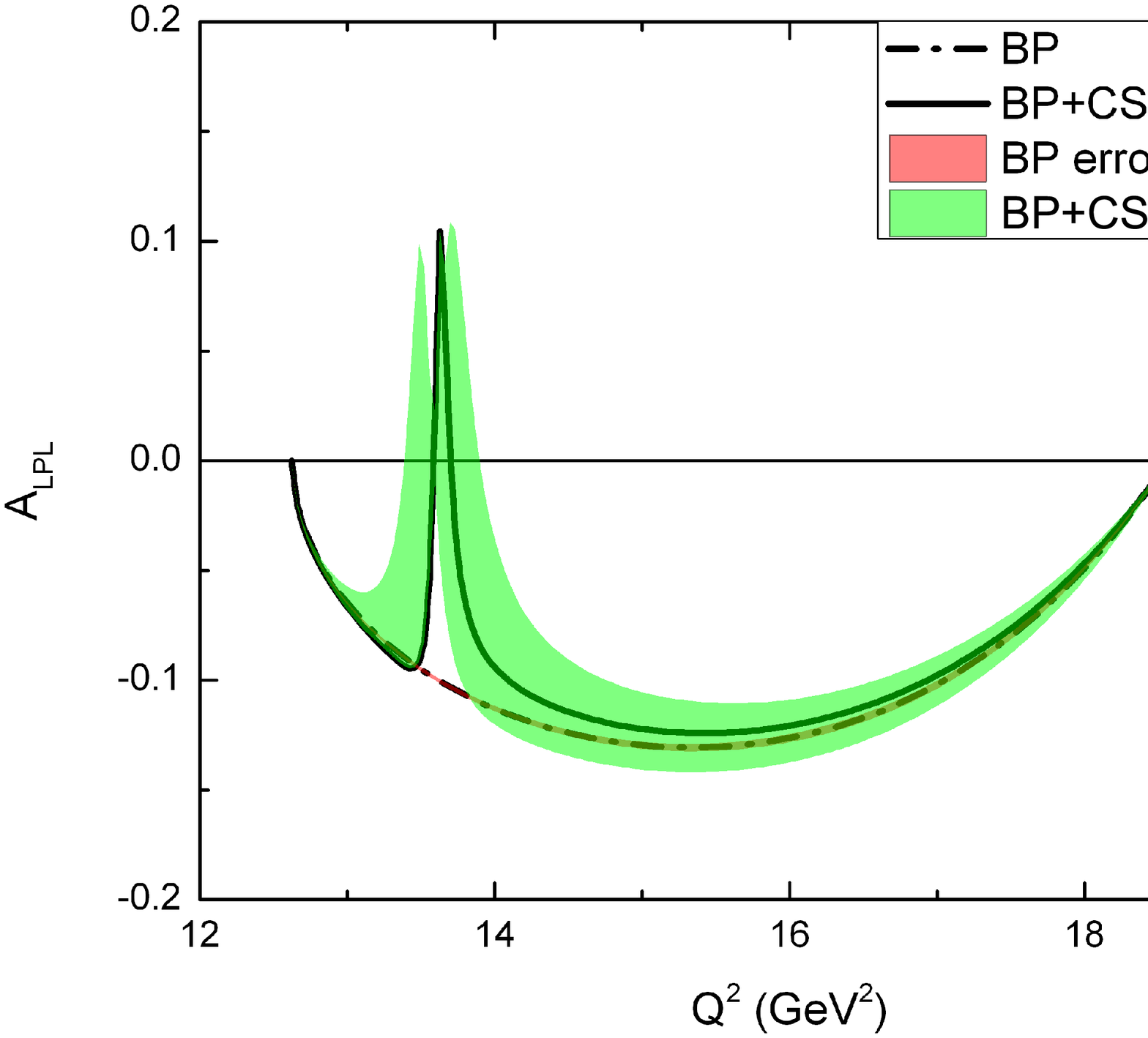}}
\subfigure[Leptonic longitudinal polarization asymmetry]{\includegraphics[width = 0.49\textwidth,height=0.4\textheight]{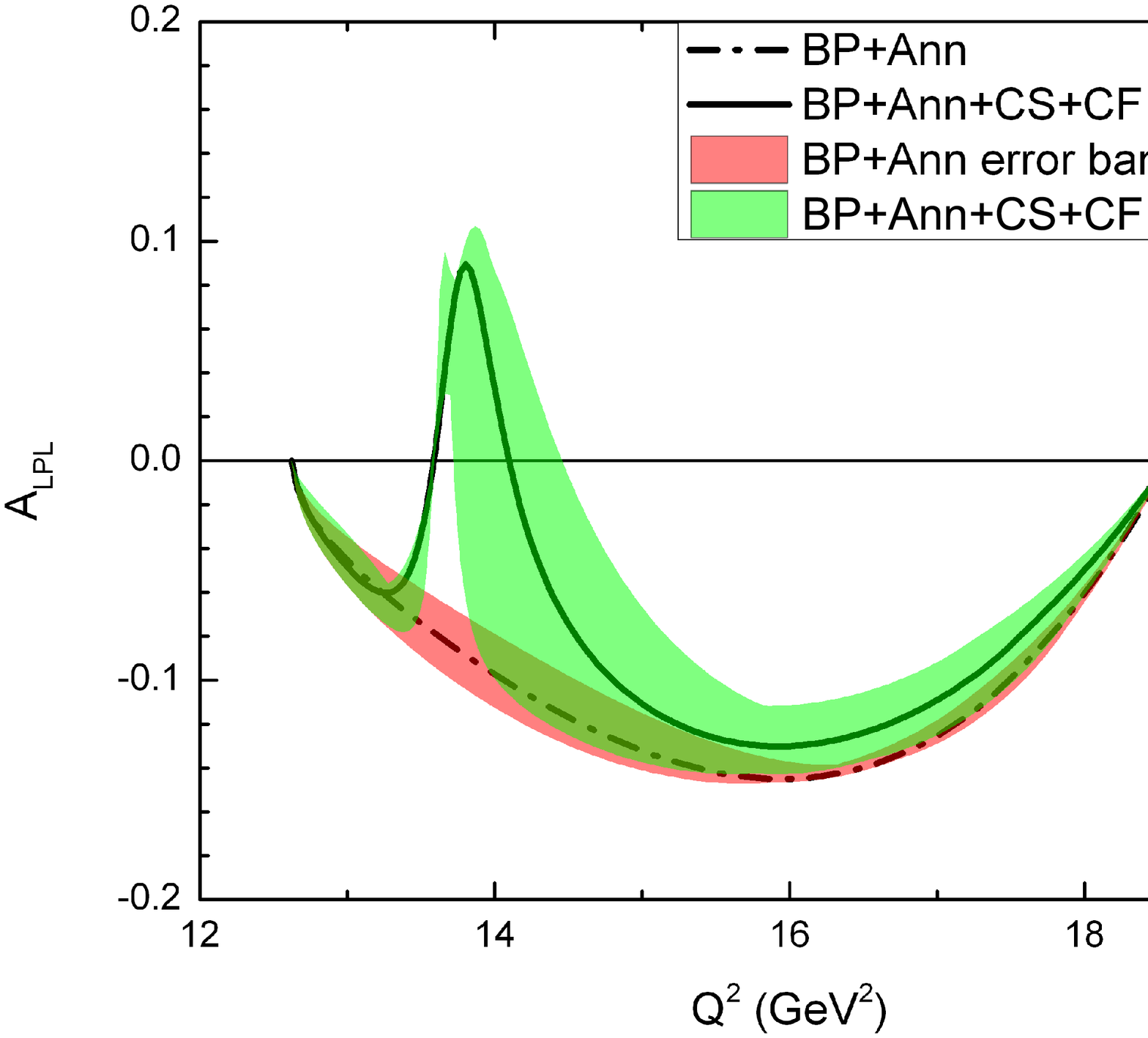}}
\caption{Observables of $B_c\rightarrow D_s \tau\bar{\tau}$.} 
\end{figure}

\begin{figure}[htbp]
\centering
\subfigure[Differential branching fraction]{\includegraphics[width = 0.49\textwidth,height=0.2\textheight]{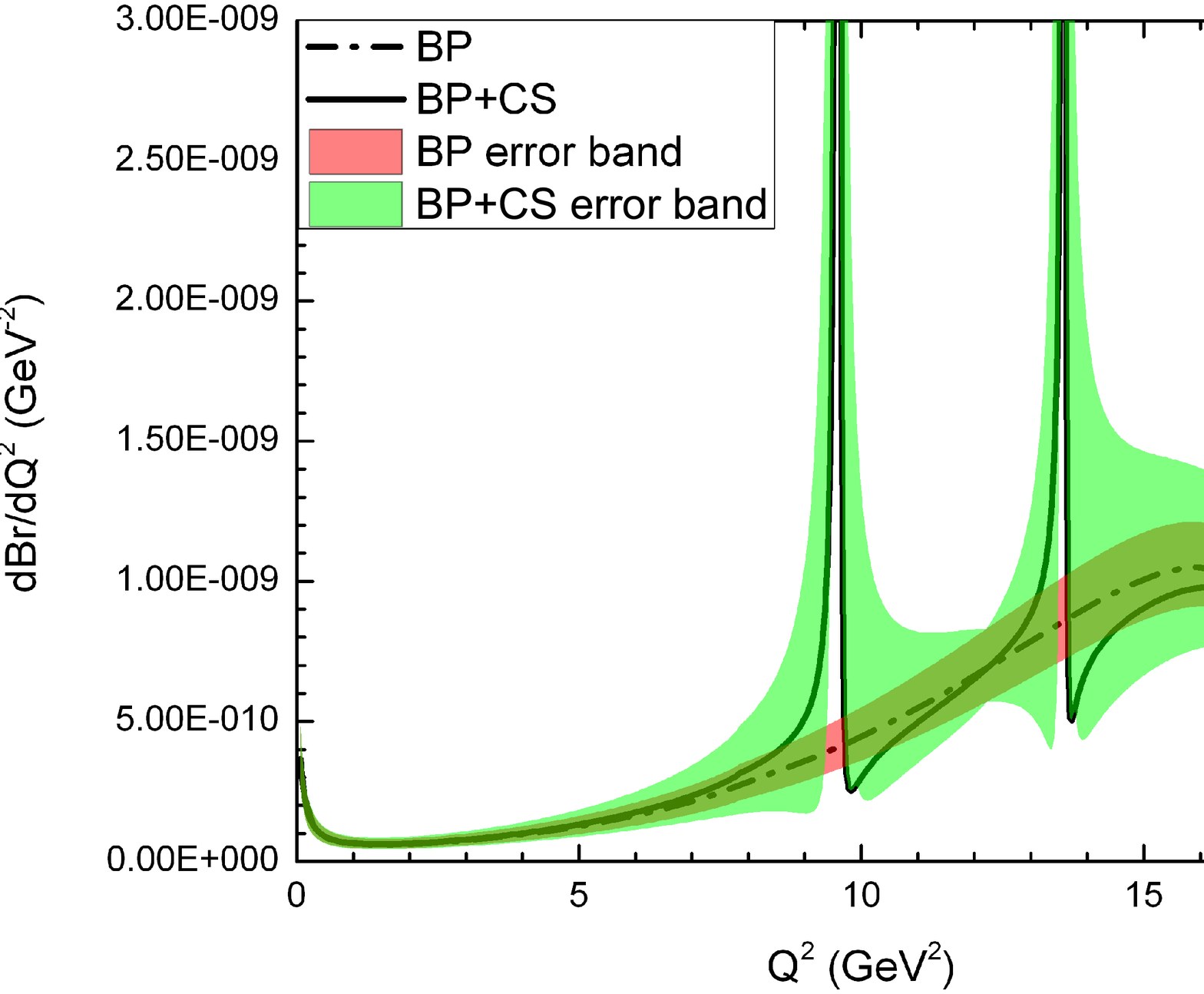}}
\subfigure[Differential branching fraction]{\includegraphics[width = 0.49\textwidth,height=0.2\textheight]{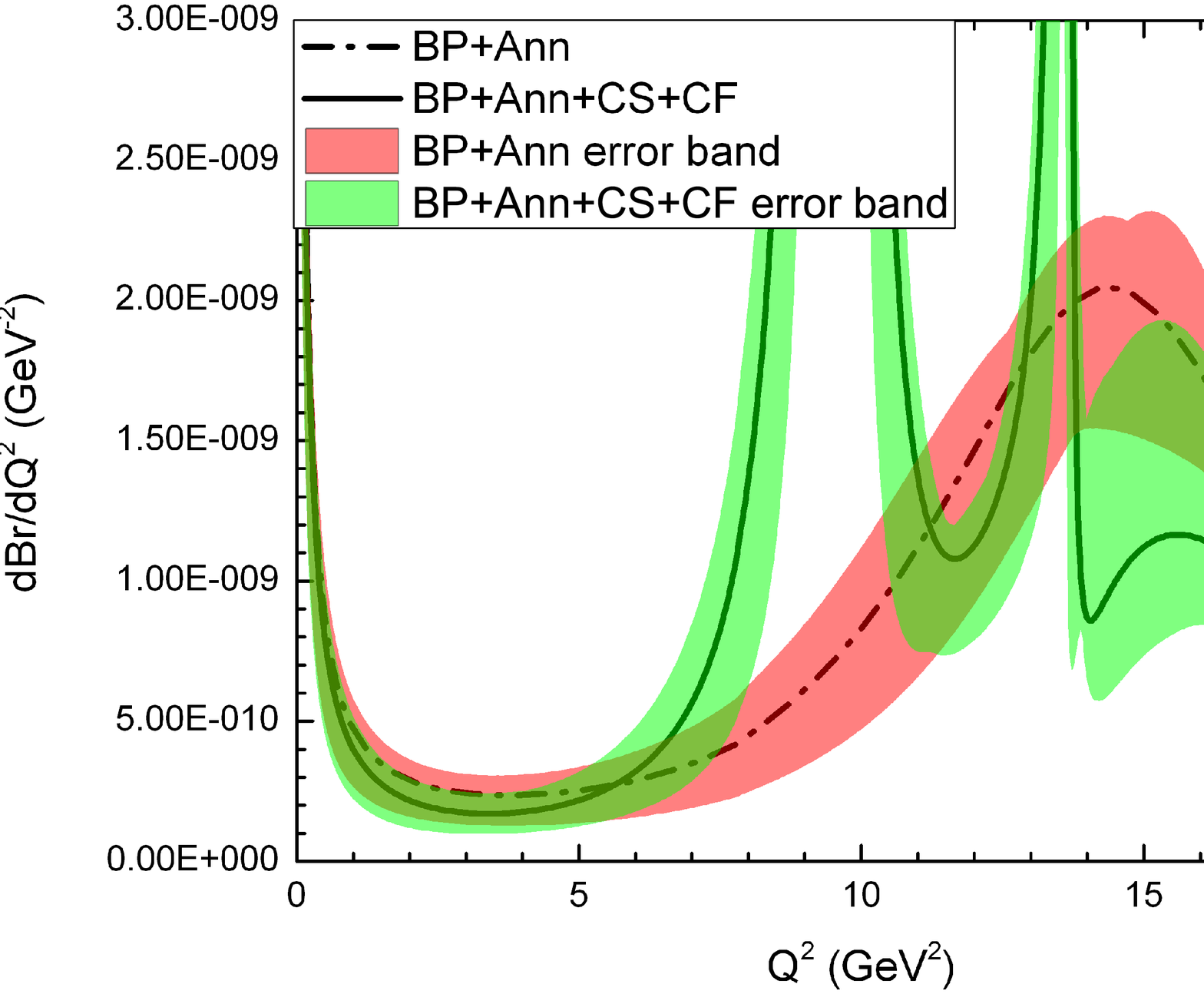}}\\
\subfigure[Forward-backward asymmetry]{\includegraphics[width = 0.49\textwidth,height=0.2\textheight]{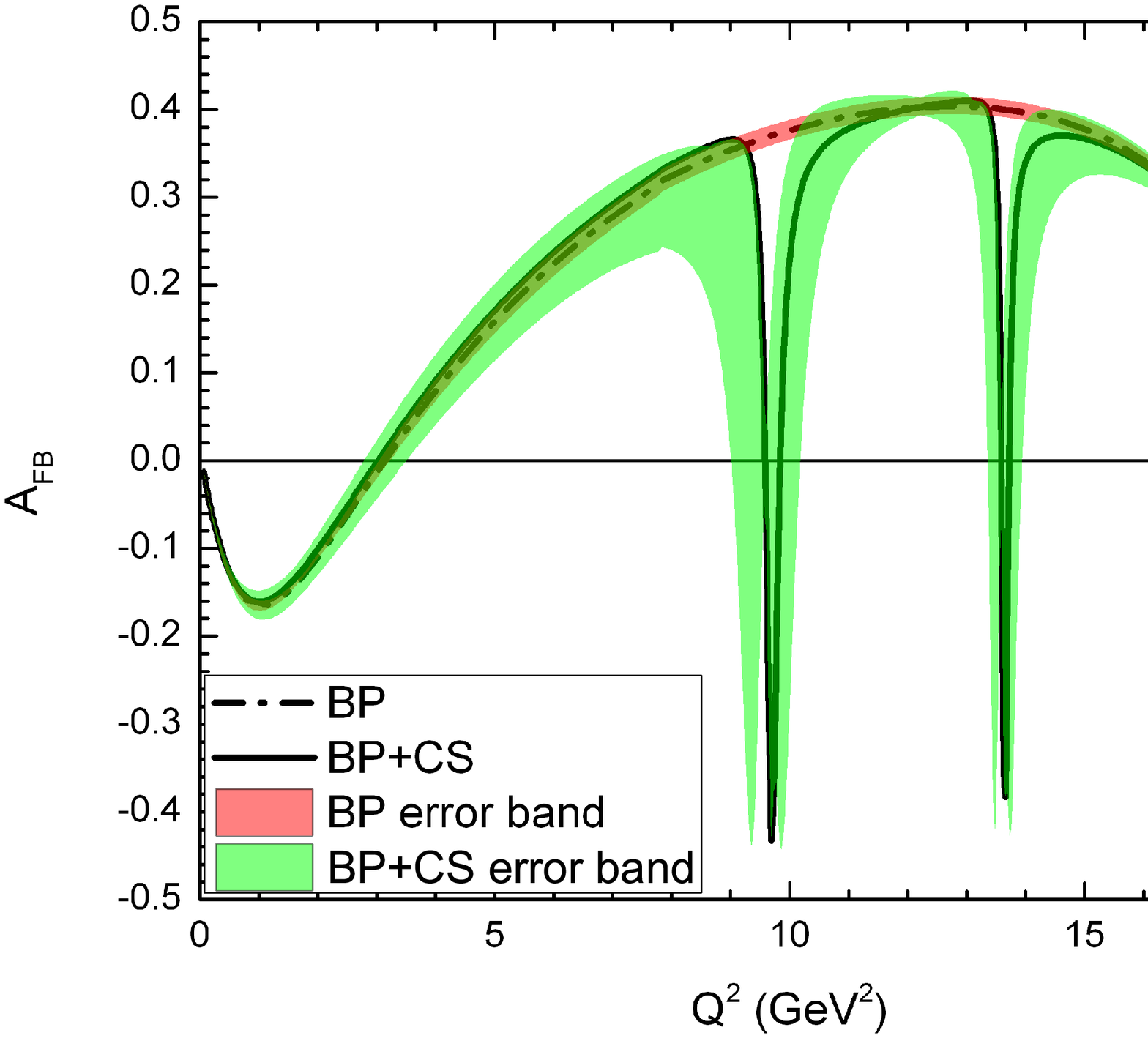}}
\subfigure[Forward-backward asymmetry]{\includegraphics[width = 0.49\textwidth,height=0.2\textheight]{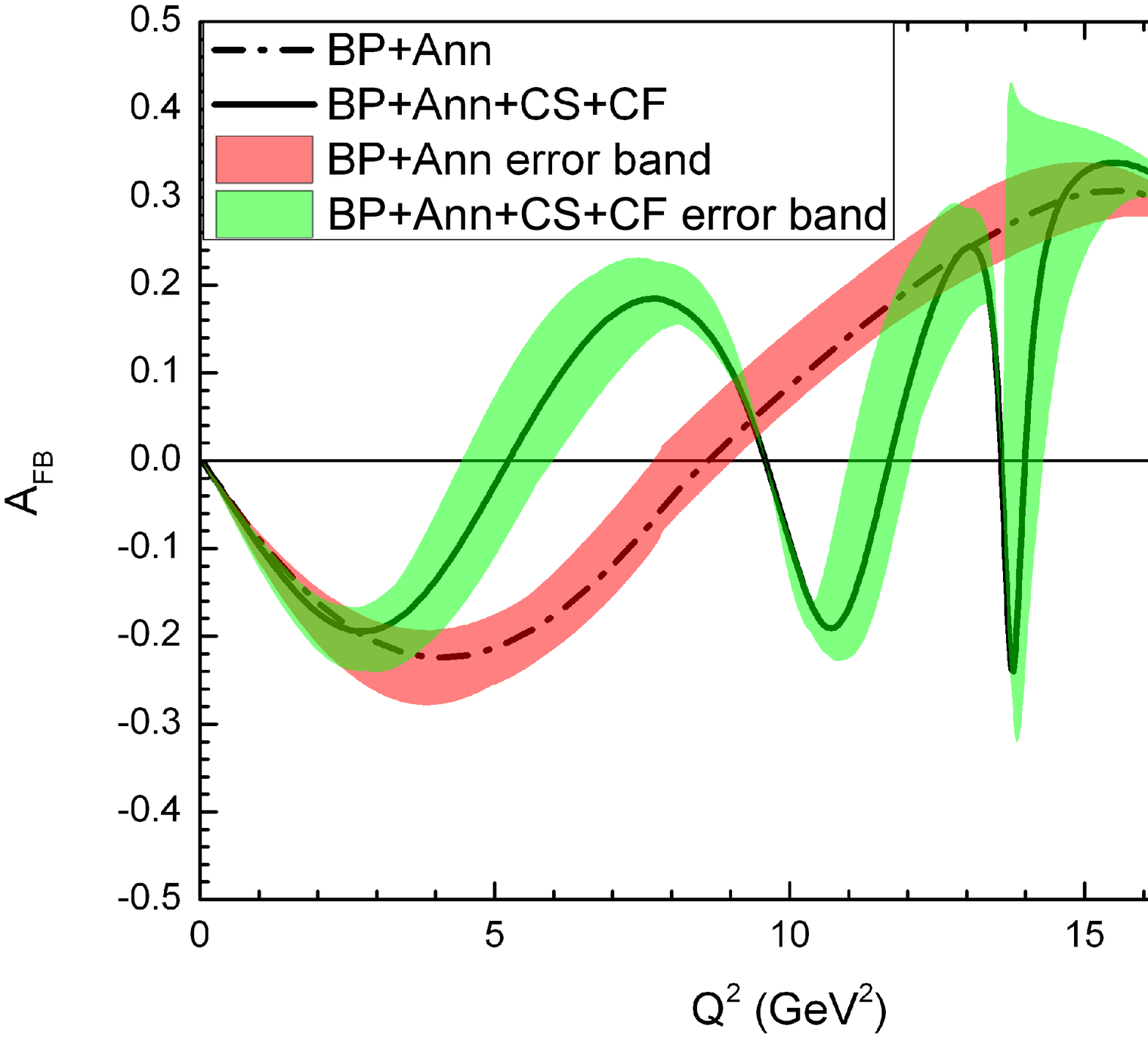}}\\
\subfigure[Longitudinal polarization of the final vector
meson]{\includegraphics[width =
0.49\textwidth,height=0.2\textheight]{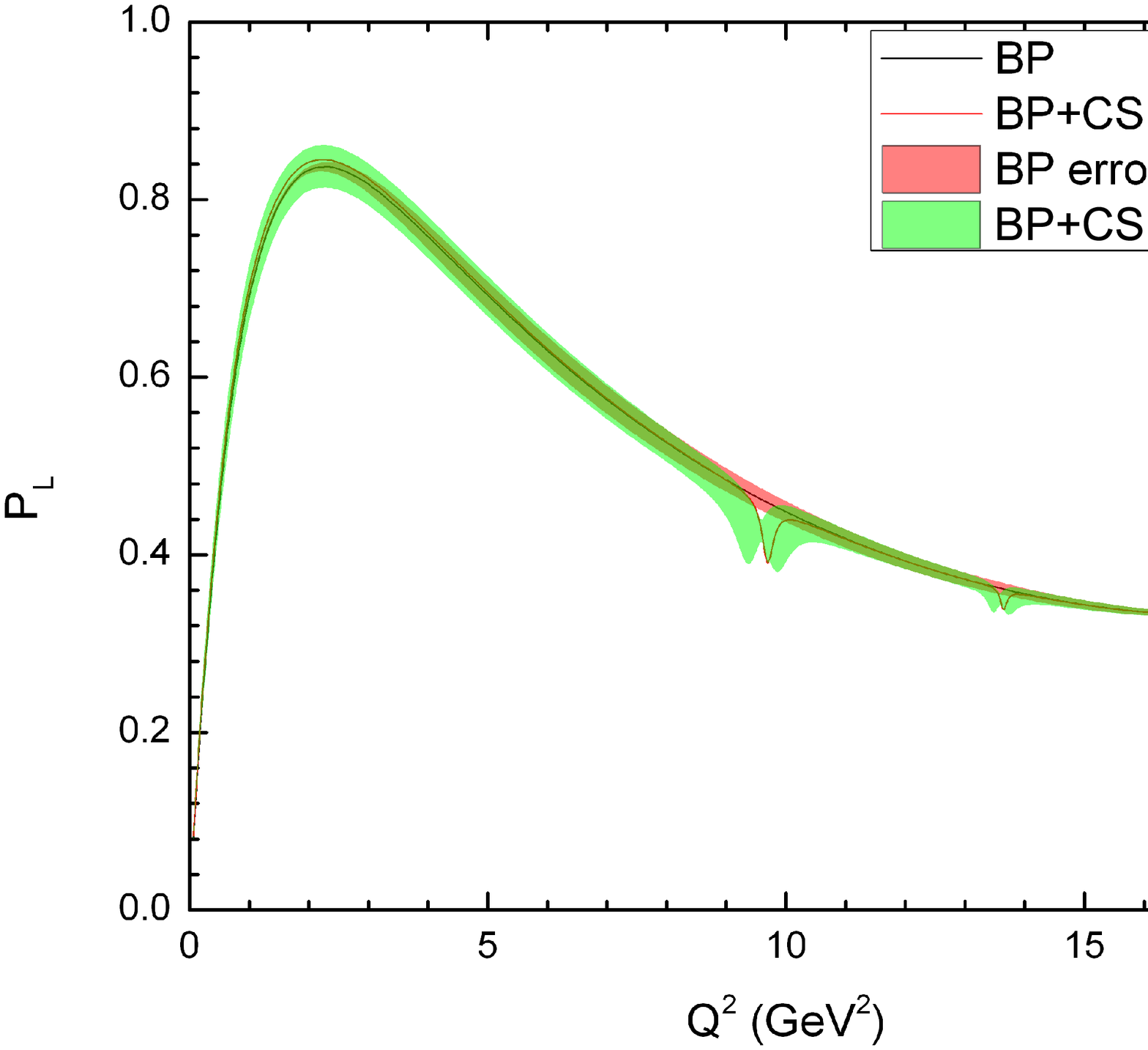}}
\subfigure[Longitudinal polarization of the final vector meson]{\includegraphics[width = 0.49\textwidth,height=0.2\textheight]{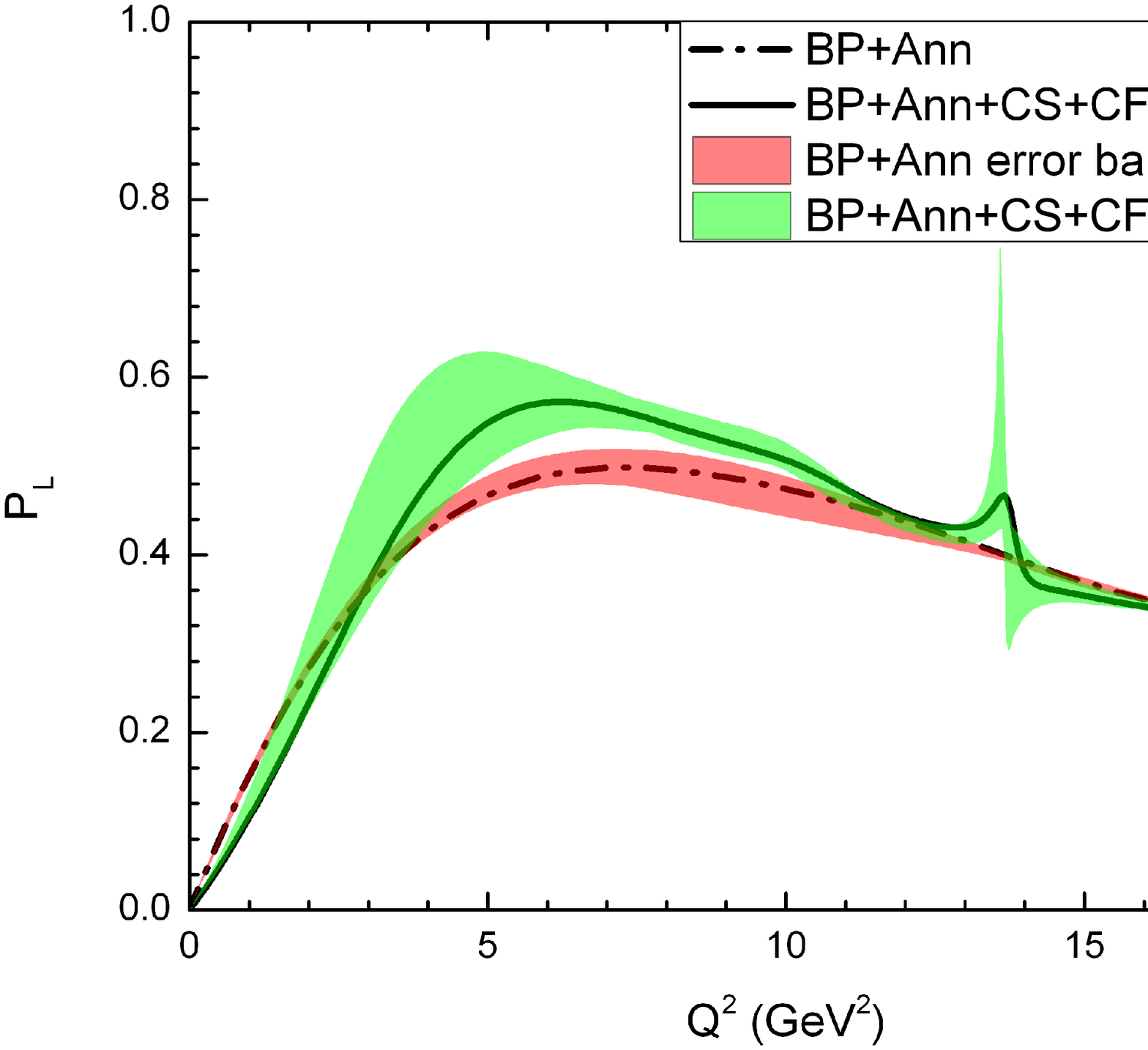}}\\
\subfigure[Leptonic longitudinal polarization asymmetry]{\includegraphics[width = 0.49\textwidth,height=0.2\textheight]{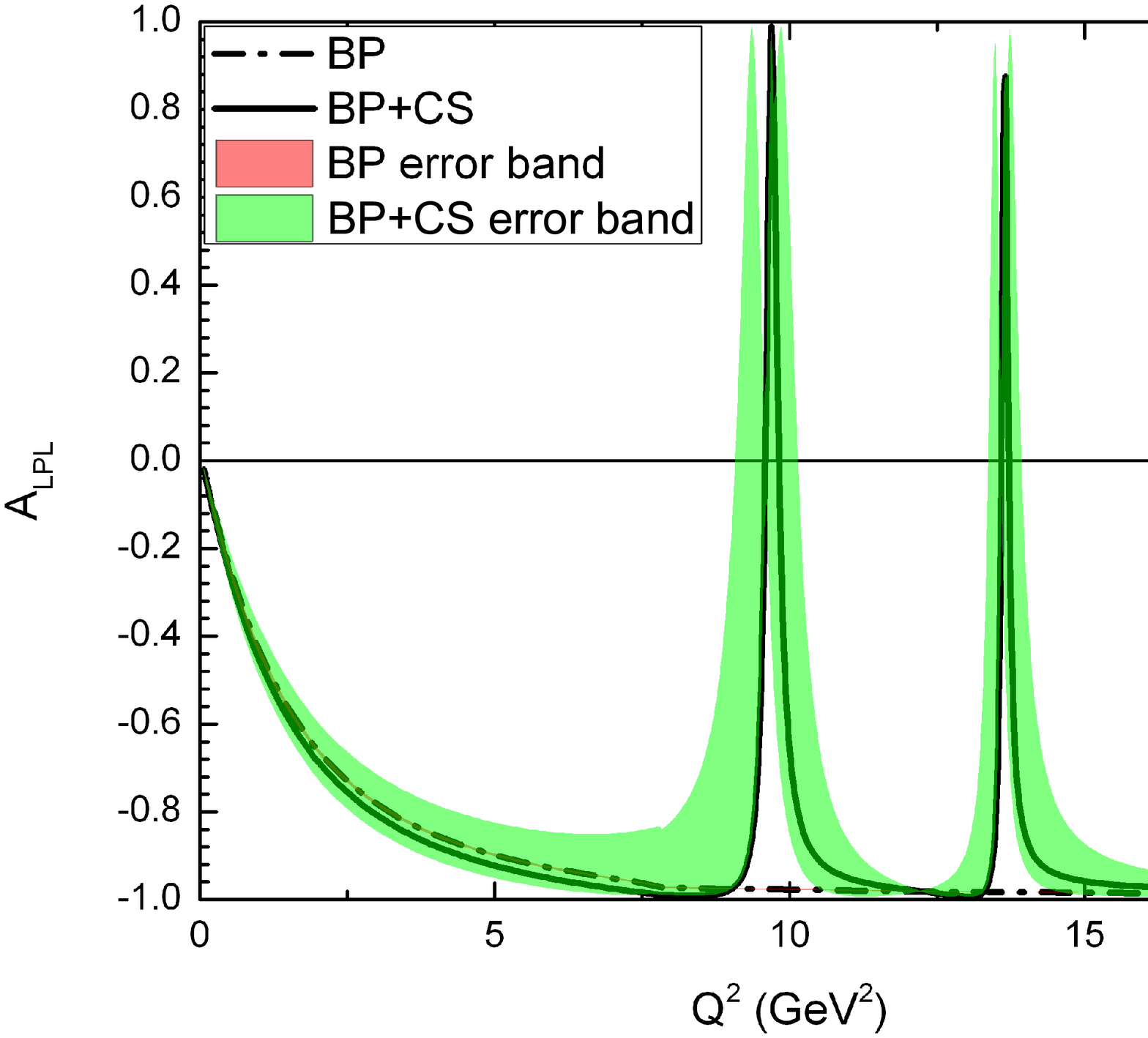}}
\subfigure[Leptonic longitudinal polarization asymmetry]{\includegraphics[width = 0.49\textwidth,height=0.2\textheight]{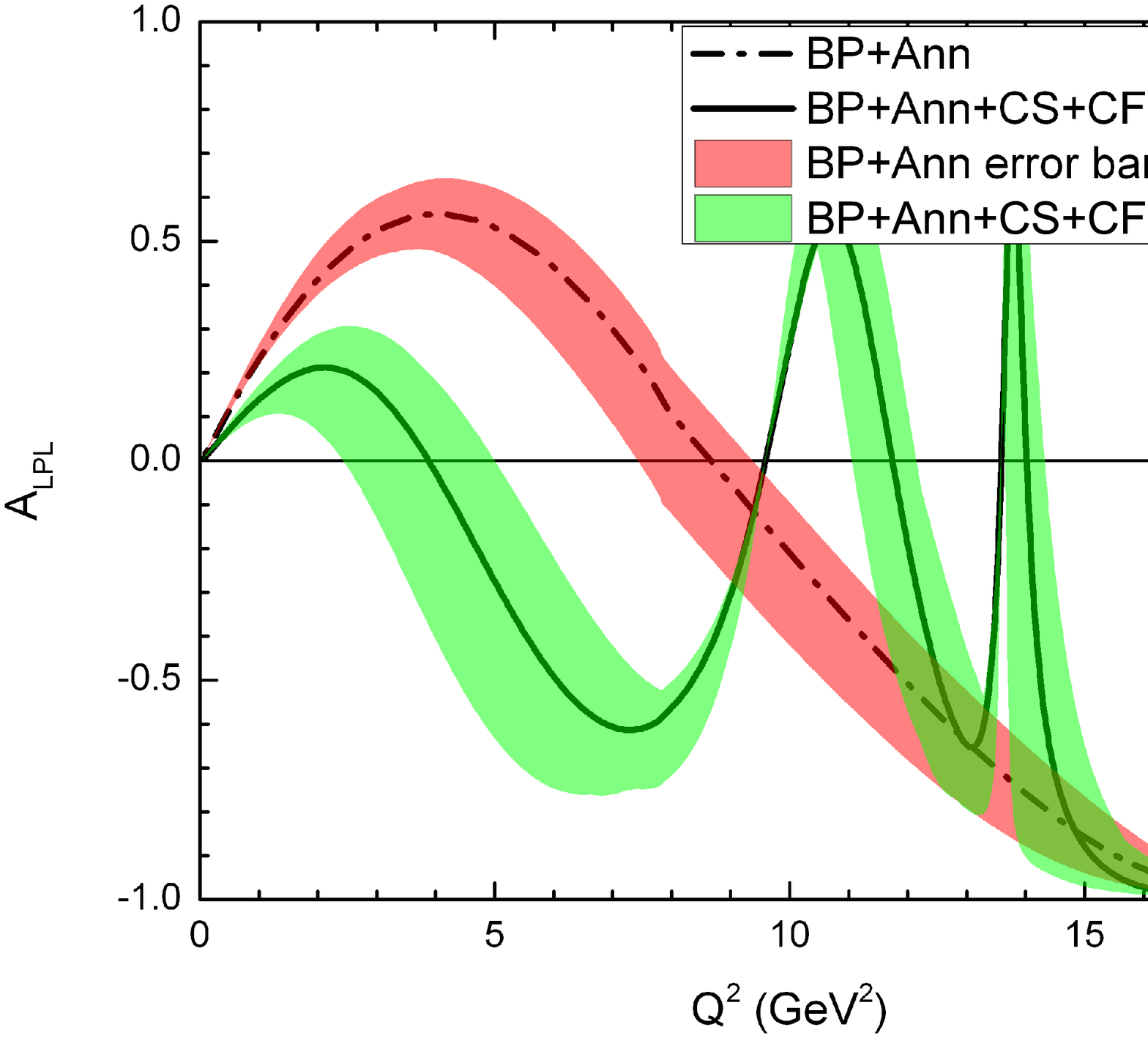}}\\
\caption{Observables of $B_c\rightarrow D^* \mu\bar{\mu}$. } 
\end{figure}

\begin{figure}[htbp]
\centering
\subfigure[Differential branching fraction]{\includegraphics[width = 0.49\textwidth,height=0.2\textheight]{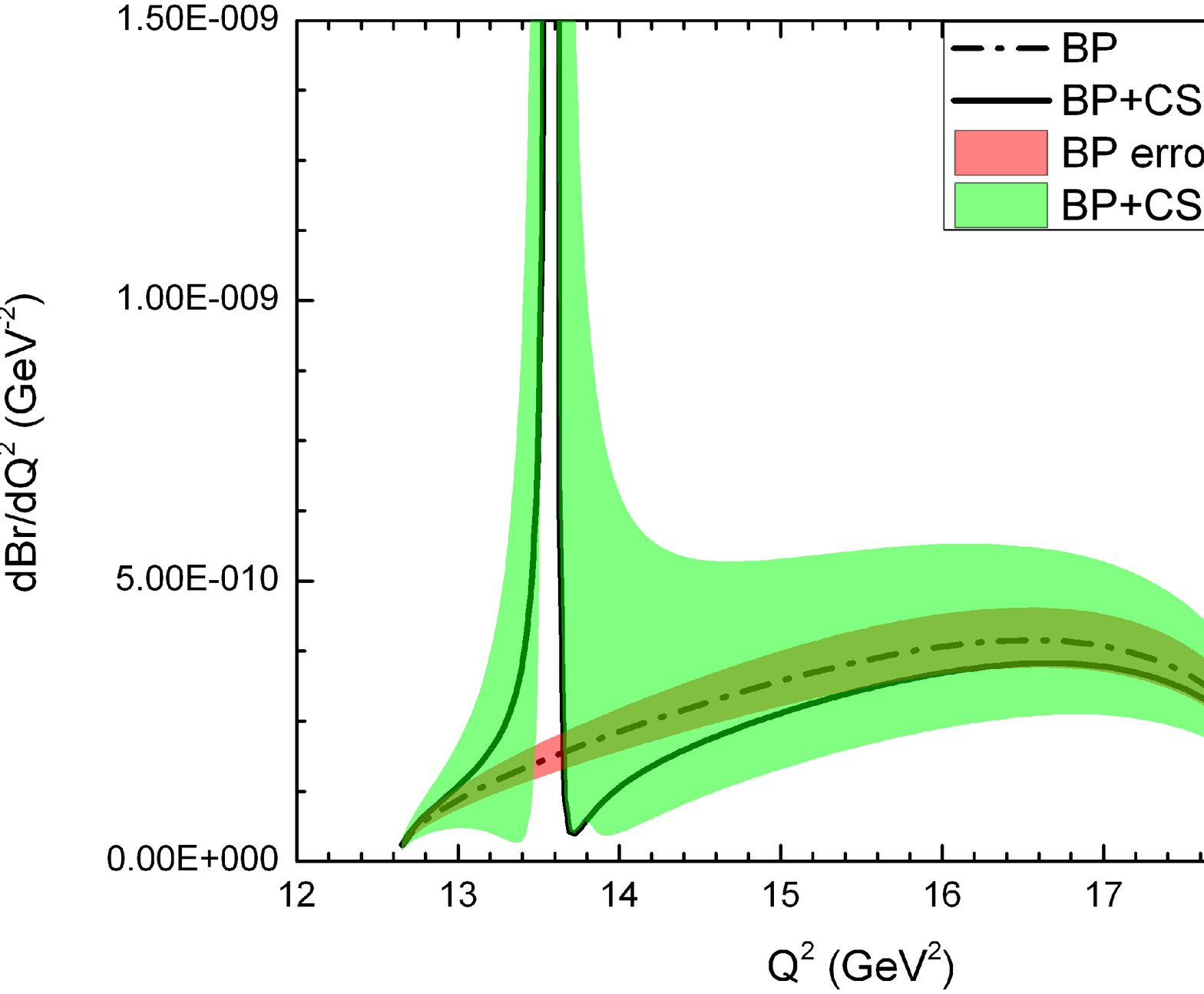}}
\subfigure[Differential branching fraction]{\includegraphics[width = 0.49\textwidth,height=0.2\textheight]{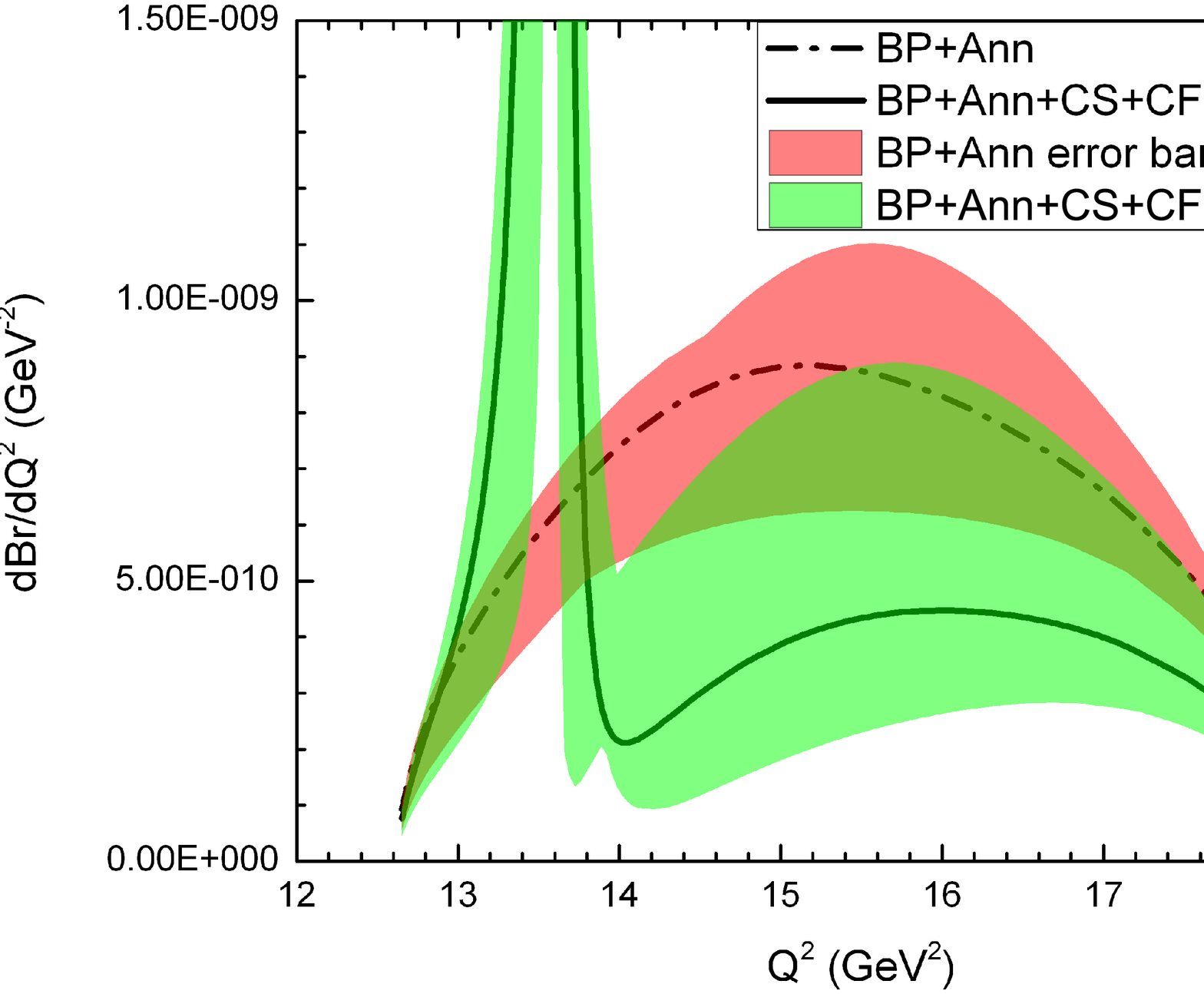}}\\
\subfigure[Forward-backward asymmetry]{\includegraphics[width = 0.49\textwidth,height=0.2\textheight]{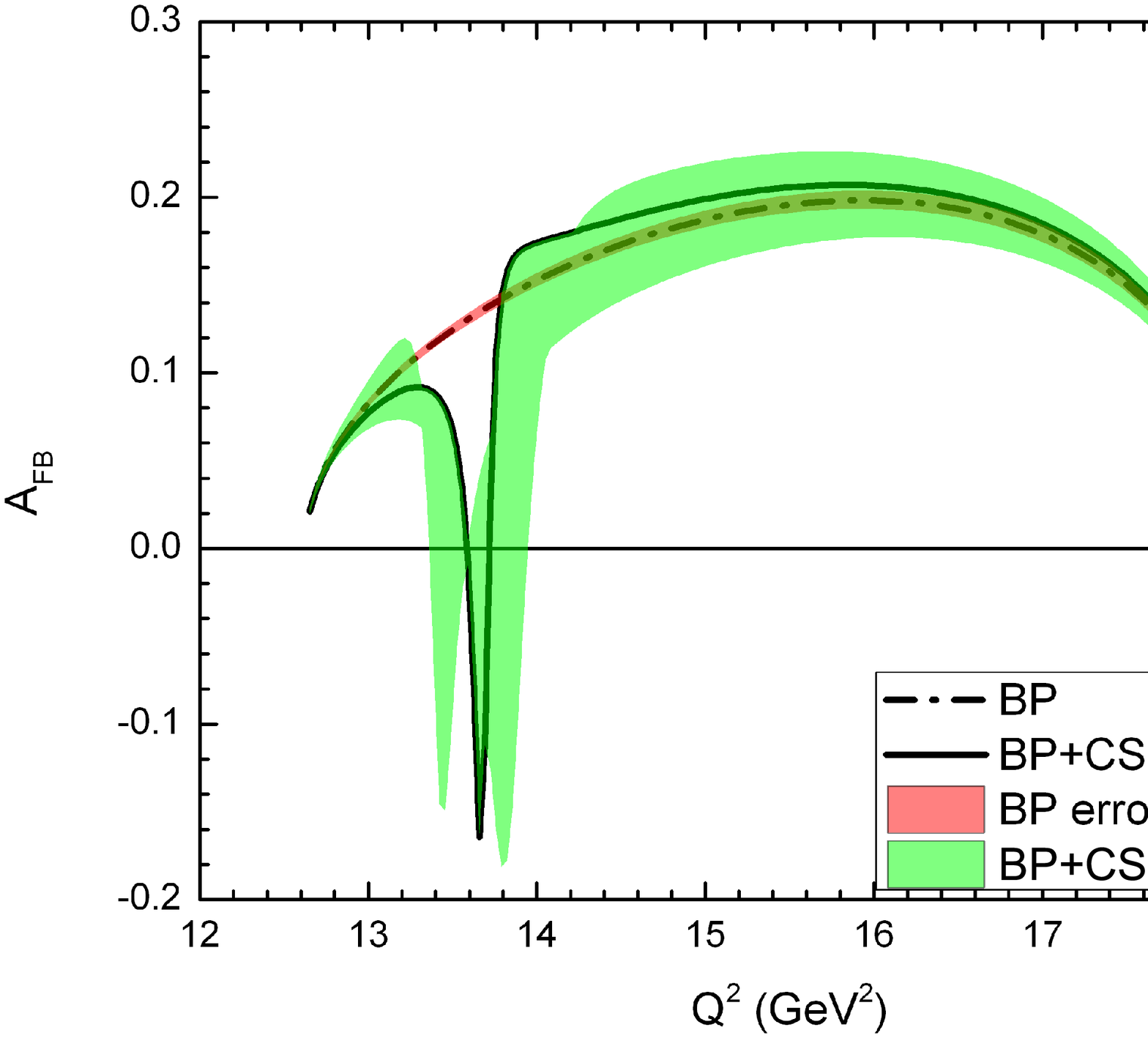}}
\subfigure[Forward-backward asymmetry]{\includegraphics[width = 0.49\textwidth,height=0.2\textheight]{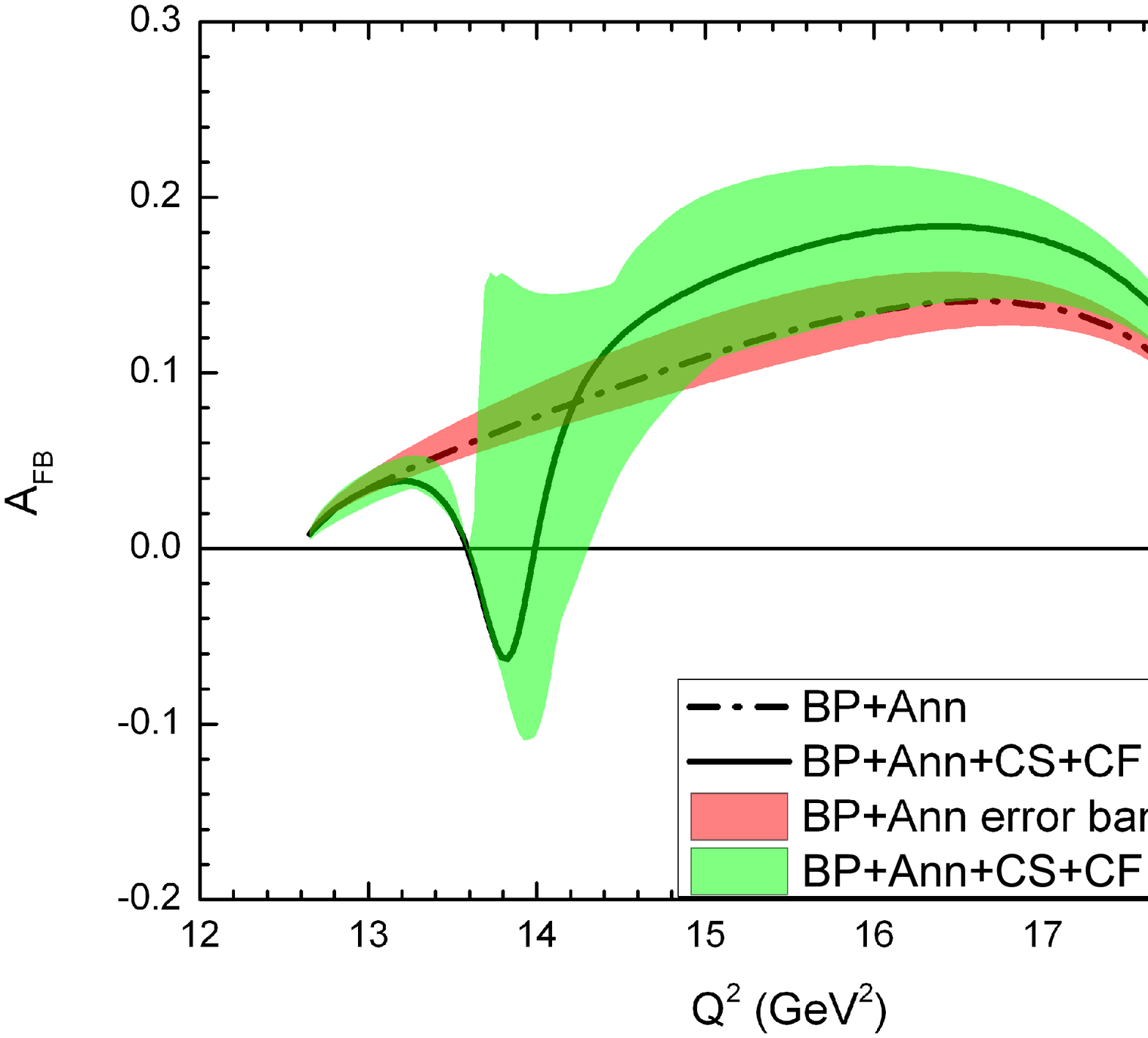}}\\
\subfigure[Longitudinal polarization of the final vector
meson]{\includegraphics[width =
0.49\textwidth,height=0.2\textheight]{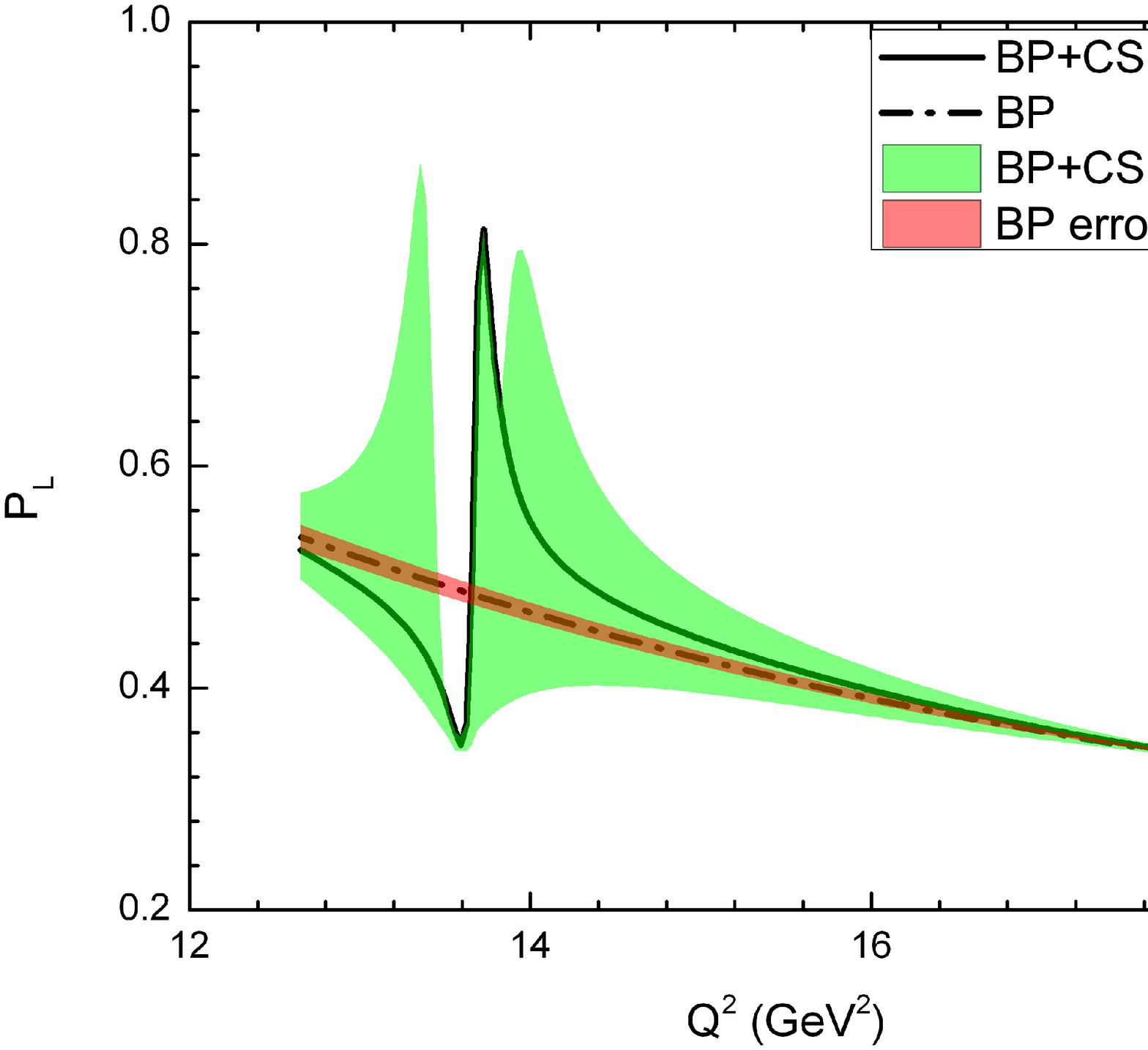}}
\subfigure[Longitudinal polarization of the final vector meson]{\includegraphics[width = 0.49\textwidth,height=0.2\textheight]{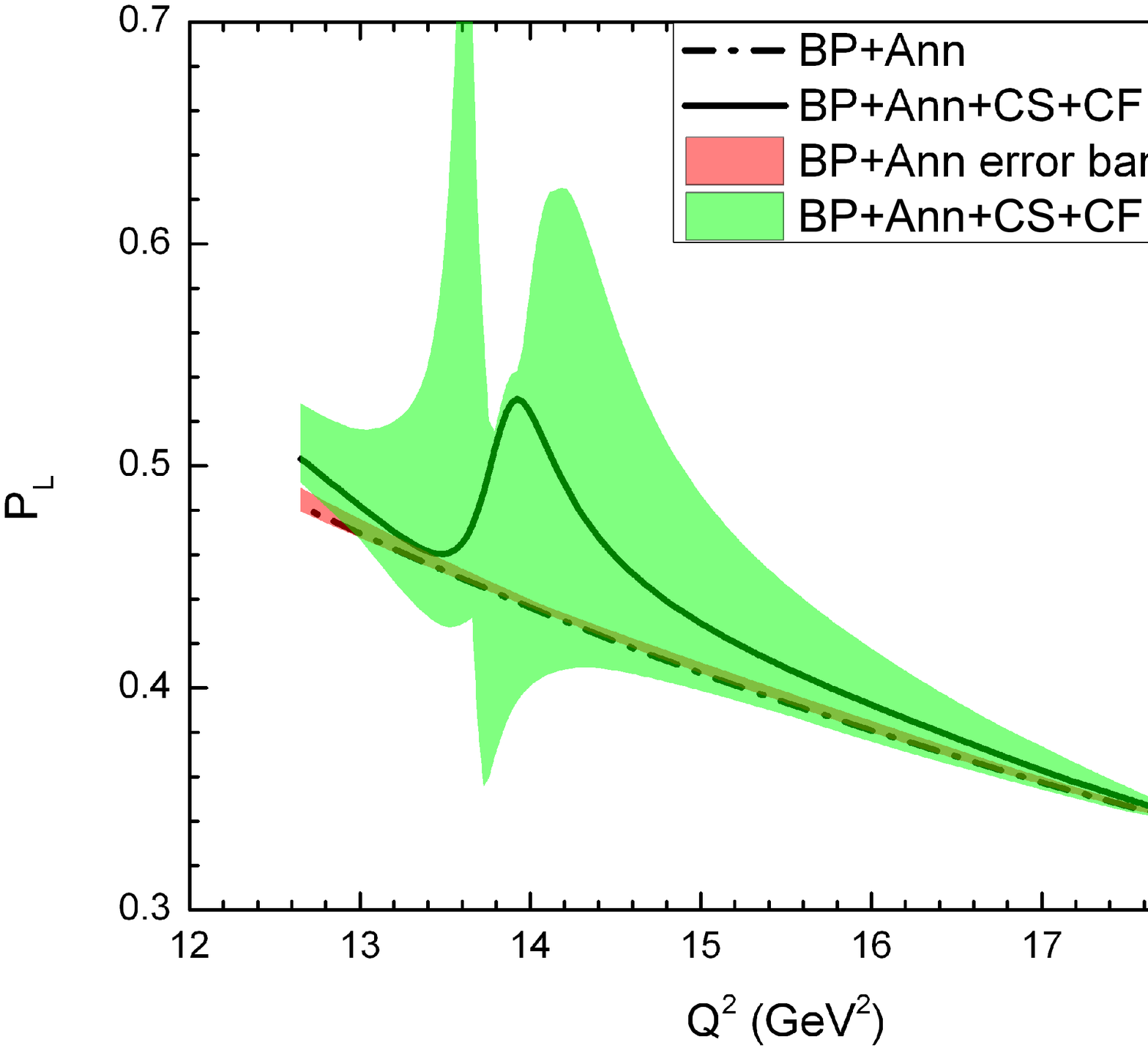}}\\
\subfigure[Leptonic longitudinal polarization asymmetry]{\includegraphics[width = 0.49\textwidth,height=0.2\textheight]{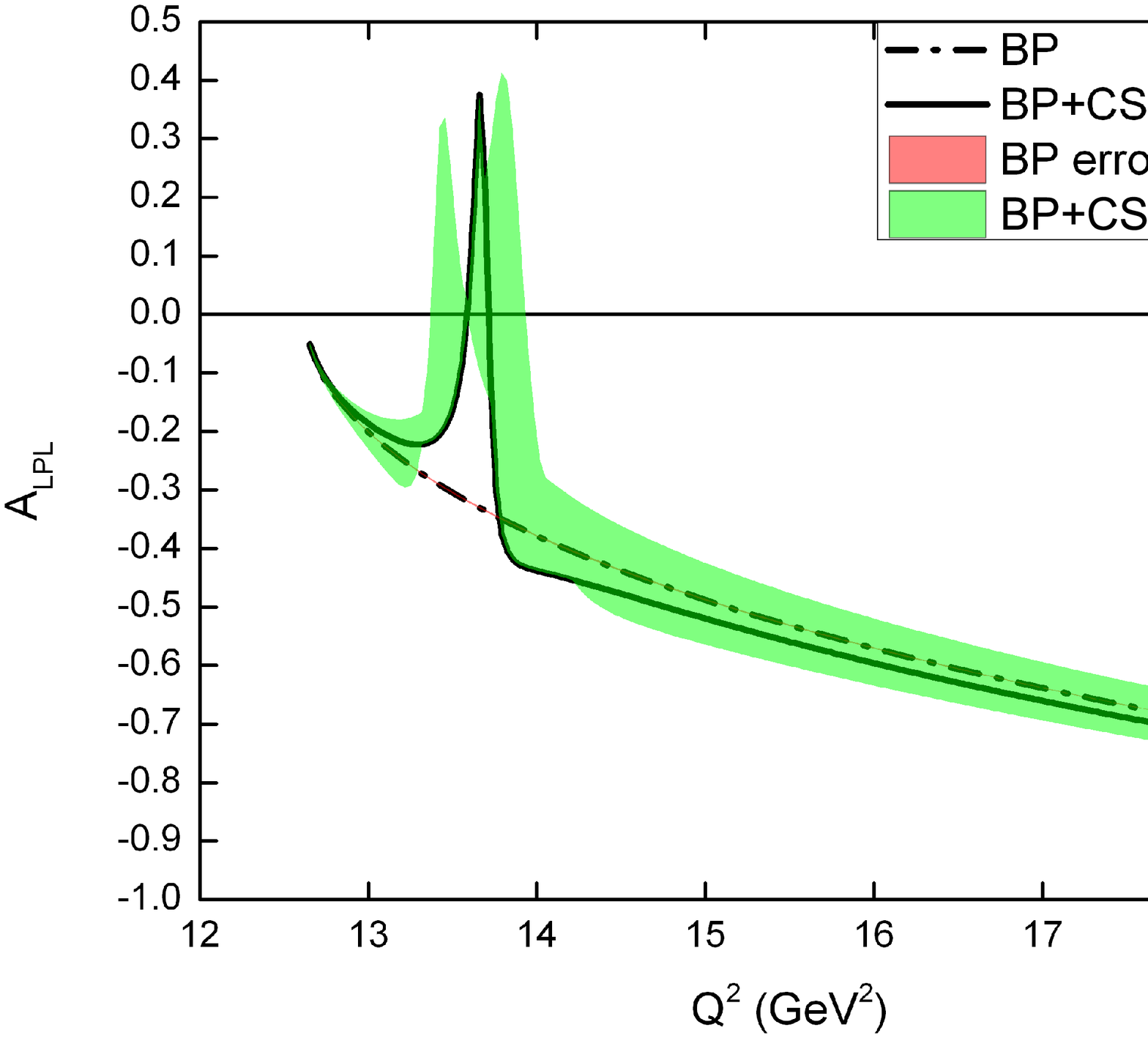}}
\subfigure[Leptonic longitudinal polarization asymmetry]{\includegraphics[width = 0.49\textwidth,height=0.2\textheight]{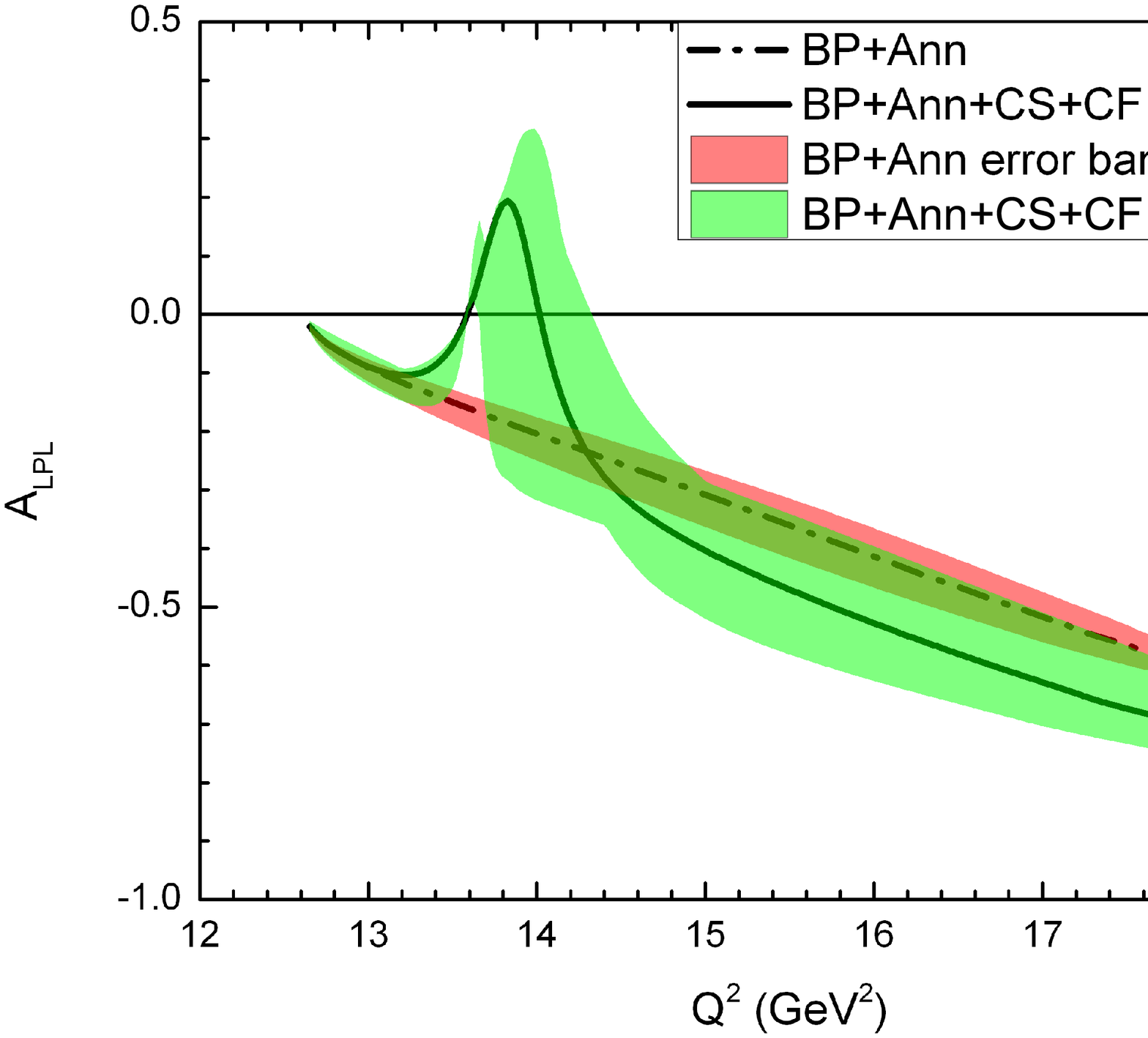}}\\
\caption{Observables of $B_c\rightarrow D^* \tau\bar{\tau}$. } 
\end{figure}

\begin{figure}[htbp]
\centering
\subfigure[Differential branching fraction]{\includegraphics[width = 0.49\textwidth,height=0.2\textheight]{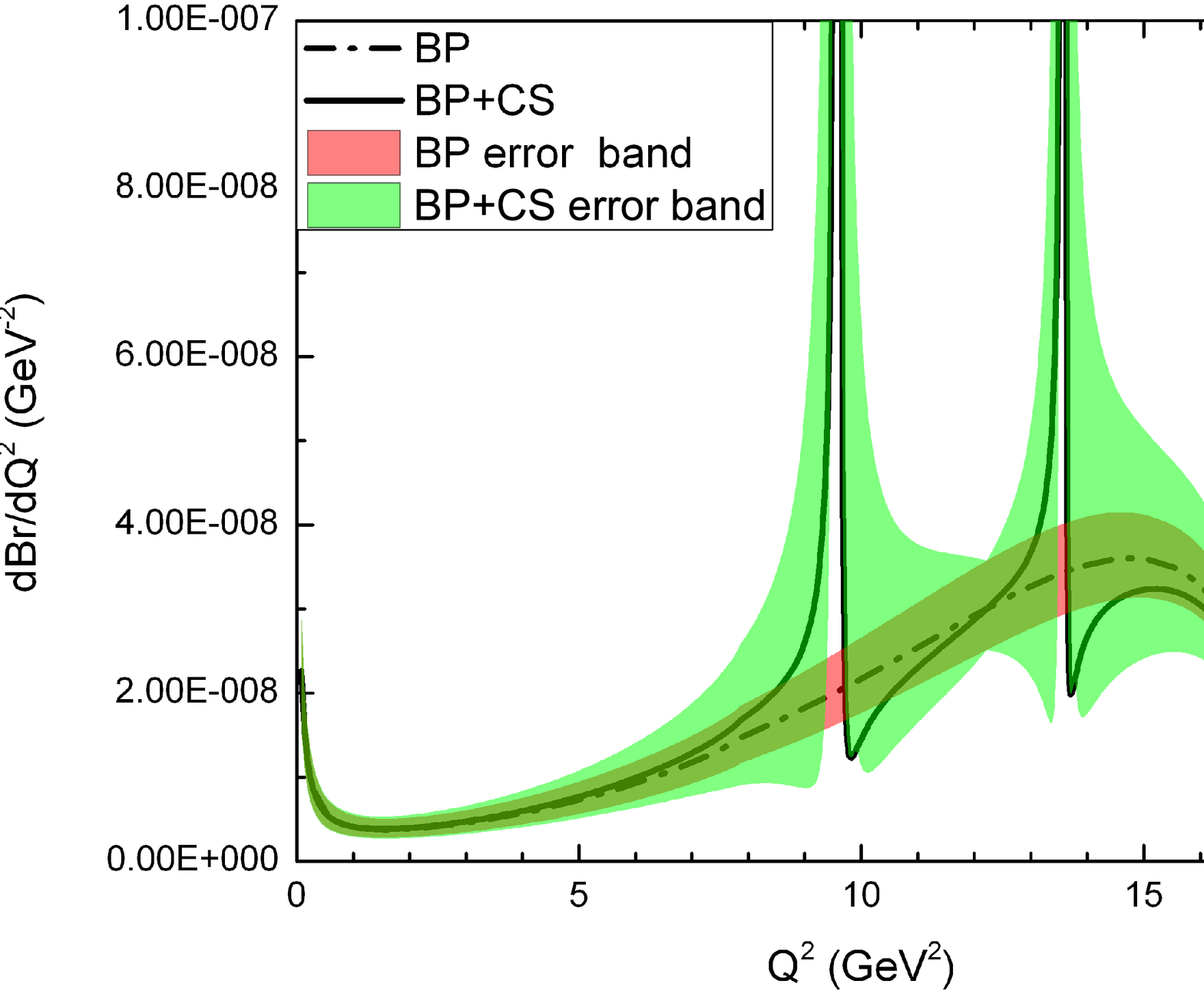}}
\subfigure[Differential branching fraction]{\includegraphics[width = 0.49\textwidth,height=0.2\textheight]{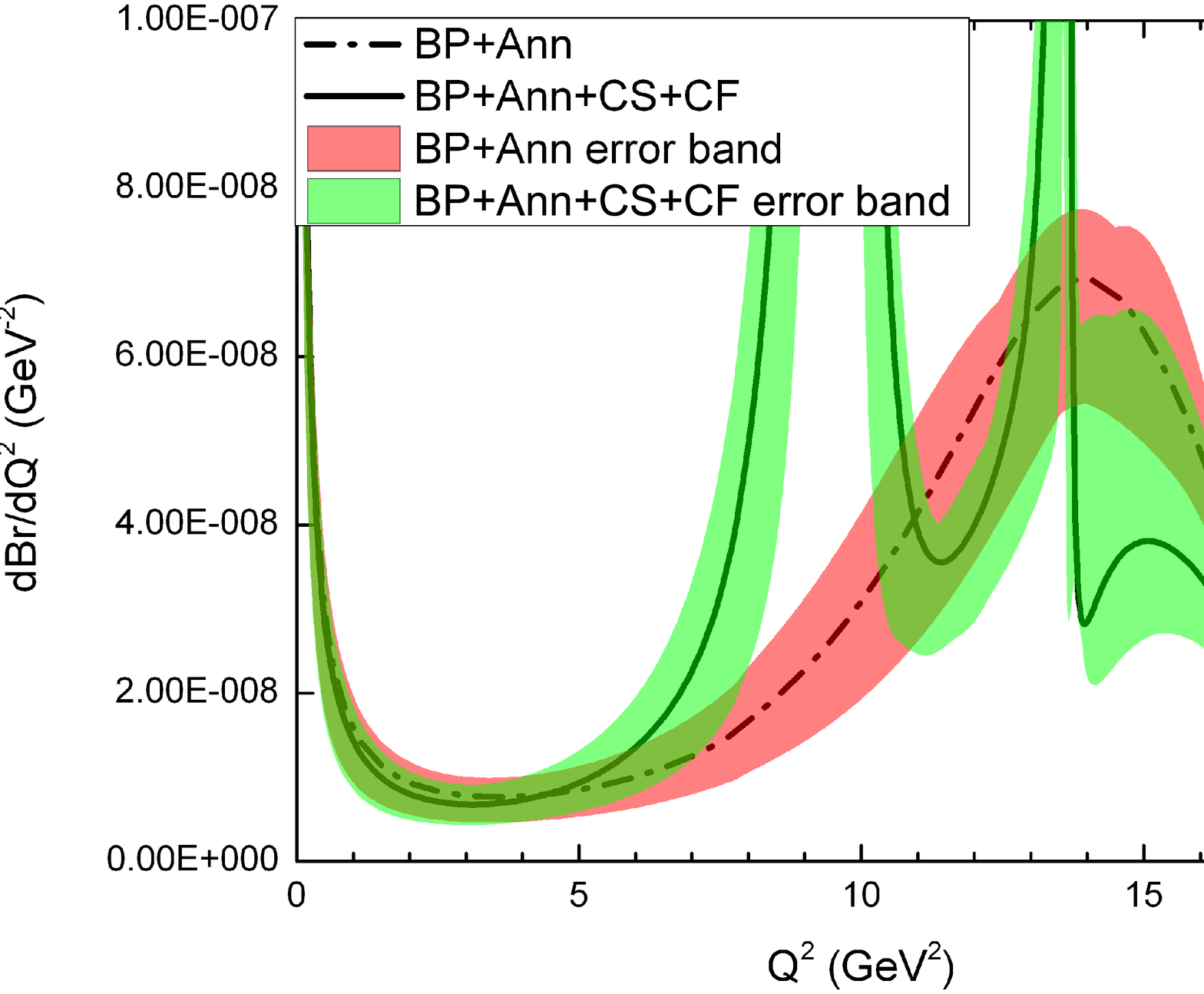}}\\
\subfigure[Forward-backward asymmetry]{\includegraphics[width = 0.49\textwidth,height=0.2\textheight]{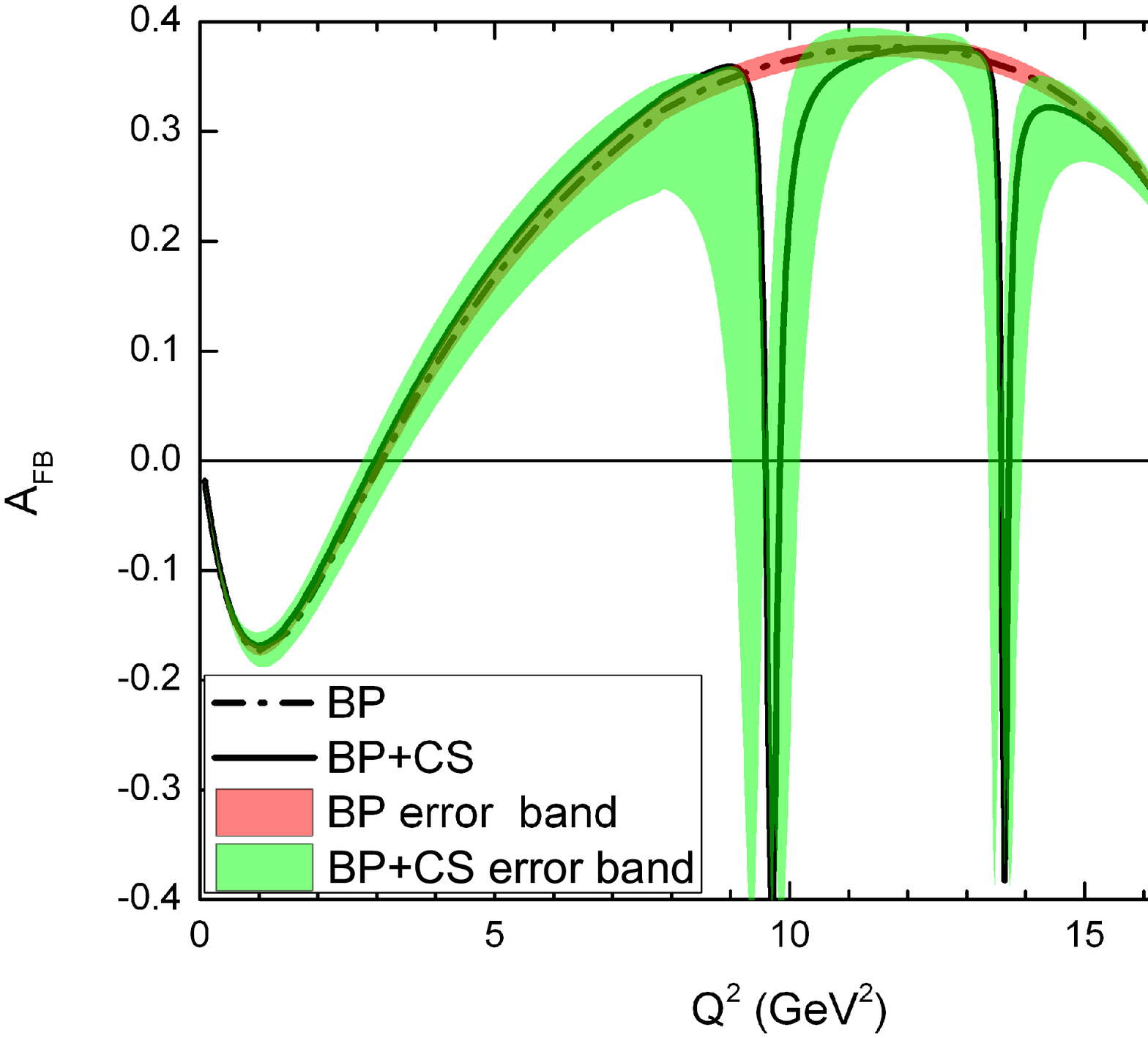}}
\subfigure[Forward-backward asymmetry]{\includegraphics[width = 0.49\textwidth,height=0.2\textheight]{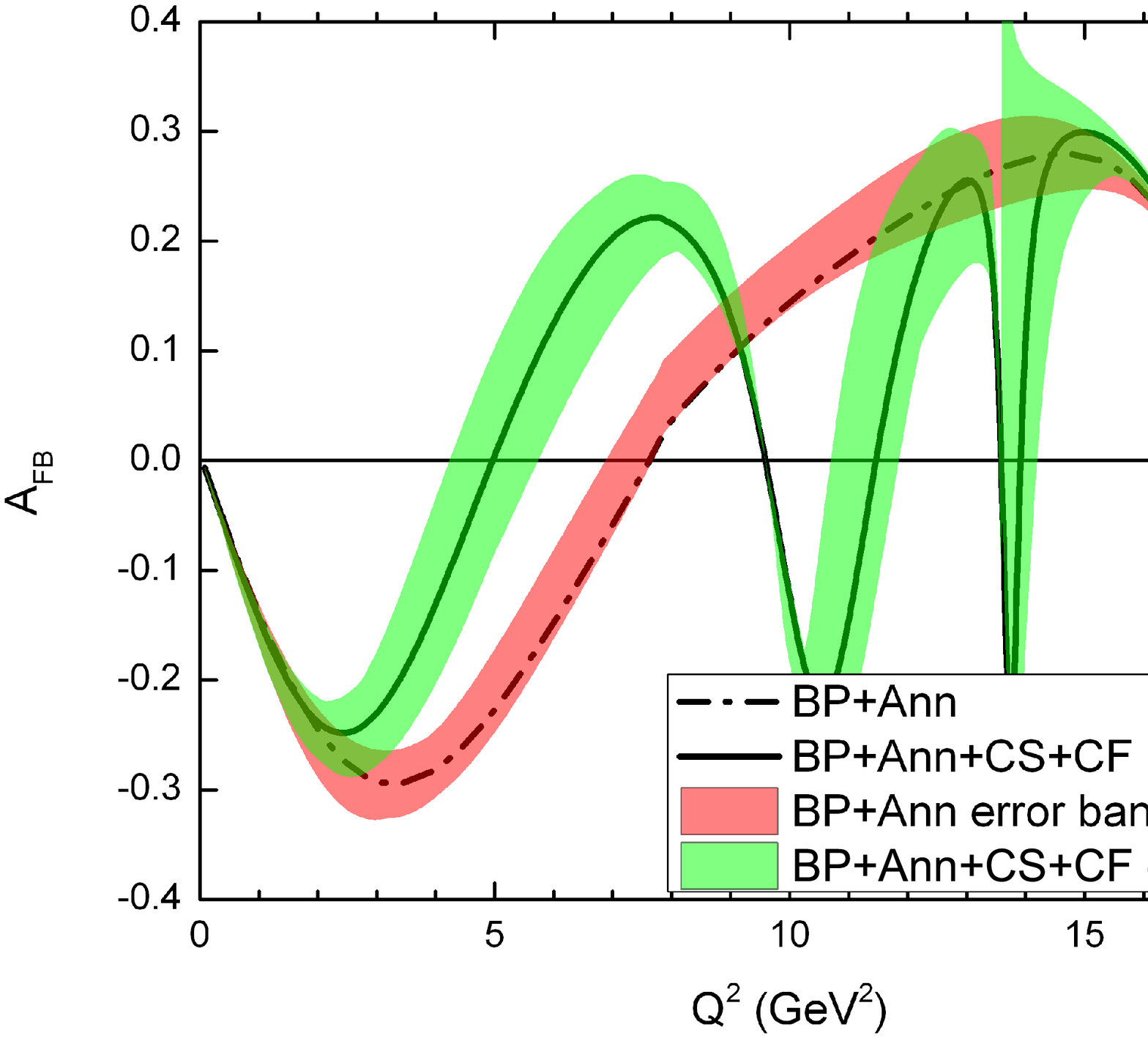}}\\
\subfigure[Longitudinal polarization of the final vector meson]{\includegraphics[width = 0.49\textwidth,height=0.2\textheight]{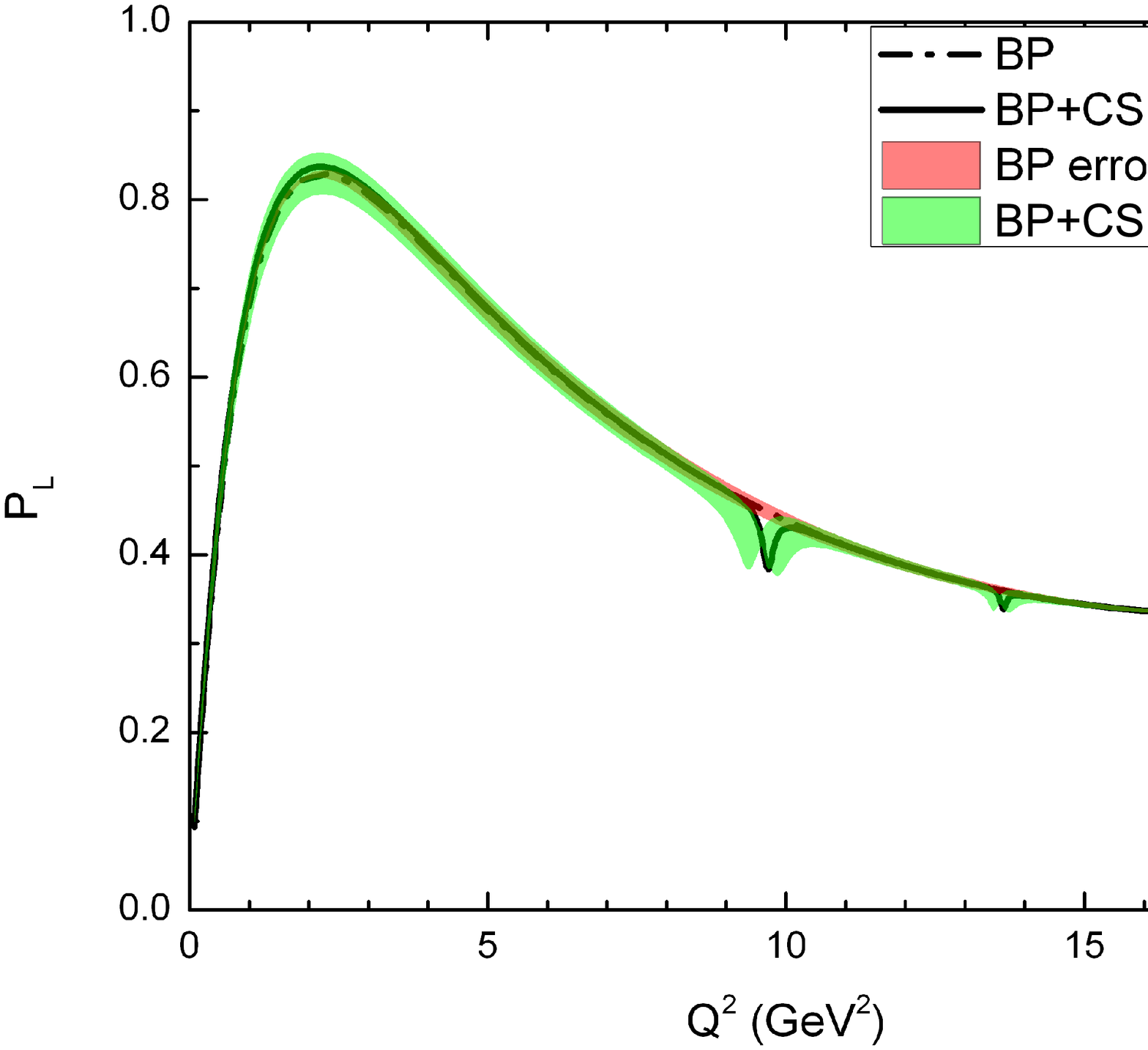}}
\subfigure[Longitudinal polarization of the final vector meson]{\includegraphics[width = 0.49\textwidth,height=0.2\textheight]{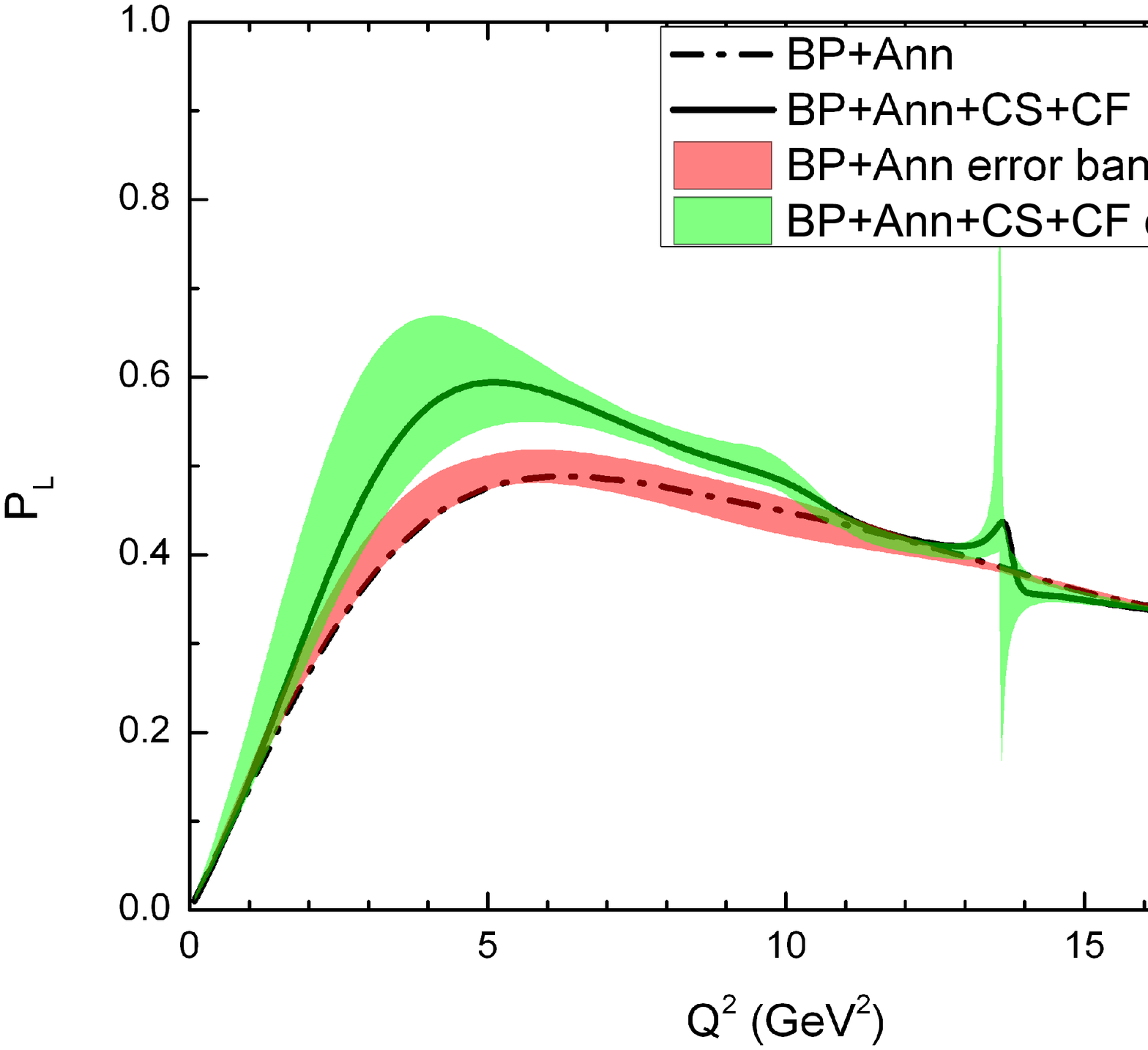}}\\
\subfigure[Leptonic longitudinal polarization asymmetry]{\includegraphics[width = 0.49\textwidth,height=0.2\textheight]{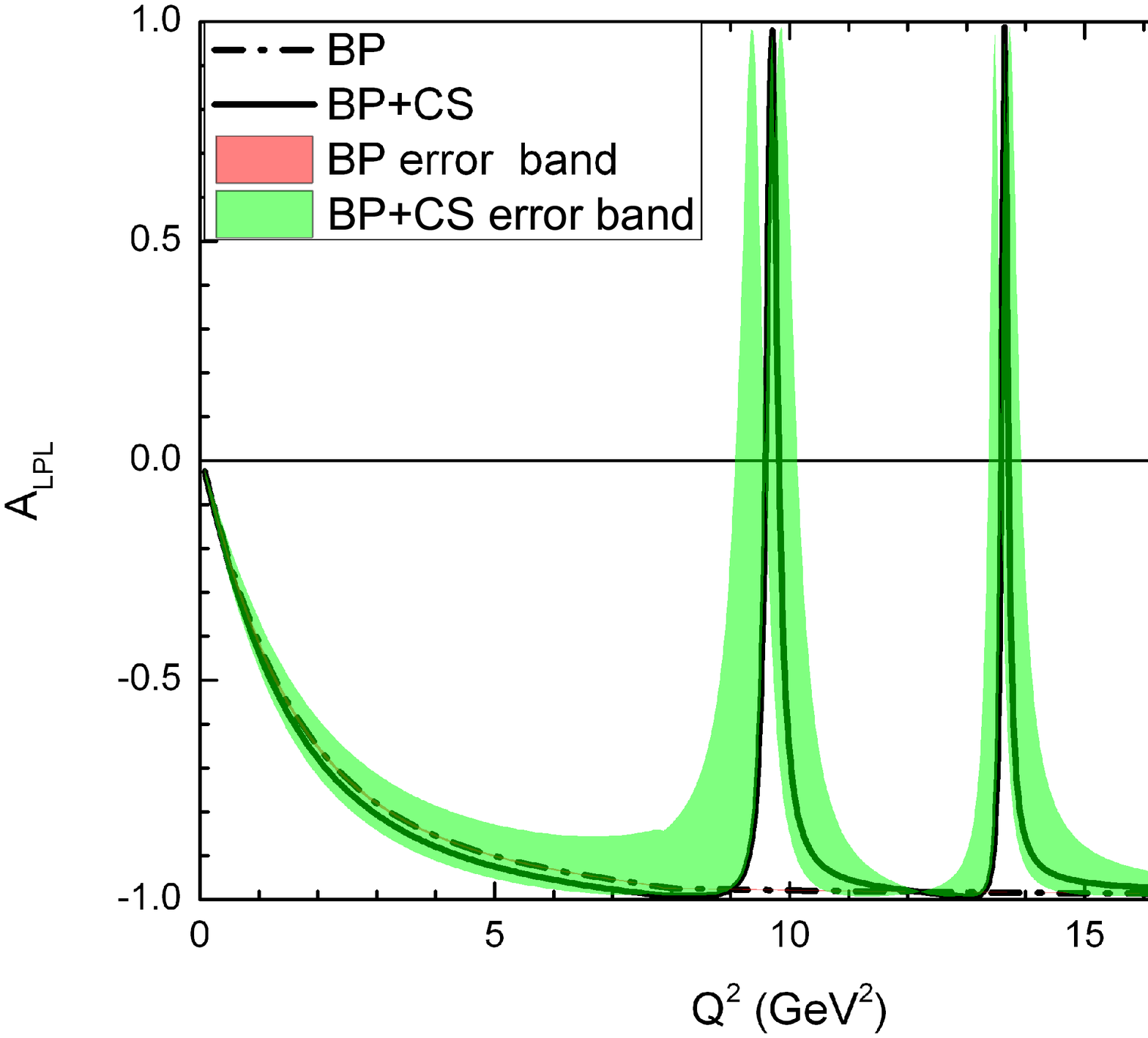}}
\subfigure[Leptonic longitudinal polarization asymmetry]{\includegraphics[width = 0.49\textwidth,height=0.2\textheight]{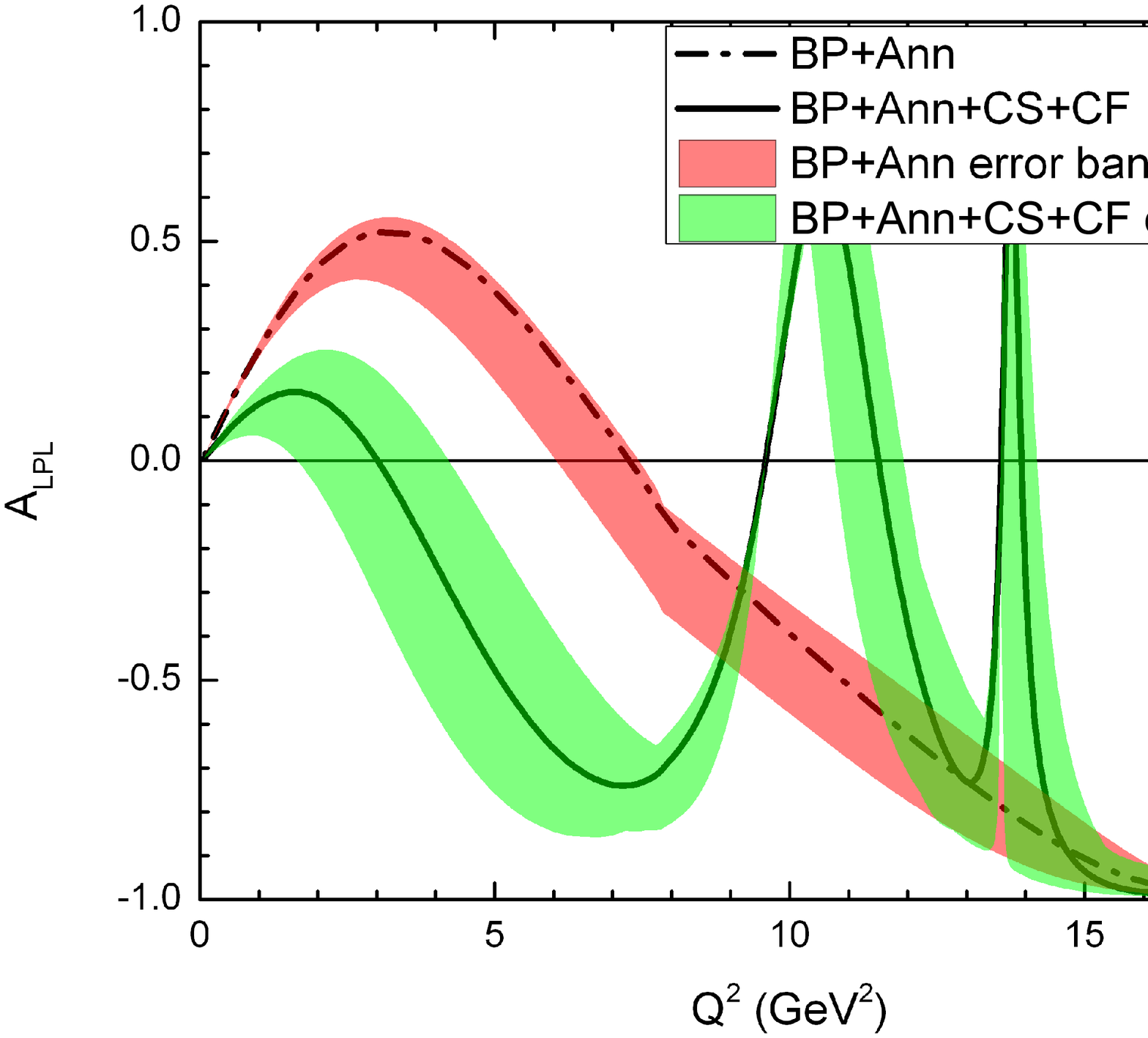}}\\
\caption{Observables of $B_c\rightarrow D_s^* \mu\bar{\mu}$.} 
\end{figure}

\begin{figure}[htbp]
\centering
\subfigure[Differential branching fraction]{\includegraphics[width = 0.49\textwidth,height=0.2\textheight]{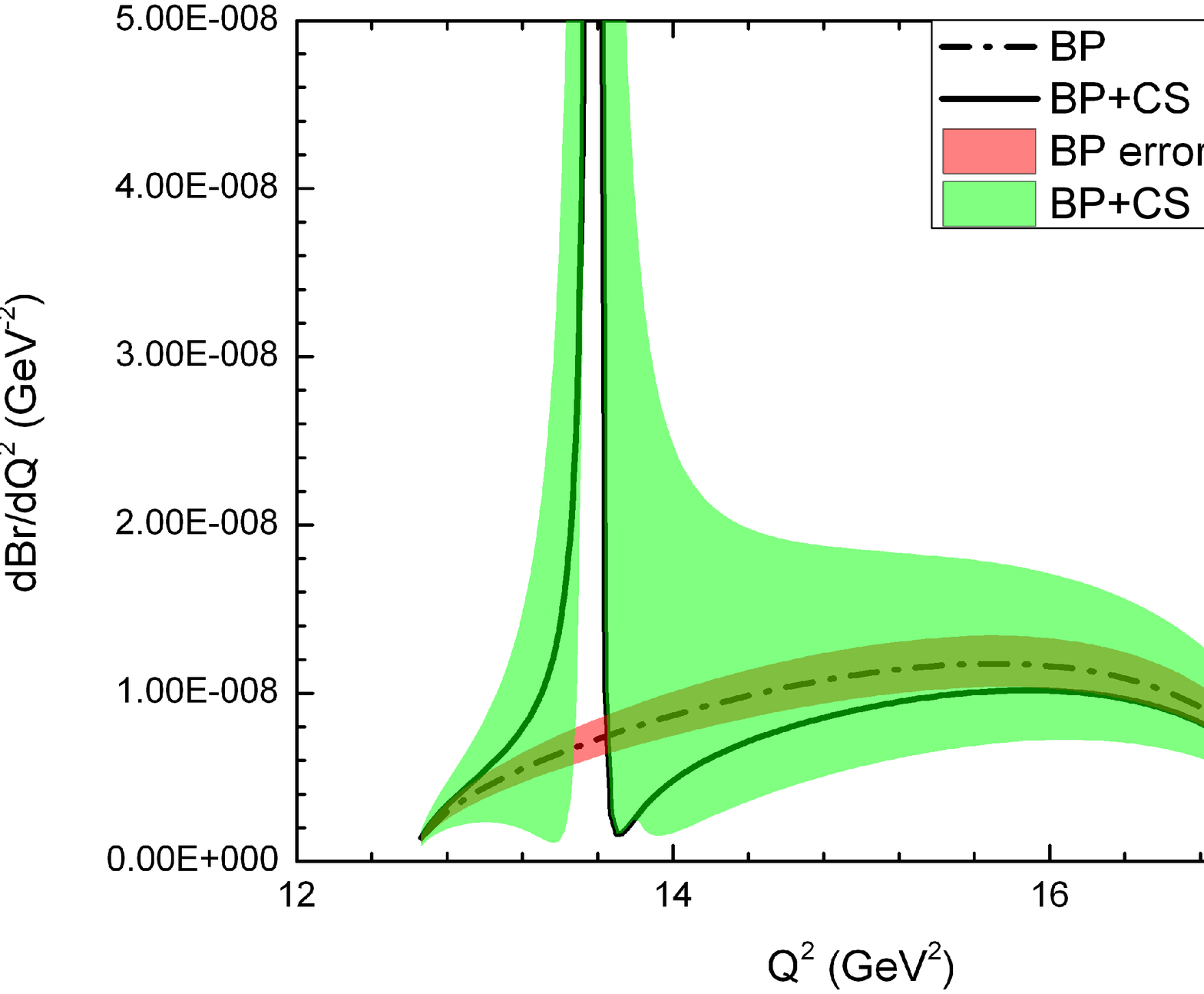}}
\subfigure[Differential branching fraction]{\includegraphics[width = 0.49\textwidth,height=0.2\textheight]{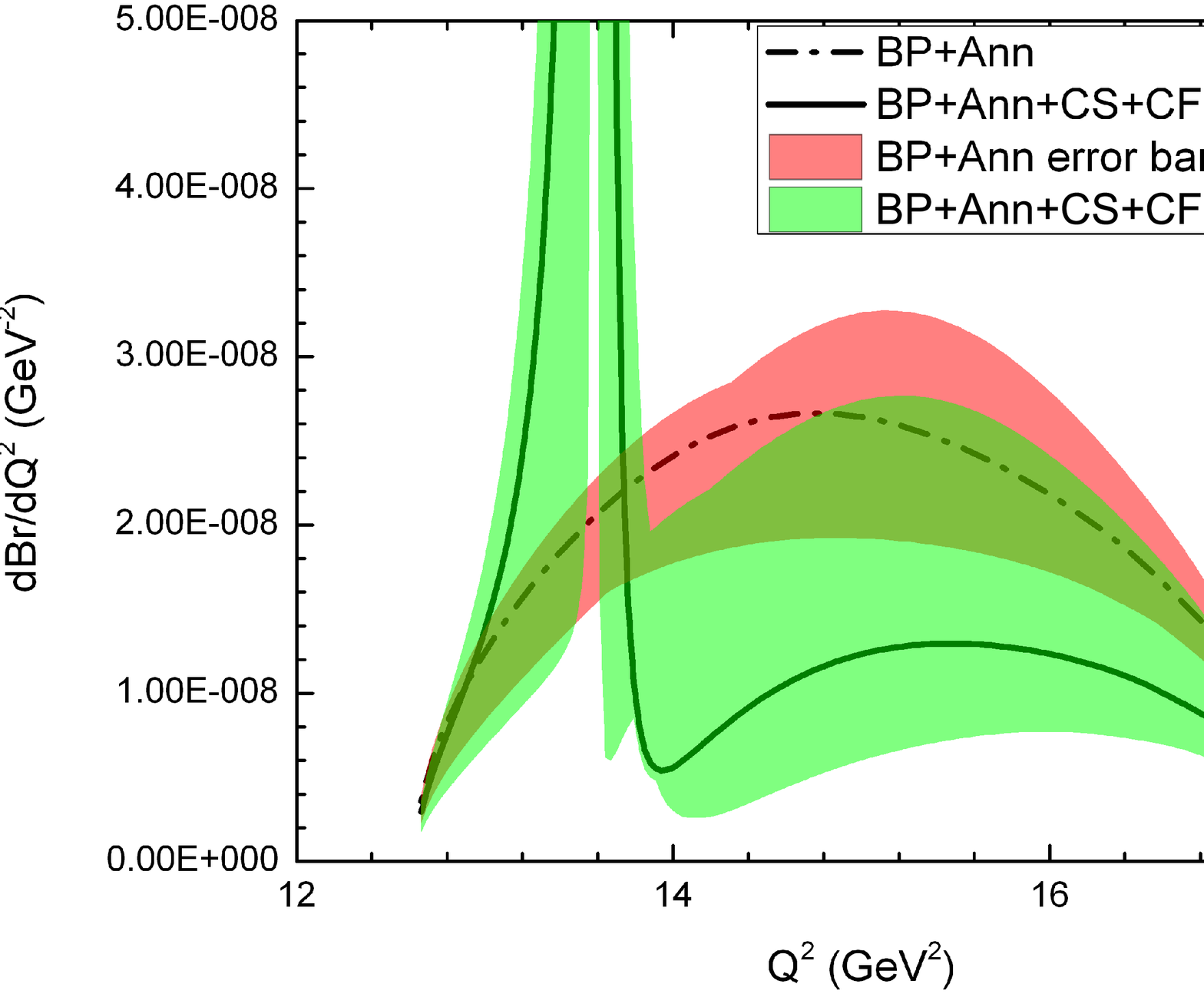}}\\
\subfigure[Forward-backward asymmetry]{\includegraphics[width = 0.49\textwidth,height=0.2\textheight]{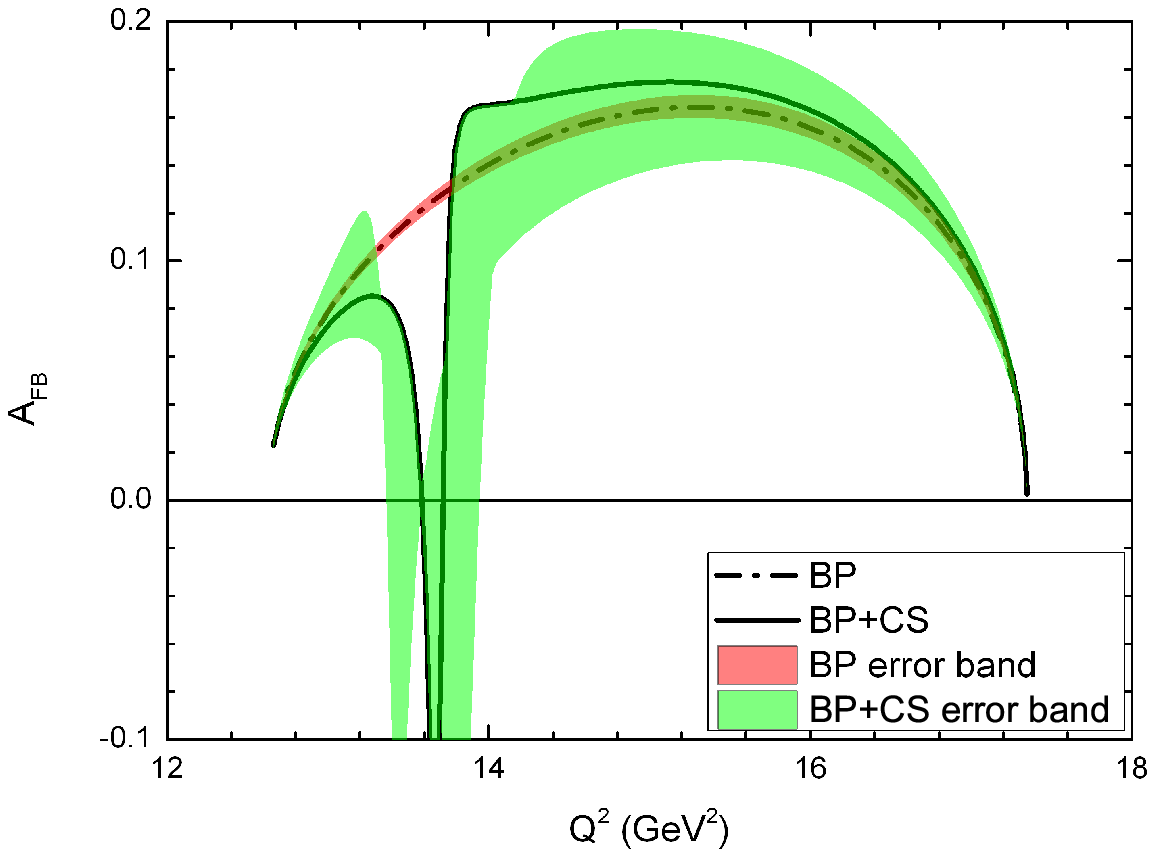}}
\subfigure[Forward-backward asymmetry]{\includegraphics[width = 0.49\textwidth,height=0.2\textheight]{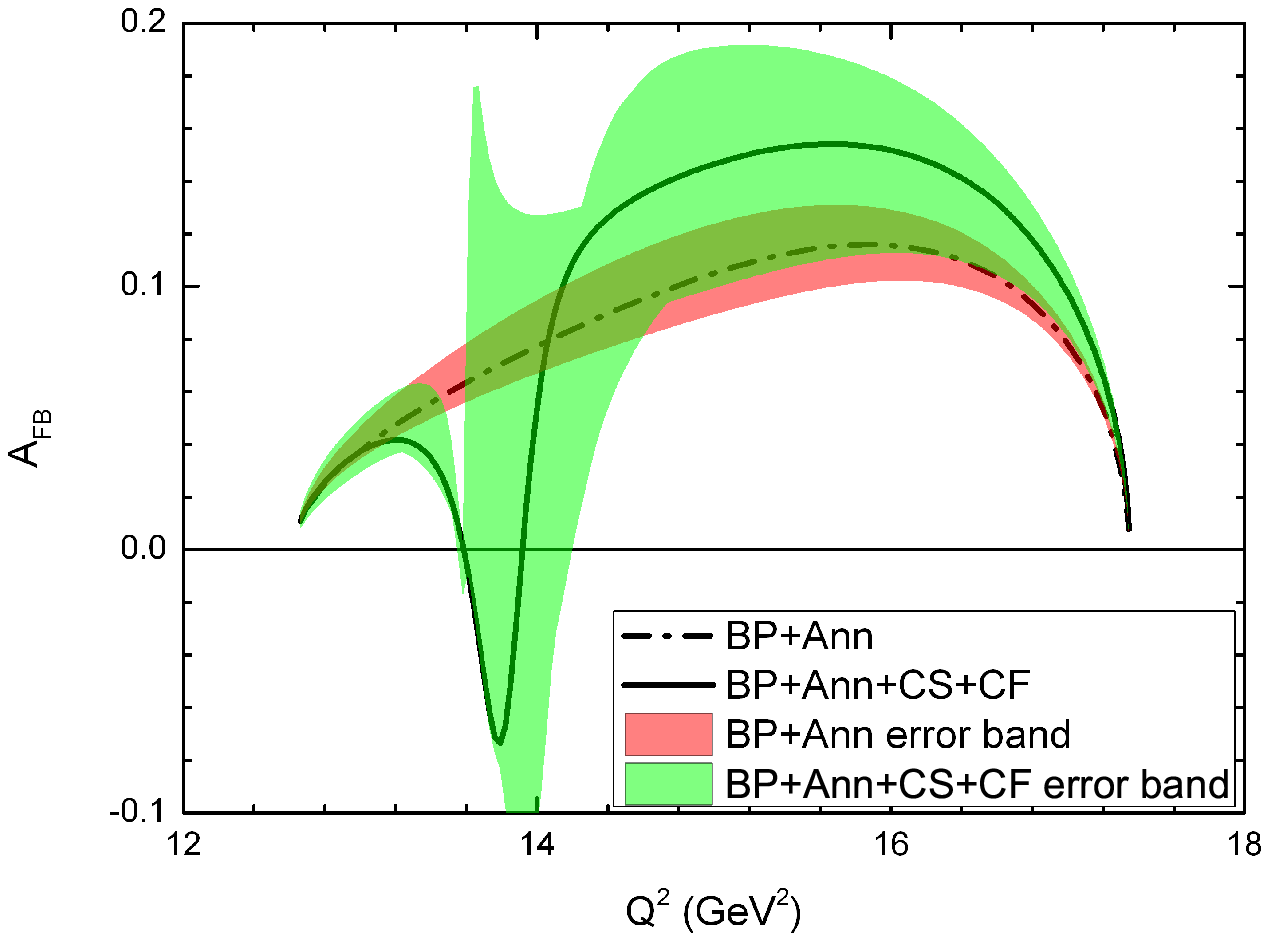}}\\
\subfigure[Longitudinal polarization of the final vector meson]{\includegraphics[width =0.49\textwidth,height=0.2\textheight]{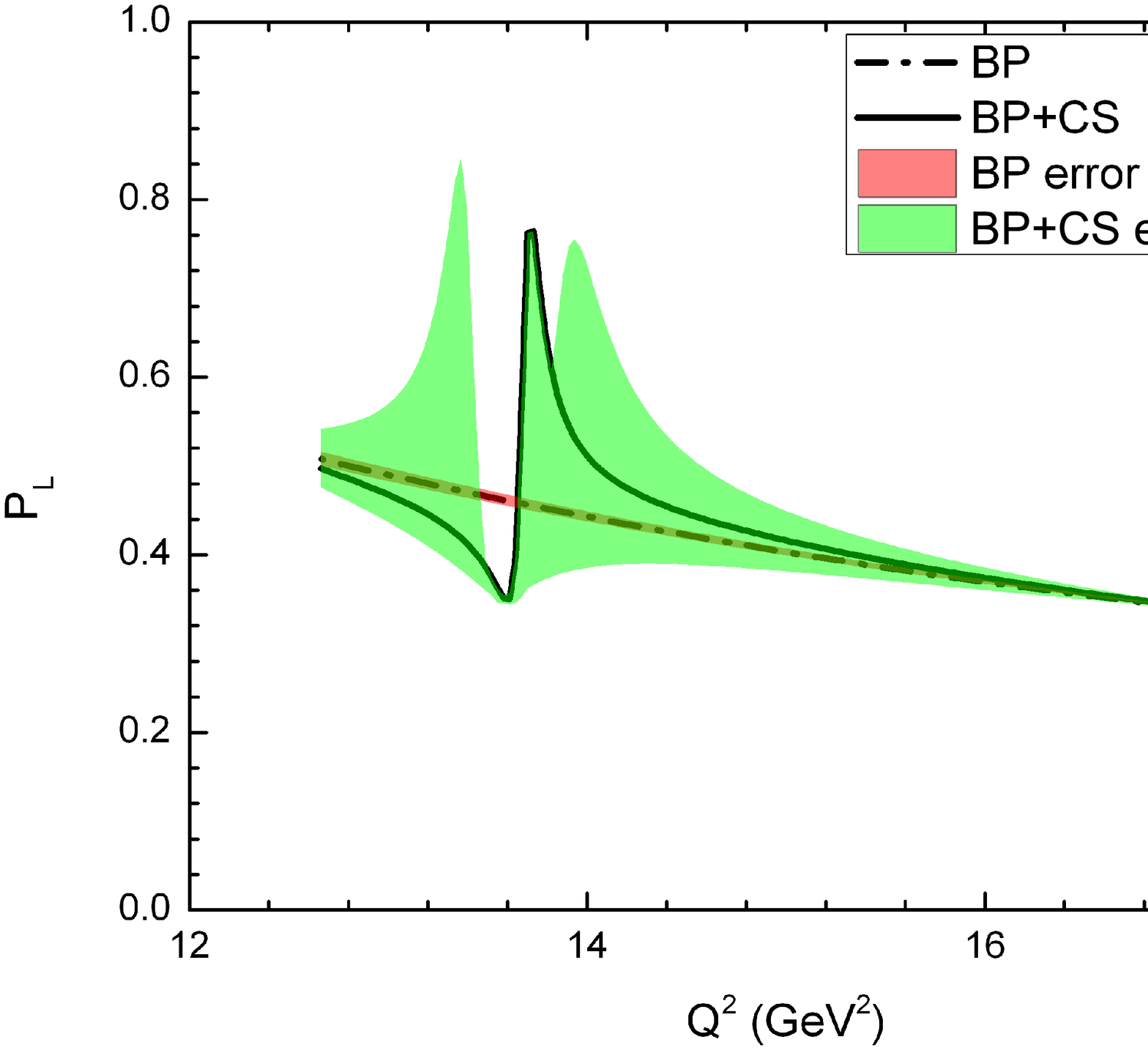}}
\subfigure[Longitudinal polarization of the final vector meson]{\includegraphics[width = 0.49\textwidth,height=0.2\textheight]{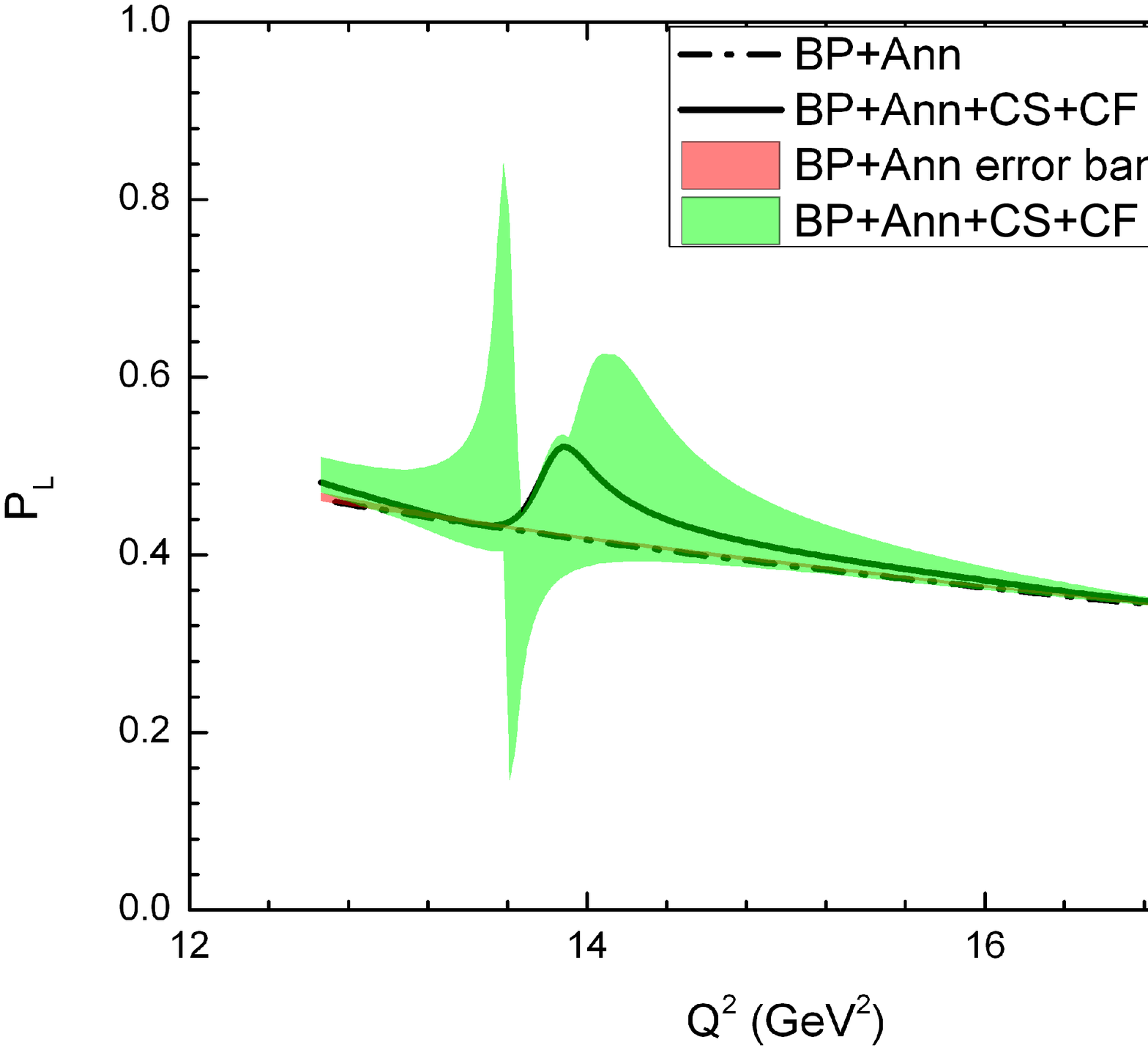}}\\
\subfigure[Leptonic longitudinal polarization asymmetry]{\includegraphics[width = 0.49\textwidth,height=0.2\textheight]{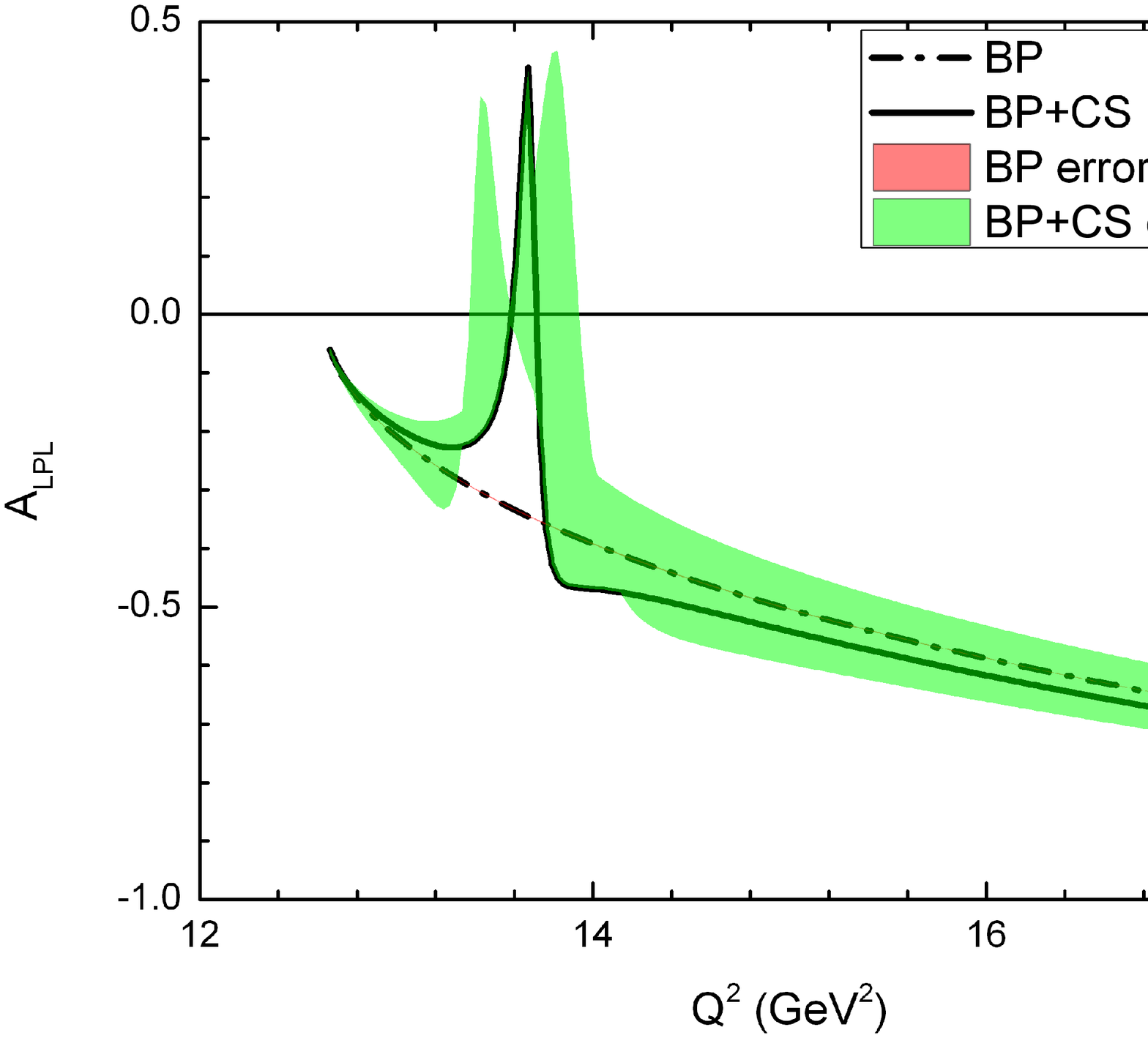}}
\subfigure[Leptonic longitudinal polarization asymmetry]{\includegraphics[width = 0.49\textwidth,height=0.2\textheight]{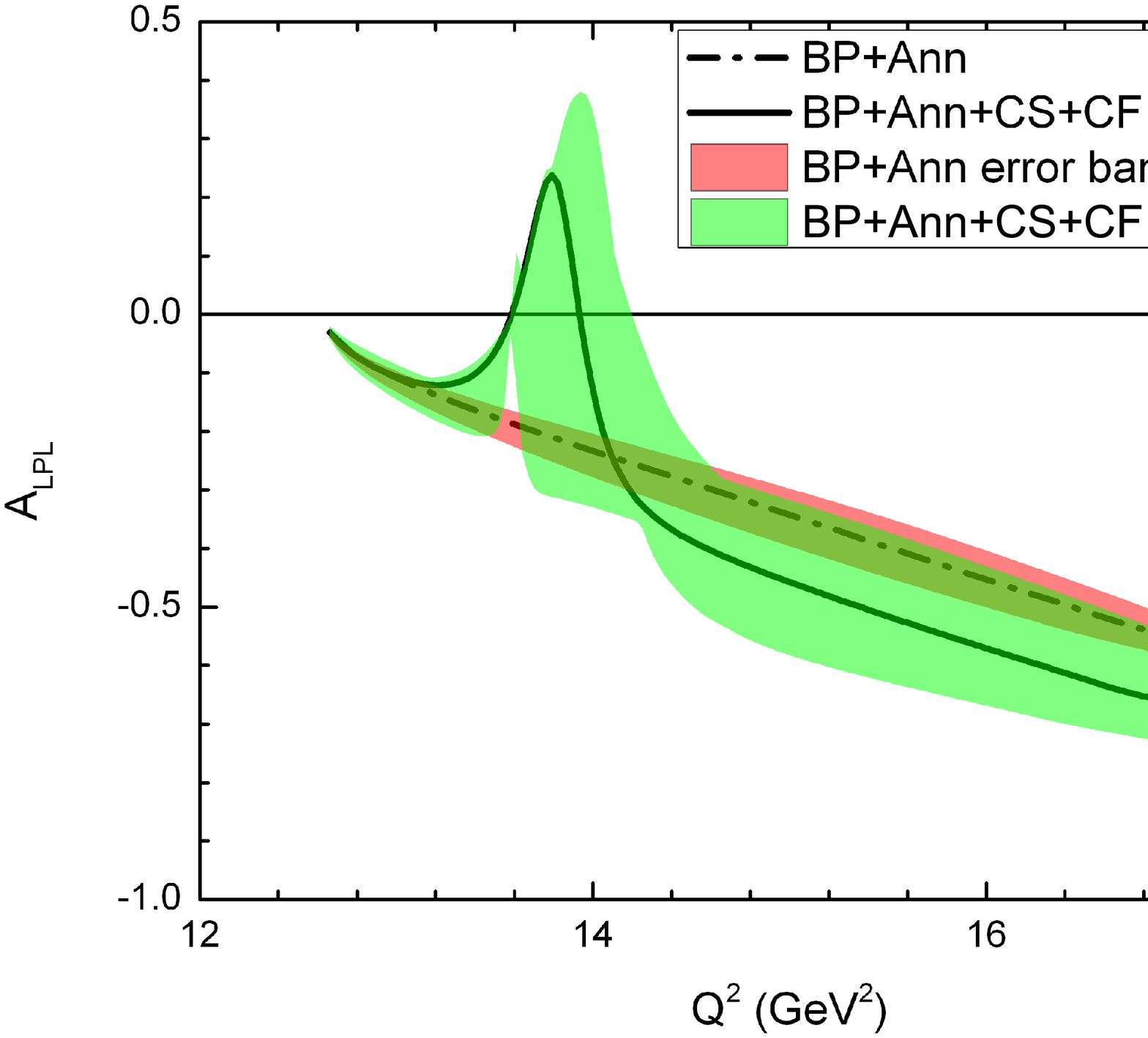}}\\
\caption{Observables of $B_c\rightarrow D_s^* \tau\bar{\tau}$.} 
\end{figure}

\end{document}